\documentclass[%
aps%
,groupedaddress%
,showpacs%
,amsfonts%
,amsmath%
,eqsecnum%
,floatfix%
,twocolumn%
]{%
revtex4%
}

\usepackage[usenames]{color}

\usepackage{epsfig}
\usepackage{graphicx}
\usepackage{mathptmx}      
\usepackage{latexsym}
\usepackage{bm}
\usepackage{verbatim}

\def\slashchar#1{\setbox0=\hbox{$#1$}
   \dimen0=\wd0 \setbox1=\hbox{/} \dimen1=\wd1
   \ifdim\dimen0>\dimen1 \rlap{\hbox to \dimen0{\hfil/\hfil}} #1
   \else  \rlap{\hbox to \dimen1{\hfil$#1$\hfil}} / \fi}

\def\tr{{\rm tr}}
\def\Tr{{\rm Tr}}
\newcommand{\MeV}{\,{\mathrm{MeV}}}
\newcommand{\GeV}{\,{\mathrm{GeV}}}
\newcommand{\SU}{{\mathrm{SU}}}
\newcommand{\U}{{\mathrm{U}}}
\newcommand{\Eq}[1]{Eq.~(\ref{eq:#1})}
\newcommand{\eq}[1]{eq.~(\ref{eq:#1})}

\newcommand{\vx}{{\bm{x}}}

\newcommand{\vp}{{\bm{p}}}

\newcommand{\ignore}[1]{}

\newcommand{\qb}{{\bar{q}}}
\newcommand{\R}{\mathbb{R}}
\newcommand{\Z}{\mathbb{Z}}

\begin{document}

\title{The Polyakov loop in various representations in the confined phase of
  QCD}

\author{E. Meg\'{\i}as}
\email{emegias@ifae.es}

\affiliation{Grup de F\'{\i}sica Te\`orica and IFAE, Departament de F\'{\i}sica, Universitat Aut\`onoma de Barcelona, Bellaterra E-08193 Barcelona, Spain}

\author{E. \surname{Ruiz Arriola}}
\email{earriola@ugr.es}

\author{L.L. Salcedo}
\email{salcedo@ugr.es}

\affiliation{Departamento de F\'{\i}sica At\'omica, Molecular y Nuclear and \\
  Instituto Carlos I de F\'{\i}sica Te\'orica y Computacional, \\ Universidad
  de Granada, E-18071 Granada, Spain.}

\date{\today}

\begin{abstract}
We analyze the expectation value of the Polyakov loop in the fundamental and
higher representations in the confined phase of QCD. We
discuss a hadronic like representation, and find that the Polyakov loop
corresponds to a partition function in the presence of a colored source,
explaining its real and positive character. Saturating the sum rules to
intermediate temperatures requires a large number of multipartonic excited
states. By using constituent or bag models, we find detailed low temperature
scaling rules which depart from the Casimir scaling and could be tested by
lattice calculations.
\end{abstract}

\sf

\pacs{11.10.Wx 11.15.-q 11.10.Jj 12.38.Lg}
\keywords{}

\maketitle

\tableofcontents

\section{Introduction}

The thermodynamics of $\SU(N_c)$ non Abelian gauge theories with or without
matter Dirac fields with $N_f$ flavors has received much attention and
interest due to a possible realization of new phases at sufficiently high
temperatures such as the quark-gluon
plasma~\cite{Shuryak:1978ij,MeyerOrtmanns:1996ea}. This is relevant in the
early stages of the universe or in the laboratory at accelerator facilities
such as SPS, RHIC and LHC~\cite{Harris:1996zx,MeyerOrtmanns:1996ea%
  ,Muller:2006ee,Fukushima:2011jc}. Indeed, the initial and unique pioneering
Quantum Chromodynamics (QCD) predictions of lattice
QCD~\cite{Kogut:1982rt,Polonyi:1984zt,Pisarski:1983ms}, now firmly established
~\cite{Aoki:2006we,Bazavov:2011nk}, of a phase transition provided a strong
motivation to search for the quark-gluon plasma phase in the Laboratory. The
phase transition from a confined and chirally spontaneous broken phase to a
deconfined and chirally symmetric phase is characterized by a steep change of
the chiral condensate $\langle \bar q q \rangle $ and the Polyakov loop
expectation values. Actually, they become true order parameters in the limits
of massless and infinitely heavy quarks respectively. In the real world the
transition temperature is defined as the inflexion point of both
observables. Both the quark condensate and the Polyakov loop require a
(multiplicative) renormalization and hence demand a delicate analysis on the
lattice.

Because of these promising expectations, and also because lattice calculations
become more difficult at finite but small temperatures, the phase transition
picture has naturally and traditionally dominated theoretical approaches and
insightful guesswork in the past. These include effective potential
methods~\cite{Megias:2003ui,Braun:2007bx,Braun:2009gm,Smith:2013msa}, and
quark models~\cite{Meisinger:1995ih,Fukushima:2003fw}. Hence, the focus of
studies and model building was placed on the understanding of the physics
around the phase transition with less attention on the low temperature range
and its detailed features.

On the other hand, a long term scrutiny over the last 30 years has culminated
revealing a by now widely accepted cross-over~\cite{Aoki:2006we} (for a review
see e.g. \cite{Fodor:2012rma}), at about a temperature of $T \approx 200 {\rm
  MeV}$ indicating the co-existence of hadronic and quark-gluon degrees of
freedom. To our knowledge, the physical mechanism how this cross-over starts
to manifest itself remains unclear (see however \cite{Jakovac:2013iua}).

This turn of the subject suggests that we may actually improve our
theoretical understanding by looking into the pre-deconfinement regime
in terms of hadronic degrees of freedom. At very low temperature, the
low lying and well known hadronic states will dominate any physical
observable, as they will effectively behave as elementary and stable
states, so that their multipartonic nature will not show up. This
effective elementarity is buttressed by the quantum virial expansion
among hadrons (including unstable
resonances)~\cite{Dashen:1969ep,Dashen:1974yy} and provides the basis
of the Hadron Resonance Gas (HRG) model. Through the assumption of
completeness of hadronic states the HRG model implements the
quark-hadron duality at finite temperature as a multicomponent gas of
non-interacting massive stable and point-like
particles~\cite{Hagedorn:1984hz}.  Such a simple model has been used
as a reference to compare with lattice calculations of the trace
anomaly and the quark condensate at low temperature, particularly
because of initially unsettled discrepancies, which finally came to an
agreement among themselves and {\em with} the HRG
model~\cite{Karsch:2003zq,Borsanyi:2010cj,Huovinen:2009yb,Borsanyi:2013bia}.
Remarkably, the disagreement still persists beyond the expected range
of validity of the HRG model. Although this gives the HRG model a
distinct arbitrating role, its validity based on microscopic arguments
has only been checked in the strong coupling limit and for heavy
quarks to lowest orders~\cite{Langelage:2010yn} or in chiral quark
models under very specific
assumptions~\cite{RuizArriola:2012wd,Megias:2013aua,Megias:2013xda}.
Recent applications of the HRG model include also the study of QCD
transport coefficients~\cite{NoronhaHostler:2012ug}, QCD in presence
of small magnetic fields~\cite{Endrodi:2013cs,Bali:2013esa}, and
nucleus-nucleus collisions at relativistic energies~\cite{Cleymans:1999st,Begun:2012rf}.

The HRG model requires using specific hadronic states, and those
listed in the PDG booklet~\cite{Beringer:1900zz} provide the standard
ones in computations of the trace anomaly for light flavors.  These
calculations exclude the exotic states on the HRG model side, although
high accuracy would be needed anyhow in order to discriminate if they
are to be {\em seen} distinctly in lattice calculations. One must in
addition make an assessment on the error of the HRG model itself. The
simple half width rule error estimate of Ref.~\cite{Arriola:2012vk}
based on the resonance character of most excited states suggests that
both lattice and the HRG model already agree within their
uncertainties. Thus, we do not view finite temperature calculations of
the trace anomaly as a viable way of unveiling the, so far, scarce
exotic states.

In a recent paper~\cite{Megias:2012kb} we have established a hadronic
representation of the Polyakov loop in the fundamental representation,
a purely gluonic but gauge invariant and hence color singlet operator,
which corresponds to the QCD partition function in the presence of a
color triplet fixed source. For the usual light $u,d,s$ quarks the low
lying part of the spectrum of such a theory can be approximated by
mesons and baryons with just one heavy quark, either $c$ or $b$.  This
partition function character guarantees the positivity and
monotonicity of the Polyakov loop expectation value at low
temperatures, a fact which is not obvious by other means but has
always been observed in lattice calculations~\footnote{Color charge
  conjugation just allows to insure reality of the expectation
  value~\cite{Dumitru:2005ng,Megias:2006df,Megias:2006ke}}. In our
view, these observations make, despite traditional reservations, the
renormalized Polyakov loop an observable in much the same way as the
renormalized chiral condensate.

The situation and prospectives for the Polyakov loop in the fundamental
representation are rather different as compared to the trace anomaly or the
quark condensate for light $u,d,s$ quarks. Firstly, there are much less listed
PDG states with one additional $c$ or $b$ quark. Thus, in order to saturate
the hadronic representation one would need rather small temperatures, which by
the leading exponential Boltzmann suppression would provide too weak a signal
in lattice calculations. However, at current available temperatures two
different lattice groups agree on this
observable~\cite{Bazavov:2011nk,Borsanyi:2010bp} so its analysis may be more
robust. Besides, the renormalized Polyakov loop turns out to be approximately
bound by the number of colors~\footnote{This is a tricky point, since after
  renormalization there is a slight violation of this bound at high
  temperatures~\cite{Gava:1981qd,Burnier:2009bk,Brambilla:2010xn}.}, which
sets a necessary validity range for the HRG model calculation. As we will
show, the presence of exotic states becomes very visible still within the
range where we expect the HRG model to work.

A recent lattice calculation has explicitly implemented the sum rule
for the Polyakov loop we have derived in a previous
work~\cite{Megias:2012kb} using a finite number of excited lattice QCD
states~\cite{Bazavov:2013yv} although large uncertainties are
displayed for $T \ge 140 {\rm MeV}$. The origin of the uncertainties
is intriguing, and no clear conclusions regarding the existence of
exotic states have been reached.

In any case, as the temperature is raised there will be manifest quark
and gluon exchange effects which we will discuss in some detail. Our
analysis will face, once more, the difficulties in making a clear cut
definition of a hadronic state out of multiparton states. We note
incidentally, that {\em all} established states listed in the PDG have
a $\bar q q $ and $qqq$ assignment in the quark model, with no missing
states but further states. This implies that in the confining regime
we expect the hadronic basis of states to be complete.

As mentioned above, most works on finite temperature have been
stimulated by the occurrence of the phase-transition. However, even if
a rapid change of order parameters takes place at some critical
temperature, the proper low temperature behavior is not guaranteed. A
prominent example is provided by chiral quark models at finite
temperature, which have been massively used and reproduced a chiral
phase transition in spite of violating low temperature requirements,
such as the $1/N_c$ suppression of finite temperature corrections as
well as chiral perturbation theory requirements (see e.g. for a
thorough discussion~\cite{Megias:2004hj,Megias:2006bn}). In quark
models the situation was mended by including the Polyakov loop
variable~\cite{Meisinger:1995ih}, which has generated a wealth of
publications~\cite{Fukushima:2003fw,Megias:2004hj,Megias:2006bn%
  ,Ratti:2005jh,Sasaki:2006ww,Ciminale:2007sr,Contrera:2007wu%
  ,Schaefer:2007pw,Ghosh:2007wy,Costa:2008dp,Mao:2009aq,Sakai:2010rp%
  ,Radzhabov:2010dd,Zhang:2010kn}. However, in most implementations of
the Polyakov loop Nambu--Jona-Lasinio (PNJL) model, a plain mean-field
approximation is applied, the goal being to describe the interplay
between the breaking of chiral and center symmetries and features of
the QCD phase diagram. Unfortunately such mean-field approximation
erases detailed information such as the Polyakov loop expectation
values in higher representations. In a recent communication we have
shown that the quantum and local nature of the Polyakov
loop~\cite{Megias:2006df,RuizArriola:2012wd} becomes indispensable in
order to make contact with the HRG model both for the partition
function as well as for the Polyakov loop as deduced in
Ref.~\cite{Megias:2012kb}. Thus, there is at present no model where i)
the HRG model is reproduced at low temperatures and ii) the
confinement-chiral crossover transition observed on the lattice is
reproduced~\cite{RuizArriola:2012wd,Megias:2012hk,Arriola:2013jxa}. In
the present paper we focus on features at low temperatures, leaving
possible extensions for the deconfined phase for future work. While at
this level of description we are only confronted with rather global
aspects of hadrons we already face ambiguities regarding color singlet
clustering inside a global color singlet state.

Ideally the sum rule for the Polyakov loop would be saturated by just
using accepted PDG states with $c,b$ as heavy quarks and $u,d,s$ as
light quarks. However, unlike the $u,d,s$ spectrum there are much less
$c,b$ states. For this reason in Ref.~\cite{Megias:2012kb} we
saturated the sum rule for the Polyakov loop using quark model spectra
for $\bar q Q$ and $qqQ$ color singlet states with one heavy $Q=c,b$
quark and the remaining light quarks $q=u,d,s$. While heavy quarks
exhibit their non-relativistic nature for the low lying states, the
fact that many excited states were needed suggested using relativistic
kinematics as the one of the Relativized Quark Model
(RQM)~\cite{Godfrey:1985xj,Capstick:1986bm}. Unfortunately, going
beyond three-particle systems, i.e. tetraquark, hybrids or pentaquark
excited states within the RQM requires assumptions about the color
structure of interactions. There is a wealth of work on multiquark
systems (see e.g.Ref.~\cite{Richard:2010ab} for a lucid summary and
references therein) but we consider that despite much progress in
recent times in quark models~\cite{Vijande:2007ix}, QCD sum
rules~\cite{Nielsen:2009uh} or lattice QCD, the theory is not on a
satisfactory state as to make unambiguous predictions on what states
should one consider into the partition function at finite
temperatures. Therefore, and following our previous work, we will
analyze independent particle models, such as the MIT bag model and
PNJL models where the problem reduces to evaluating degeneracies of
singlet multiquark and multigluon states, which will be refereed to as
multiparton states for short.

However, even if a sufficient number of PDG states were available there
still remains the problem on {\em what} states should be used. This
brings us to the issue of completeness of the hadronic spectrum. There
are currently no redundancy of states in the PDG as compared to the
quark model assignment. For instance, the concerns on the
proliferation of the new $X,Y,Z$ states poses a serious theoretical
question which have been spelled out~\cite{Carames:2012th} and can be
traced to the identification of states in terms of constituents. One
of the advantages of our approach is that the correct counting of
singlet multiparton states is guaranteed, and the discussion on
under-- or over-- completeness of states is shifted to the concept of
singlet cluster irreducibility and the corresponding effective
elementarity~\cite{rajaraman1979elementarity}.

The generalization of the previous discussion to other representations
besides the fundamental one is straightforward, and in the present
paper we want to analyze the Polyakov loop in the lowest higher
representations. Unfortunately, the extraction of the spectrum from
hadronic systems seems even less obvious so we will provide some
initial estimates by using specific models with quark and gluon
degrees of freedom.

For instance, if the color source is in the adjoint representation we may
imagine that it corresponds to a heavy gluon or two heavy quarks coupled
adjointwise.  Heavy quarks exhibit their non-relativistic nature for the low
lying hadronic states. However, we expect important relativistic corrections
in the higher part of the energy spectrum. Thus, the models that we will be
using embody relativistic kinematics for the light degrees of freedom.

Casimir scaling is one of the features which is suggested by lowest
orders in perturbation theory in QCD and still holds
non-perturbatively on the lattice numerically~\footnote{Violations of
  Casimir scaling in perturbation theory have been recently
  reported~\cite{Anzai:2010td}.}. The most studied example is given by
the string tension.  This scaling has been advocated in
Ref.~\cite{Gupta:2007ax} within the study of renormalized Polyakov
loops in many representations. Casimir scaling has also been observed
on the lattice \cite{Mykkanen:2012ri} in pure $\SU(N_c)$ gauge
theories at several values of $N_c$ {\em above} the phase transition
for the renormalized Polyakov loop. In the present paper we provide
alternative scaling patterns which differ from the Casimir scaling
ones and apply at low temperatures {\em below} the phase transition.

QCD is characterized among other things by quark-gluon confinement of
the physically observable hadronic states, which exhibit a finite
energy gap with the vacuum. For non strange hadrons there are two main
gaps: $m_\pi$ and $m_\rho$. This allows for a clear separation at low
temperatures where thermal effects are just due to a pion gas in a
defined temperature range. Beyond these gaps, hadron states start to
pile up with large multiplicities, eventually suggesting a Hagedorn
spectrum which is not manifested in the thermodynamics of QCD. Thus,
we expect that by looking into violations of quark-hadron duality at
finite temperature, we may learn about the mysterious mechanisms of
deconfinement at the lowest possible temperatures.  As a general rule
we find an expansion in the number of constituents a suitable tool to
discriminate the effective elementarity of hadrons at low
temperatures.

The manuscript is organized as follows. In Sec.~\ref{sec:2} we
introduce the hadron resonance gas model for the Polyakov loop based
on generic QCD arguments. We apply this formalism in
Sec.~\ref{sec:ind-part.model} within an independent multiparticle
picture, and state the basis of the expansion in the number of
constituents. These results will be applied later in concrete models,
in particular the Polyakov constituent quark model in
Sec.~\ref{sec:cqm}, and the bag model in Secs.~\ref{sec:3} and
\ref{sec:4}. We present in Sec.~\ref{sec:cqm.c} some low temperature
scaling relations of the Polyakov loop. Finally we conclude with a
discussion of our results and an outlook towards possible future
directions in Sec.~\ref{sec:conclusions}. Some technical details and
further numerical results for the bag model are collected in the
appendices.

\section{Hadron resonance gas model for the Polyakov loop}
\label{sec:2}

In this section we elaborate on the derivation of the sum rule for
the Polyakov loop in a general irreducible representation of the
color $\SU(N_c)$ gauge group in terms of singlet hadronic states.
Also $N_c$ is kept arbitrary here.

\subsection{The Polyakov loop and the hadron resonance gas}
\label{sec:2.A}

Let $\cal{H}$ denote the Hilbert space of all possible configurations of
(dynamical) quarks, antiquarks and gluons.  As is known in gauge theories, the
functional integration over $A_0$ takes care of projecting onto the {\em
  physical} subspace of states which are color singlet at every point $\vx$,
$\cal{H}_{\rm singlet} \subset \cal{H}$. That is, the QCD partition function
(we do not include a chemical potential in this work) is
\begin{equation}
Z_{\rm QCD} = \Tr_{\cal{H}_{\rm singlet}} ( e^{-\beta H_{\rm QCD}} )
= \Tr_{\cal{H}} (e^{-\beta H_{\rm QCD}} P_{\cal{H}_{\rm singlet}} )
,
\end{equation}
where $H_{\rm QCD}$ denotes the QCD Hamiltonian, $\beta=1/T$ is the inverse
temperature and
\begin{equation}
P_{\cal{H}_{\rm singlet}} = \int \prod_{\vx} U(\Omega(\vx)) d\Omega(\vx)
,
\end{equation}
is the projector onto $\cal{H}_{\rm singlet}$, the space that contains the
physical states made of quarks, antiquarks and gluons. Here $U(\Omega)$ is the
unitary operator representing the element $\Omega$ of $\SU(N_c)$ acting on the
fields at given point, and $d\Omega$ is the $\SU(N_c)$ Haar measure.

Other subspaces of $\cal{H}$ can be explored by introducing static
color charges at different points. Specifically, the subspace which is
in the antifundamental representation at $\vx_0$ and singlet
everywhere else, can be investigated by adding a fundamental source at
$\vx_0$, such as an infinitely heavy quark sitting at that point. The
source polarizes the system since dynamical quarks, antiquarks, and
gluons collaborate to neutralize the source in order to have a color
singlet everywhere. In the confining phase the polarization takes
place through nearby dynamical particles which screen the source at
distance $r\sim T/\sigma$, where $\sigma$ is the string tension.
Other irreducible representations (irreps) of $\SU(N_c)$ can be
considered as well by using different color sources, e.g., by adding
the appropriate combination of heavy quarks and/or antiquarks to form
the given representation. Note that the color sources represent heavy
enough particles so that they can be given a well-defined position,
are at rest and have no other active degree of freedom apart from
color (even the spin and flavor states of the source can be
disregarded due to heavy-quark symmetry).

Mathematically, the projector on a given irrep $\mu$ can be written as
\cite{Tung:1985bk}
\begin{equation}
P_\mu = n_\mu \int \chi_\mu(\Omega^{-1}) \, U(\Omega) \, d\Omega 
\label{eq:2.3}
\end{equation}
where $n_\mu$ denotes the dimension of the representation, $\chi_\mu(\Omega)$
denotes the character of the element $\Omega$ in the irrep $\mu$, i.e., the
trace of $U(\Omega)$ when $U$ falls in that irrep. Therefore, the QCD
partition function when a color charge in the irrep $\mu$ is sitting at
$\vx_0$ would be
\begin{equation}
Z_{{\rm QCD},\mu} = 
\Tr_{\cal{H}} \left[e^{-\beta H_{\rm QCD}}
\int
\chi_\mu(\Omega(\vx_0))  
\prod_{\vx} U(\Omega(\vx)) \, d\Omega(\vx)
\right]
.
\label{eq:2.5}
\end{equation}
Note (i) that our convention is to call $\mu$ to the irrep of the
static source, and so the polarized system of dynamical quarks,
antiquarks and gluons itself is in the conjugate irrep $\bar\mu$ at
$\vx_0$ (and in color singlet at any other point). So we have applied
$P_{\bar\mu}$, and used $\chi_{\bar\mu}(\Omega^{-1})=
\chi_\mu(\Omega)$, with $\Omega=\Omega(\vx_0)$ in \Eq{2.3}. And (ii),
$Z_{{\rm QCD},\mu}$ is actually the partition function divided by
$n_\mu$. Just one of the $n_\mu$ color-degenerated states is
counted. This counting is automatically obtained by the coupling of
the heavy source and the dynamical system to form a color
singlet~\cite{Luscher:2002qv,Jahn:2004qr}. The (infinite) mass of the
source is excluded from the partition function.

In the Euclidean formulation of the gauge theory, the local gauge rotation
$\Omega$ is realized by the Polyakov loop, i.e., the gauge covariant operator
defined as
\begin{equation}
\Omega(\vx) = P\, e^{\, i \int_0^\beta A_0(x) \, dx_0}
,
\end{equation}
where $P$ indicates path ordering and $A_0$ and $x$ are Euclidean. Thus
\Eq{2.5} can be rewritten as the ratio of two partition functions, to match
the usual definition of expectation value of the Polyakov loop in the irrep
$\mu$:
\begin{equation}
 L_{{\rm QCD},\mu}(T)
:=
\langle \chi_\mu(\Omega) \rangle_{\rm QCD}
=
\frac{Z_{{\rm QCD},\mu}}{Z_{\rm QCD}}
.
\label{eq:2.6}
\end{equation}

A different normalization is also found in the literature, namely, the
normalized trace, $\langle \chi_\mu(\Omega)/n_\mu \rangle$. Here we use
directly the trace, since it is more directly related to a true partition
function (see \Eq{2.7} below).

It should be noted that the formal definitions based on projecting
exactly at some $\vx_0$ are intrinsically UV divergent. The subsequent
renormalization leaves a self-energy ambiguity that translates into a
factor $e^{\beta \,C_\mu}$ in $L_{{\rm QCD},\mu}(T)$, $C_\mu$ being an
arbitrary energy scale~\footnote{For uniformity, we always include the
  singlet in the set of irreps, with $L_{{\rm QCD},{\bf 1}}(T)=1$ and
  $C_{\bf 1}=0$. No ambiguity in this case.}. In the static $q\qb$
potential, this corresponds to the ambiguity in fixing the origin of
energies of the potential. If instead one introduces a heavy quark and
subtracts its large mass at the end, this leaves a similar finite
ambiguity.

The Polyakov loop expectation value in any irrep $\mu$ can be computed in the
lattice formulation. In order to estimate this quantity in the confining
phase, we make an assumption paralleling that of the hadron resonance gas
model for the partition function, namely, we neglect non-confining
interactions. This approximation is used as follows. The heavy (as opposed to
dynamical) color source will be screened by forming a heavy hadron with the
dynamical quarks and gluons.  That heavy hadron will stay anchored at $\vx_0$
and of course it will interact with other dynamical hadrons present in the
resonance gas, e.g., through nuclear forces mediated by meson exchange.  We
retain the confining forces that give rise the heavy hadron but neglect the
corrections from non-confining ones. A detailed study of their contributions
is beyond of the scope of this work. When such residual interactions are
neglected, the dynamical hadrons decouple and form the ordinary unpolarized
hadron resonance gas. Therefore their contribution in $Z_{{\rm QCD},\mu}$ is
just to produce a factor equal to $Z_{{\rm QCD}}$ that cancels with the
denominator in \Eq{2.6}.  These considerations lead us to the following
approximate sum rule \cite{Megias:2012kb},
\begin{equation}
L_{{\rm QCD},\mu}(T) \approx 
\sum_i g_i \, e^{-\beta \Delta_i}
.
\label{eq:2.7}
\end{equation}
Here $i$ denotes each of the heavy hadron states at rest obtained by combining
the static source in the irrep $\mu$ with the dynamical quarks, antiquarks and
gluons in the irrep $\bar{\mu}$, $g_i$ is the degeneracy and $\Delta_i$ is the
mass of the heavy hadron excluding the mass of the heavy source. Except for
color, the source is completely inert and does not contribute to the mass nor
to the degeneracy of the state. Obviously, the approximate sum rule in
\Eq{2.7} should break down at temperatures beyond the confining regime, as the
same statement holds for hadron resonance gas model itself. The sum rule is
expected to work better at low temperatures, but still being only
approximated, due to the simplifying assumptions introduced in its derivation.

Of course, these considerations hold not only in QCD but also in gluodynamics
and we treat both cases together. In gluodynamics ``hadron'' would refer to a
glueball, and only triality trivial irreps would produce a non vanishing
Polyakov loop expectation value in the confining phase. While in the particular case
of the Polyakov loop in the fundamental representation the central symmetry $\Z(N_c)$ has played a
key role to characterize the deconfinement transition, this symmetry leaves 
unconstrained some higher dimensional representations (e.g., the adjoint representation).

Before leaving this section, let us note that there is some
ambiguity as to exactly which states should be included in the sum
rule \Eq{2.7}. The problem is as follows, let $V$ be the spatial
neighborhood of the static color source with the dynamical
constituents (quarks, antiquarks, gluons) producing the
screening~\footnote{In what follows, by constituents we mean the
  dynamical quarks, antiquarks or gluons forming the heavy hadron. We
  do not necessarily mean constituent quarks or gluon in the technical
  sense. In the bag model, the constituents are current quarks and
  gluons. Constituent quarks are considered in Sec. \ref{sec:cqm}.}.
For instance, in a bag model such region would be the bag cavity.  The
procedure of just adding constituents in $V$, to form color singlets
with the source, and computing the resulting spectrum, will certainly
produce states which are spurious. Namely, states composed of a
genuine heavy hadron plus one or more ordinary dynamical hadrons. A
prime example is obtained when the irrep is precisely the singlet one,
$\mu= {\bm 1}$. Clearly, in this case all states are spurious and they
would produce a non trivial value for $\langle \chi_{\bm 1}(T)\rangle$
when $1$ is the correct result in this case. In order to remove the
spurious states, one prescription is to include just configurations of
constituents which are {\em color irreducible}, that is, those in
which all constituents are needed to screen the source, without
additional constituents forming a color singlet by themselves. One
estimate of $\langle \chi_\mu(T)\rangle$ is thus
\begin{equation}
L_\mu(T)
:=
\sum_{i, \, {\rm irred}} g_i \, e^{-\beta \Delta_i}
.
\label{eq:2.7a}
\end{equation}
In particular the correct normalization $L_{{\bm 1}}(T)=1$ is ensured.  It is
interesting that tetraquark configurations are always reducible; when non
confining interactions are switched off they split into two mesons.

An alternative prescription will also be considered. We have argued that, in
the absence of purely hadronic interactions, the dynamical hadrons in the
numerator of \Eq{2.6} decouple, producing just the partition function of the
hadron resonance gas (which, in the same approximation, coincides with the
denominator). However, strictly speaking one would obtain a hadron gas with a
hole corresponding to the removal of the spatial region $V$. This implies that
the cancellation with the denominator would not be exact. Instead, we would
obtain the ratio between the contribution to $Z_{{\rm QCD},\mu}$ in $V$ and
the contribution to the hadron gas in $V$. Assuming that $V$ does not strongly
depend on $\mu$, this implies a $\mu$-independent, but $T$-depending,
ambiguity in the normalization which can be settled by the requirement
$L_{{\bm 1}}(T)=1$. Let us denote by $Z_\mu(T)$ the sum over all
configurations in $V$ (color irreducible or not), then the obvious procedure
to achieve the correct normalization for a source $\mu$ is to take the ratio
\begin{equation}
\tilde{L}_\mu(T) := \frac{Z_\mu(T)}{Z_{{\bm 1}}(T)}
,
\qquad
Z_\mu(T)
:=
\sum_{i, \,{\rm all}} g_i \, e^{-\beta \Delta_i}
.
\label{eq:2.8}
\end{equation}
This is another estimate of $\langle \chi_\mu(T)\rangle$.

The two definitions just given, $L_\mu(T)$ and $\tilde{L}_\mu(T)$, are not
identical in concrete models. In $Z_\mu(T)$ there are genuine states plus
dynamical hadrons of the hadron gas that just happen to pass by the region $V$
and are spurious. Intuitively, the division by the singlet sum in \Eq{2.8}
would corresponds to remove precisely those spurious dynamical hadrons. As we
will show below, in actual models the estimates $L_\mu(T)$ and
$\tilde{L}_\mu(T)$ do coincide in an expansion in the number of dynamical
constituents, up to and three constituents, but differ in general when four or
more constituents are involved.  Moreover, in that expansion,
$\tilde{L}_\mu(T)$ may give negative weights to some configurations. This is
by itself not a reason to reject the estimate because those configurations can
never produce a net negative result, for any choice of parameters (since
$Z_\mu(T)$ is always positive) but the picture is certainly cleaner if just
the irreducible configurations are retained, as in $L_\mu(T)$.

\subsection{Some generic considerations on the Polyakov loop and its renormalization}

The set of expectation values of the Polyakov loop in the different
configurations can be collected in a generating function. Any square
integrable class function, i.e., invariant under the similarity
transformations $\Omega\to \Omega_1 \Omega \Omega_1^{-1}$, of the compact
color $\SU(N_c)$ group can be Fourier expanded in terms of irrep
characters. Use of the orthonormality and completeness relations of the
characters
\begin{eqnarray}
\int d\Omega \, \chi^*_\nu(\Omega)  \chi_\mu(\Omega) &=& \delta_{\mu\nu}
,
\nonumber \\ 
 \sum_\mu \chi_\mu(\Omega_1) \chi^*_\mu(\Omega_2) &=& \delta(\Omega_1,\Omega_2)
,
\nonumber \\ 
\chi_\mu(1) &=& n_\mu
,
\label{eq:2.1}
\end{eqnarray}
($\delta(\Omega_1,\Omega_2)$ is the Dirac delta distribution
corresponding to the invariant group measure) allows us to write a
generalized Fourier decomposition of a square integrable function on
the group manifold
\begin{equation}
L_{{\rm QCD}}(\Omega,T) = \sum_\mu \chi^*_\mu(\Omega) L_{{\rm QCD},\mu}(T)
,
\label{eq:2.9}
\end{equation}
so that the corresponding Fourier coefficients are given by 
\begin{equation}
L_{{\rm QCD},\mu}(T) = \int d\Omega \, \chi_{\mu}(\Omega) \, L_{{\rm QCD}}(\Omega,T)
,
\end{equation}
and in particular, for the singlet irrep,
\begin{equation}
1 = \int d\Omega \, L_{{\rm QCD}}(\Omega,T)
.
\end{equation}

These functions are known in limiting cases. At $T=0$, $L_{{\rm QCD},\mu}=
\delta_{\mu,{\bf 1}}$, and $L_{{\rm QCD}}(\Omega) = 1$. On the other hand, at
$T=\infty$, $L_{{\rm QCD},\mu} = n_\mu$, and $L_{{\rm QCD}}(\Omega) =
\delta(\Omega,1)$. (In gluodynamics the system can choose some other central
element of $\SU(N_c)$, with equivalent dynamics.)

The point that we want to make here is that $L_{{\rm QCD}}(\Omega,T)$
is not only normalized but real and non negative and in fact it is (almost) a
proper probability density on the $\SU(N_c)$ group manifold, although
not free from renormalization ambiguities~\footnote{In general,
  probability density functions are no coordinate invariant, due to
  the Jacobian. $L_{{\rm QCD}}(\Omega,T)$ is the probability relative
  to the natural Haar measure. So this function is a scalar defined on
  $\SU(N_c)$. Ambiguities come from renormalization choices.}. Note
that $L_{{\rm QCD}}(\Omega,T)$ is a class function as a consequence of
gauge invariance. For $\SU(3)$ this implies that this and similar
functions depend on two coordinates, rather than on the full eight
coordinates of the group.

Let us consider the bare quantities, as obtained on a lattice, and let us
indicate them by a label $b$:
\begin{equation}
L^b_{{\rm QCD}}(\Omega,g,N_t) = \sum_\mu \chi^*_\mu(\Omega) L^b_{{\rm QCD},\mu}(g,N_t)
.
\label{eq:2.11}
\end{equation}
Here, $N_t$ indicates the lattice size in the time direction (the spatial
directions are assumed to be sufficiently large), and $g$ is the lattice
coupling. The lattice spacing $a$ (times a fixed scale $\Lambda$) is a
(numerically) known and well defined function of $g$. Similarly, we can also
introduce the related quantities $Z^b_{{\rm QCD},\mu}(g,N_t)$ and $Z^b_{{\rm
    QCD}}(\Omega,g,N_t)$. From their very definition, and positivity of the
Hilbert space, it follows that $L^b_{{\rm QCD},\mu}(g,N_t)$ or $Z^b_{{\rm
    QCD},\mu}(g,N_t)$ are real and non negative, since they appear as averages
of projector operators.

On the other hand, using the orthonormality of the characters in the bare
version of \Eq{2.5}, we can write
\begin{eqnarray}
&&
Z^b_{{\rm QCD}}(\Omega,g,N_t)
= 
\label{eq:2.13}\\
&&~~
\Tr_{{\cal{H}}^b} \left[e^{-\beta H^b_{\rm QCD}}
U^b(\Omega)
\int
\prod_{\vx\not=\vx_0} U^b(\Omega(\vx)) \, d\Omega(\vx)
\right]
.
\nonumber
\end{eqnarray}
The interpretation of this formula is very instructive: by adding the
character $\chi_\mu(\Omega)$ and integrating over $\Omega$ one recovers the
(unnormalized) character expectation value. And this is exactly the same
procedure one applies when computing the expectation value in lattice using
Monte Carlo. In other words, the quantity $L^b_{{\rm QCD}}(\Omega,g,N_t)$ is
just the {\em probability distribution of the random variable $\Omega$} which
is sampled in the lattice simulations. Consequently this quantity is
normalized, real and non negative definite.

Our point is that both quantities (schematically)
\begin{equation}
Z_\mu \sim \frac{1}{n_\mu} \langle P_{\bar\mu}  \rangle
,\quad
Z(\Omega) \sim \langle U(\Omega) \rangle
\label{eq:Z(Om)},
\end{equation}
are non negative. A useful point of view here is that of considering functions
defined on $\SU(N_c)$, $\psi(\Omega)$, and the corresponding $\psi_\mu$ as
wavefunctions of a state $|\psi\rangle$ in the Hilbert space
$L^2(\SU(N_c),d\Omega)$, in two conjugate bases \cite{Kogut:1974ag}
\begin{equation}
\psi(\Omega) = \langle \Omega|\psi\rangle ,
\quad
\psi_\mu = \langle \mu|\psi \rangle ,
\quad
\chi_\mu(\Omega) = \langle \mu | \Omega \rangle
.
\end{equation}
In this view, $Z^b_\mu$ and $Z^b(\Omega)$ are wavefunctions of the same state
in the two bases
\begin{equation}
Z^b_\mu = \langle \mu | Z^b \rangle
, \quad
Z^b(\Omega) =  \langle \Omega | Z^b \rangle
.
\end{equation}
In general, positivity of the components of a vector in one basis says nothing
on the positivity of the components in a different basis. The fact that
$Z(\Omega) \sim \langle U(\Omega) \rangle$ is also positive follows from the
fact that, after reintroduction of $A_0$ in the functional integral, the
measure is still positive definite, and so $\Omega$ is a proper random
variable.

This observation opens the possibility to analyses of the Polyakov loop
alternative
to the usual ones. Namely, instead of computing each expectation value of
$\chi_\mu(\Omega)$ separately, one could consider say, a Bayesian approach to
reconstruct the distribution $L^b_{{\rm QCD}}(\Omega,g,N_t)$ directly from the
Monte Carlo sampling data. We do not dwell on this point here. We just mention
that the positivity of $L^b_{{\rm QCD}}(\Omega,g,N_t)$ implies some
theoretical bounds on the expectation values $\langle \chi_\mu(\Omega) \rangle^b
= L^b_{{\rm QCD},\mu}(g,N_t)$: the characters are not positive definite
(except the singlet one), and a very large component of just one of them in the
sum \Eq{2.11} would not be consistent with positivity of the generating
function.

Let us briefly comment on the renormalization problem, from the point of view
of $Z^b(\Omega)$. In the irrep basis $|\mu \rangle$ the renormalization is
just multiplicative \cite{Polyakov:1980ca,Gupta:2007ax}
\begin{equation}
L^b_{{\rm QCD},\mu}(g,N_t)
=
z_\mu(g)^{N_t}L^r_{{\rm QCD},\mu}(T)
.
\label{eq:2.15}
\end{equation}
The non trivial statement here is that, for given $\mu$, $z_\mu$ is just a
function of $g$ (or equivalently, of the lattice spacing $a$) whereas the
renormalized expectation value is only a function of $T=1/(a N_t)$. The
dependence $z_\mu(g)^{N_t}= e^{-\beta \Sigma^b(g)}$ indicates that the bare
static source contains a divergent self-energy that has to be
renormalized. The function $z_\mu(g)$ is unique up to a multiplicative
constant (additive in $\Sigma^b$). We can regard, $L^b_{{\rm QCD},\mu}$ and
$L^r_{{\rm QCD},\mu}$ as the wavefunctions of $|L^b_{{\rm QCD}}\rangle$ and
$|L^r_{{\rm QCD}}\rangle$ in the basis $|\mu\rangle$. On the other hand,
$z(g)$ can be regarded as an operator that is purely multiplicative in that
basis. So in a basis-independent way
\begin{equation}
|L^b_{{\rm QCD}}(g,N_t)\rangle 
=
\hat{z}(g)^{N_t} |L^r_{{\rm QCD}}(T) \rangle
.
\end{equation}
The operator $\hat{z}(g)$ is no longer multiplicative in the basis
$|\Omega\rangle$, instead it defines a real and symmetric
function~\footnote{Note that this is not a generic function of the two
  arguments $\Omega_1$ and $\Omega_2$. Rather it contains the same information
  as the single argument function $z(\Omega,g)= \sum_\mu
  z_\mu(g)\chi_\mu(\Omega)$. For an Abelian group $z(\Omega_1,\Omega_2,g) =
  z(\Omega_1\Omega_2^{-1},g)$.}
\begin{equation}
z(\Omega_1,\Omega_2,g) =  \langle \Omega_1 |\hat{z}(g) |\Omega_2\rangle
= \sum_\mu z_\mu(g) \chi^*_\mu(\Omega_1) \chi_\mu(\Omega_2)
.
\label{eq:2.17}
\end{equation}
The action of $z(\Omega_1,\Omega_2,g)$ is that of a convolution. Each new
temporal layer in the lattice introduces a new convolution that tends to
flatten the distribution of $\Omega$,
\begin{equation}
L^b_{{\rm QCD}}(\Omega,g,N_t)
=
 \int \prod_{n=1}^ {N_t} \left[ d\Omega_n z(\Omega_{n-1},\Omega_n,g) \right]
\,
L^r_{{\rm QCD}}(\Omega_{N_t},T)
,
\end{equation}
(with $\Omega_{n=0}=\Omega$).
The convolution property
\begin{equation}
\int d\Omega_1 \, z(\Omega_1,\Omega_2,g) = 1
,
\end{equation}
also holds due to $z_{\bf 1}(g)=1$.

As is known, all the $L^b_{{\rm QCD},\mu}$ on the lattice (except the singlet)
tend to zero in the continuum limit, in any phase of QCD, while $L^r_{{\rm
    QCD},\mu}$ remains finite. The factor $z_\mu(g)$ in \Eq{2.15} tends to
further quench $L^b_{{\rm QCD},\mu}$ as each new temporal layer is added
(except for the singlet irrep). In terms of the $\Omega$ distribution,
$L^r_{{\rm QCD}}(\Omega)$ remains finite (retains a non trivial structure)
while $L^b_{{\rm QCD}}(\Omega)$ tends to be flatter and flatter in the
continuum limit. The interpretation of $z(\Omega_1,\Omega_2,g)$ as a
convolution suggests that this function should be, not only real, but also non
negative, and this would put some restrictions on the allowed values of the
renormalization factors $z_\mu(g)$.

Throughout this discussion we have used an atypical definition of a factor $z$
that passes from the {\em renormalized} quantity to the {\em bare} one, e.g.,
\Eq{2.15}, while the opposite point of view is normally adopted. As said, the
bare quantities tend to vanish or flatten as the cutoff is removed. In the
renormalized quantities this effect is avoided by means of increasingly large
renormalization factors $z_\mu^{-1}$. The operation $\hat{z}^{-1}$ produces an
{\em anti-convolution}, rather than a true convolution (flattening), and
likely the sum over irreps similar to that in \Eq{2.17}, but with
$z_\mu(g)^{-1}$, would not converge at all. For instance, within the Casimir
scaling conjecture, $z_\mu(g) \sim z_{\bf 3}(g)^{C_2(\mu)/C_2({\bf 3})}$, the
power raises very rapidly with $\mu$ producing an exponential rate while,
presumably, the characters change polynomially. This implies that, although
$L^b_{\rm QCD}(\Omega)$ truly defines a proper probability density on
$\SU(N_c)$ the same needs not be strictly correct for its renormalized version
$L_{\rm QCD}(\Omega)$ as the anti-convolution might require regions where this
function is negative or singular. This is certainly the case in the unconfined
phase where, for any choice of renormalization condition, $\langle \chi_{\bf
  3}(\Omega)\rangle_{\rm QCD}$ gets larger than $3$ at high enough
temperatures \cite{Gava:1981qd}.

Finally, we would like to briefly comment on the intrinsic
renormalization ambiguity due to the source self-energy, which
introduces a factor $e^{\beta C_\mu}$ in the Polyakov loop expectation
value. At high temperatures such term becomes irrelevant as its effect
decreases much rapidly than the perturbative tail \cite{Gava:1981qd},
that is, a privileged value of $C_\mu$ cannot be selected by means of
a perturbative calculation. Nevertheless, the analysis carried out in
\cite{Megias:2005ve} of the lattice data in
\cite{Kaczmarek:2002mc,Kaczmarek:2005ui} for the Polyakov loop in the
fundamental representation in gluodynamics and QCD, shows that the
perturbative prediction is attained at a rate $e^{-b/T^2}$ (for
certain constant $b$) as the temperature increases. Corrections to the
perturbative result of the type $e^{c/T}$, which would be dominant,
are not seen.  (Similar $O(1/T^2)$ corrections have also been noted
\cite{Pisarski:2006yk,Megias:2009mp,Megias:2009ar} in the lattice data
for the pressure and the trace anomaly
\cite{Boyd:1996bx,Cheng:2007jq,Lucini:2012gg}. However, in principle,
these quantities differ from the Polyakov loop in that they are not
subject to renormalization ambiguities.)  Obviously, the lack of
$O(1/T)$ corrections in the Polyakov loop at high temperatures does
not hold for just any choice of source self-energy, and this suggests
the existence of preferred choices of $C_\mu$. In \cite{Megias:2007pq}
it was argued that $O(1/T)$ terms are removed by the Cornell form of
the quark-antiquark potential $a/r+c+\sigma r$ with $c=0$, and also by
dimensional regularization, which will never introduce new
dimensionful constants (like $c$) in addition to
$\Lambda_{\mathrm{QCD}}$ (that fixes the string tension $\sigma$ and
the constant $b$ above \cite{Megias:2007pq}). The odd mass-dimension
of $c$ is also problematic, perturbatively and non perturbatively as
it cannot be related to any available condensate.  In any case this
point certainly deserves further study.

\section{Independent parton models}
\label{sec:ind-part.model}

In this section we model the partition function in the presence of a
color source in any irrep of the $\SU(3)$ group by using an
independent parton picture. We will also show how the same results can
be found by explicitly introducing the Polyakov loop as a dynamical
variable.

\subsection{Hamiltonian}

A direct application of the HRG model for \Eq{2.7} in the case of
higher representations has several limitations: i) there may be not
enough observed states to saturate the sum rule at a given
temperature, ii) most of the states are presumably unstable resonances
or finite size bound states, iii) we may incur into double counting if
the states are not constructed from some underlying quark-gluon
dynamics. So, in view of these general limitations we will resort to
specific models.

We will focus now on some general results which are deduced within an
independent multiparticle picture. This includes the Polyakov
constituent quark model and the bag model with a fixed-radius
cavity. The MIT bag is not strictly an independent particle model as
the energy is not simply additive, but it is very close to it and the
same machinery can be adapted to this case with some suitable
modifications.

As we have shown in the previous section, in order to compute the
Polyakov loop in a given color group representation $\mu$ one may
either use directly \Eq{2.7} or, alternatively, use the Polyakov loop
distribution (as in \Eq{2.13}). The two approaches are summarized in
\Eq{Z(Om)} and are discussed separately below. The first approach is
more suited to calculations with a small number of constituents and
allows to identify color reducible and irreducible contributions
separately. This is not possible in the second approach but it allows
to treat any number of constituents. Both approaches are equivalent
and, as it is shown below, the way $\Omega$ acts on the partons,
namely, through the operator $U(\Omega)$ in \Eq{2.5}, is just standard
minimal coupling. In the independent particle model we consider,
$\Omega$ takes a common value through the confining region (e.g. the
bag cavity in the bag model), so this is similar to a Hartree
approximation. Several models used in the literature and to be
discussed below belong or can be taken into this category.

We will thus assume the Hamiltonian to be given by
\begin{equation}
H= h_q + h_{\bar q}+ h_g 
=
\sum_{\alpha,c} \epsilon_\alpha a_{\alpha,c}^\dagger a_{\alpha,c} 
,
\end{equation}
where $a_{\alpha,c}$ and $a^\dagger_{\alpha,c}$ are partonic annihilation and
creation operators which have bosonic or fermionic character depending on
whether they are gluons or quarks respectively. $c$ indicates the color label
of the single-particle state and $\alpha$ refers to all other labels (type of
particle, flavor, angular momentum, etc). For short we refer to $\alpha$ as
the {\em spin-flavor} state. Thus, the general mass formula of a multiparticle
state is
\begin{eqnarray}
\Delta= \sum_\alpha n_\alpha \epsilon_\alpha
,
\end{eqnarray}
$n_\alpha$ being the occupation number of $\alpha$. This provides a
generalized shell model picture, familiar from mean field studies in nuclear
and atomic physics. The Fock space is built from the
multiparton and a generic state can be expanded as
\begin{eqnarray}
 | \psi\rangle = \sum_{n,m,k}  \psi_{n,m,k} | q^n \bar q^m g^k  \rangle 
\end{eqnarray}
The only requirement is that the total state be a color singlet. The
one body nature of the model simplifies matters tremendously, but
still enables to envisage some interesting features.

\begin{table*}[t]
\caption{Number of color states for different configurations $(q^{n_q},
  \qb^{n_\qb}, g^{n_g})$ and different irreps of $\SU(3)$.  In the first row the
  irreps are denoted by their dimension. In the second row they are denoted by
  their Young tableau.  We include the possible irreps for up to three
  constituents, i.e., $n_q+n_\qb+n_g\le 3$. For pairs of conjugate irreps,
  only one member of the pair has been included. The notation ``$1^n$''
  indicates that the $n$ constituents are in $n$ different spin-flavor states,
  e.g., a configuration $\alpha\beta\gamma$ for $n=3$. Likewise ``12''
  indicates a configuration $\alpha\beta\beta$, etc. For each entry, the total
  of the sum gives the number of $\SU(3)$ multiplets produced of the given
  type (most zero values have been omitted). The first (second) term in the
  sum is the number of irreducible (reducible) color configurations (see
  text).}
\label{tab:1}
\begin{tabular}{ccccccccccccc}
\hline 
& {\bf 1} & {\bf 3} & {\bf 6} & {\bf 8} & {\bf 10} & {\bf 15$^\prime$} 
& {\bf  15} & {\bf 24} & {\bf 27} & {\bf 35} & {\bf 42} & {\bf 64}
\cr
$q$ $\qb$ $g$ & [~] & [1] & [2] & [2,1] & [3] & [4]
& [3,1] & [4,3] & [4,2] & [5,1] & [5,2] & [6,3] 
\cr
\hline 
$(0,0,0)$ & 1+0 & & & & & & & & & & & 
\cr
$(0,1,0)$ & & 1+0 & & & & & & & & & & 
\cr
$(0,0,1)$ & & & & 1+0 & & & & & & & & 
\cr
$(0,1^2,0)$ & & & 1+0 & & & & & & & & & 
\cr
$(0,2,0)$ & & & 0+0 & & & & & & & & & 
\cr
$(1^2,0,0)$ & & 1+0 & & & & & & & & & & 
\cr
$(2,0,0)$ & & 1+0 & & & & & & & & & & 
\cr
$(0,0,1^2)$ & 0+1 & & & 2+0 & 1+0 & & & & 1+0 & & & 
\cr
$(0,0,2)$ & 0+1 & & & 1+0 & 0+0 & & & & 1+0 & & & 
\cr
$(1,1,0)$ & 0+1 & & & 1+0 & & & & & & & & 
\cr
$(0,1,1)$ & & 1+0 & & & & & 1+0 & & & & & 
\cr
$(1,0,1)$ & & & 1+0 & & & & & & & & & 
\cr
$(0,1^3,0)$ & 0+1 & & & 2+0 & 1+0 & & & & & & & 
\cr
$(0,12,0)$ & 0+1 & & & 1+0 & 0+0 & & & & & & & 
\cr
$(0,3,0)$ & 0+1 & & & 0+0 & 0+0 & & & & & & & 
\cr
$(1^3,0,0)$ & 0+1 & & & 2+0 & & & & & & & & 
\cr
$(12,0,0)$ & 0+1 & & & 1+0 & & & & & & & & 
\cr
$(3,0,0)$ & 0+1 & & & 0+0 & & & & & & & & 
\cr
$(0,0,1^3)$ & 0+2 & & & 5+3 & 4+0 & & & & 6+0 & 2+0 & & 1+0 
\cr
$(0,0,12)$ & 0+1 & & & 2+2 & 2+0 & & & & 3+0 & 1+0 & & 1+0 
\cr
$(0,0,3)$ & 0+1 & & & 0+1 & 1+0 & & & & 1+0 & 0+0 & & 1+0 
\cr
$(1,1^2,0)$ & & 0+2 & & & & & 1+0 & & & & & 
\cr
$(1,2,0)$ & & 0+1 & & & & & 0+0 & & & & & 
\cr
$(1^2,1,0)$ & & & 1+0 & & & & & & & & & 
\cr
$(2,1,0)$ & & & 1+0 & & & & & & & & & 
\cr
$(0,1^2,1)$ & & & 2+0 & & & & & & & & & 
\cr
$(0,2,1)$ & & & 1+0 & & & & & & & & & 
\cr
$(1^2,0,1)$ & & 2+0 & & & & & 2+0 & 1+0 & & & & 
\cr
$(2,0,1)$ & & 1+0 & & & & & 1+0 & 0+0 & & & & 
\cr
$(0,1,1^2)$ & & 2+1 & & & & 1+0 & 4+0 & 2+0 & & & 1+0 & 
\cr
$(0,1,2)$ & & 1+1 & & & & 0+0 & 2+0 & 1+0 & & & 1+0 & 
\cr
$(1,0,1^2)$ & & & 3+0 & & & & & & & & & 
\cr
$(1,0,2)$ & & & 1+0 & & & & & & & & & 
\cr
$(1,1,1)$ & 0+1 & & & 2+1 & 1+0 & & & & 1+0 & & & 
\cr
\hline
\end{tabular}\\
\end{table*}

\subsection{Multiparton states}
\label{sec:3.B}

In general each state in the sum \Eq{2.7} is a multiparticle state
composed of quarks, antiquarks and gluons, $q^{n_q}\qb^{n_\qb}g^{n_g}$,
occupying certain energy levels, with color state coupled to the irrep
$\bar{\mu}$. In order to compute the degeneracy $g_i$ it is sufficient to know
how many times the irrep $\bar{\mu}$ appears in the product of $n_q$ ${\bf
  3}$'s, ~$n_\qb$ ${\bf \bar{3}}$'s, and $n_g$ ${\bf 8}$'s. These multiplicities
are given in Table \ref{tab:1} for the irreps that appear up to a total of
three constituents, i.e., $n_q+n_\qb+n_g\le 3$. Irreps that can be obtained
from their conjugate by exchanging quarks with antiquarks, are omitted. In
this counting, each constituent is in a well-defined spin-flavor state. The
case of several spin-flavor states forming a degenerated energy level is
considered below.  The multiplicity can be reduced when two or more
constituents are in the same spin-flavor state, as the color wavefunctions may
no longer be linearly independent. So the values for the various cases $1^2$
($\alpha\beta)$ and $2$ ($\alpha\alpha$), or $1^3$ ($\alpha\beta\gamma$), $12$
($\alpha\beta\beta$), and $3$ ($\alpha\alpha\alpha$), are given. The size of
the table would increase quickly if more constituents were considered. We have
checked the table with the results to be obtained in the Sec. \ref{sec:3.G}
by integration on the Polyakov loop. The correct dimensions are also
verified. For instance, the configuration $(2,0,1)$ has dimension $3\times 8 =
24$ ($3$ color antisymmetric states of two quarks, and $8$ gluon states). The
same number is obtained from one ${\bf 3}$, one ${\bf 15}$, and one ${\bf
  \bar{6}}$. The latter is taken from $(0,2,1)$ which gives one ${\bf 6}$.

\subsection{Reducible and irreducible color configurations}
\label{sec:3.C}

In Table \ref{tab:1} we distinguish between {\em reducible} and {\em
  irreducible} color configurations. To illustrate these concepts, consider a
static source in the irrep $\mu={\bf 8}$, i.e., the adjoint
representation. And let us consider its screening by three gluons in three
different spin-flavor states $\alpha\beta\gamma$. The product ${\bf
  8}\otimes{\bf 8}\otimes{\bf 8}$ produces $8$ adjoint irreps.  The color
wavefunctions of the three gluons and the source can be represented by means
of four linearly independent Hermitian traceless matrices $A$, $B$, $C$ and
$S$. One can construct nine invariant products: six invariants of the form
$\tr(ABCS)$, plus five further permutations of $ABC$, and three invariants of
the form $\tr(AB)\tr(CS)$, plus two further permutations of $ABC$. The fully
symmetric sum of the six permutations of $\tr(ABCS)$ equals the symmetric sum
of the three permutations of $\tr(AB)\tr(CS)$. The other combinations are
linearly independent, producing an eight-dimensional vector space
\cite{Coleman:1988bk}. We say that the three configurations of the type
$\tr(AB)\tr(CS)$ are reducible.  In them the source is screened by just a
subset of the constituents: in $\tr(AB)\tr(CS)$, $C$ forms a singlet with $S$
while $AB$ form another singlet by themselves. In general, we say that a color
configuration is {\em reducible} when it contains a {\em non-empty subset of
  constituents forming a singlet by themselves}. In the three gluon example,
the reducible configurations span a well-defined three-dimensional
subspace. This is indicated in Table \ref{tab:1} by the notation ``$5+3$'',
corresponding to $5$ irreducible color configurations and $3$ reducible
ones. Obviously, up to three constituents, only the irreps ${\bf 1}$, ${\bf
  3}$, ${\bf \bar{3}}$ and ${\bf 8}$ can have reducible color
configurations. By definition, a color singlet source is always in a reducible
color configuration (except in the trivial case of no screening
particles). The reducible configurations can be further classified, according
to whether the proper subset of constituents forming a color singlet are
themselves reducible or not, but such analysis will not be pursued here.
The explicit construction of irreducible configurations is discussed in Appendix
\ref{app:A}.

In the previous example of three gluons screening an adjoint source, a
reducible configuration indicates that just one of the gluons is
confined to the source, whereas the other two are actually forming a
glueball, that is, a color singlet state that is not forced to remain
near the source by confining interactions. The glueball can be moved
far apart from the source using only a bounded amount of energy,
rather than an energy increasing linearly with the separation, as in
the case of confining forces. In view of our previous argument of
keeping just confining forces, to estimate the Polyakov loop
expectation value, it seems more natural to include only irreducible
color configurations in the hadronic spectrum of the
system-plus-source. The concept of reducible color configuration
subspace, and so the dimension of the irreducible subspace, seems to
be a well-defined mathematical object, nevertheless, a possible
caveat should be noted here. The QCD Hamiltonian conserves color, but
nothing seems to prevent this Hamiltonian from coupling irreducible to
reducible configurations. This would imply that irreducibility is not
preserved by the QCD dynamics; a confined configuration could tunnel
to a non-confining one, and the former would be unstable against
decay, depending on the available phase space. For few constituents it
is often the case that all color configurations are irreducible,
depending on the irrep considered, so that the previous objection
would not apply. However, the QCD Hamiltonian could connect these
configurations to virtual ones with more constituents (adding sea
quarks, antiquarks and gluons) making them unstable. In any case, for
each given irrep, a certain absolute minimum of constituents is always
required to screen the source. What is not clear is whether the number
of these states is ever increasing or most of them are really
unstable. In what follows, we just assume that the irreducible
configurations are the relevant set to be included in the spectrum in
\Eq{2.7}.

\begin{table}[t]
\caption{As in Table~\ref{tab:1} for a source in the fundamental
  representation and four constituents (pentaquark).}
\label{tab:2}
\begin{tabular}{ccccccccccccc}
\hline
$q^3 \qb$ &
$(1^3,1,0)$ & $(12,1,0)$ & $(3,1,0)$ 
\cr
& $0+3$ & $0+2$ & $0+1$
\cr
\hline
$ \qb^4 $ &
$ (0,1^4,0) $ & $ (0,1^22,0) $ & $ (0,2^2,0) $ & $ ( 0, 13, 0) $ & $ (
0,4,0)$
\cr
& $0+3$ & $0+2$ & $0+1$ & $0+1$ & $0+0$
\cr
\hline
$ q \qb^2 g $ & 
$ (1,1^2,1) $ & $ (1,2,1) $ 
\cr
& $0+4$ & $0+2$
\cr
\hline
$ q^2 g^2 $ & 
$ (1^2,0,1^2) $ & $ (2,0,1^2) $ & $ (1^2,0,2) $ & $ (2,0,2) $  
\cr
& $5+1$ & $2+1$ & $2+1$ & $1+1$
\cr
\hline
$ \qb g^3 $ & 
$ (0,1,1^3) $ & $ (0,1,12) $ & $ (0,1,3) $
\cr
& $5+5$ & $2+3$& $0+2$
\cr
\hline
\end{tabular}\\
\end{table}

There are no irreducible configurations of the type $q\qb^2$ (i.e.,
$(1,1^2,0)$ or $(1,2,0)$) to screen a static source in the fundamental
representation. This suggests that tetraquark configurations $Q q\qb^2$ can be
separated into two mesons and would not be genuine hadronic states to be
included in the computation. On the other hand, configurations $\qb g$
(i.e. $(0,1,1)$), not included in a calculation with just quarks and
antiquarks but no gluons, gives such genuine contribution in a similar range
of energies as the tetraquark. For completeness, we have also looked at
pentaquark configurations for the fundamental representation, $Q c^4$ ($c$
being any constituent) for the five allowed configurations, $q^3\qb$, $\qb^4$,
$q \qb^2 g$, $q^2 g^3$ and $\qb g^3$. The results are presented in
Table~\ref{tab:2}. The pure pentaquark states (no gluons) are reducible, and
irreducible color configurations need the presence of at least two gluons. So
these contributions will be suppressed at low temperatures, but will become
rapidly relevant as the temperature is raised due to their large degeneracy.

\subsection{Degeneracies of the simplest configurations}
\label{sec:3.D}

Table~\ref{tab:1} gives the multiplicity of each color irrep assuming that the
levels are not degenerated.  In general each level is degenerated and of
course, it is more efficient to take this into account in carrying out the sum
over states in \Eq{2.7}. Our strategy will be to make an ordered list of the
single particle levels, with its degeneracy, till some cutoff value, and sum
over the various configurations of constituents. We include up to three
constituents because the number of states increases very rapidly as more
particles are added. The treatment including any number of constituents is
attempted in the next section.

Let $(a_1\le \cdots \le a_n)$ be the ordered set of levels of the $n$
constituents present in a given state. As said we take $n\le 3$. Let
$(\gamma_1,\ldots,\gamma_n)$ be their degeneracies (color excluded), and
$(c_1,\ldots,c_n)$ the particle types (quark, antiquark, or gluon). Note
that the information on degeneracies and particle types is already contained
in the label $a_i$.

Particles in different levels will certainly be in different
spin-flavor states, but particles in the same level may or may not be
in the same spin-flavor state, if the degeneracy of the level is
larger than one. So, in order to use Table~\ref{tab:1} it is necessary
to know the number of fillings of configurations of the type
$(1^{n_1},2^{n_2},\ldots)$ when each of the $\sum_k k n_k$ particles
can take any of $\gamma$ degenerated states~\footnote{E.g., a
  configuration $(1^3,2^2)$ is one of the type
  $\alpha\beta\gamma\delta\delta\epsilon\epsilon$.}.  This number is
given by the combinatorial symbol
\begin{eqnarray}
\left(\begin{matrix} \gamma \cr n_1,n_2,\ldots \end{matrix}\right)
&:=&
\left(\begin{matrix} \gamma \cr n_1 \end{matrix}\right)
\left(\begin{matrix} \gamma -n_1\cr n_2 \end{matrix}\right)
\left(\begin{matrix} \gamma -n_1-n_2\cr n_3 \end{matrix}\right)
\cdots \nonumber \\
&=&
\frac{\gamma!}{(\gamma-\sum_k n_k)!\prod_k n_k!}
.
\end{eqnarray}
The symbols are completely symmetric with respect to the $n_k$, and vanishing
values of $n_k$ can be dropped. Nevertheless, for the sake of clarity, we take
the convention of making explicit all $n_k$ in their natural order (namely,
increasing $k$), till the last non null one.

For $n=1$ the degeneracy of the states is
\begin{equation}
g_i(a_1) = \left(\begin{matrix} \gamma_1 \cr 1 \end{matrix}\right) 
d^\mu_{(1)}(c_1)
.
\end{equation}
Here $\mu$ is the irrep of the source and $d^\mu_{(1)}$ is the appropriate
entry in Table~\ref{tab:1}, for the given irrep ($\mu$) and type of particle
($c_1$). Of course, in the present case, the combinatorial number is just
$\gamma_1$. For $n=2$, there are two cases,
\begin{eqnarray}
g_i(a_1 < a_2) &=&
\left(\begin{matrix} \gamma_1 \cr 1 \end{matrix}\right)
\left(\begin{matrix} \gamma_2 \cr 1 \end{matrix}\right)
d^\mu_{(1^2)}(c_1,c_2)
\\
g_i(a_1 = a_2) &=&
\left(\begin{matrix} \gamma_1 \cr 2 \end{matrix}\right)
d^\mu_{(1^2)}(c_1,c_2)
+
\left(\begin{matrix} \gamma_1 \cr 0,1 \end{matrix}\right)
d^\mu_{(2)}(c_1,c_2)
.
\nonumber
\end{eqnarray}
Finally, for $n=3$
\begin{eqnarray}
g_i(a_1 < a_2 < a_3) &=&
\left(\begin{matrix} \gamma_1 \cr 1 \end{matrix}\right)
\left(\begin{matrix} \gamma_2 \cr 1 \end{matrix}\right)
\left(\begin{matrix} \gamma_3 \cr 1 \end{matrix}\right)
d^\mu_{(1^3)}(c_1,c_2,c_3)
\nonumber \\
g_i(a_1 = a_2 < a_3) &=&
\left(\begin{matrix} \gamma_1 \cr 2 \end{matrix}\right)
\left(\begin{matrix} \gamma_3 \cr 1 \end{matrix}\right)
d^\mu_{(1^3)}(c_1,c_2,c_3) \nonumber \\ 
&&+
\left(\begin{matrix} \gamma_1 \cr 0,1 \end{matrix}\right)
\left(\begin{matrix} \gamma_3 \cr 1 \end{matrix}\right)
d^\mu_{(12)}(c_1,c_2,c_3)
\nonumber \\
g_i(a_1 < a_2 = a_3) &=&
\left(\begin{matrix} \gamma_1 \cr 1 \end{matrix}\right)
\left(\begin{matrix} \gamma_2 \cr 2 \end{matrix}\right)
d^\mu_{(1^3)}(c_1,c_2,c_3) \nonumber \\ 
&&+
\left(\begin{matrix} \gamma_1 \cr 1 \end{matrix}\right)
\left(\begin{matrix} \gamma_2 \cr 0,1 \end{matrix}\right)
d^\mu_{(12)}(c_1,c_2,c_3)
\nonumber \\
g_i(a_1 = a_2 = a_3) &=&
\left(\begin{matrix} \gamma_1 \cr 3 \end{matrix}\right)
d^\mu_{(1^3)}(c_1,c_2,c_3) \nonumber \\ 
&&+
\left(\begin{matrix} \gamma_1 \cr 1,1 \end{matrix}\right)
d^\mu_{(12)}(c_1,c_2,c_3) \nonumber \\ 
&&+
\left(\begin{matrix} \gamma_1 \cr 0,0,1 \end{matrix}\right)
d^\mu_{(3)}(c_1,c_2,c_3)
.
\end{eqnarray}

Obviously, these formulas apply equally well whether one chooses to
use the total number of color configurations or just the number of
irreducible ones in the coefficients $d^\mu$. For future reference, we
quote in Table~\ref{tab:3} the degeneracies of the states obtained
within both schemes when just one level is assumed for each type of
constituent, with degeneracies $\gamma_q$, $\gamma_\qb$ and
$\gamma_g$, respectively. The table shows results for the irreps ${\bf
  1}$ and ${\bf 8}$ up to three constituents, and four constituents
for ${\bf 3}$. Some bag model information to be used
in Sec.~\ref{sec:3} is also displayed.

In Table \ref{tab:3} we also express the same degeneracies (polynomials in
$\gamma_q$, $\gamma_\qb$ and $\gamma_g$), in terms of Young tableaux. This
exposes the symmetry properties of the corresponding wavefunctions. For
instance, the pentaquark configuration $(\bar{q}^4)_{\bf\bar{3}}$, requires a
color configuration with symmetry $[2,1^2]$, fermion statistics then requires
a dual spin-flavor configuration $[3,1]$. This tableau is filled regularly
with labels from $1$ to $\gamma_\qb$. We denote the number of such fillings
(namely $\frac{1}{8}\gamma_\qb (\gamma_\qb^2-1) (\gamma_\qb+2)$) as
$[3,1]_\qb$. In general, for a given tableau $Y$, we use $Y_q$, $Y_\qb$, and
$Y_g$ to denote the dimension of the irrep $Y$ of $\SU(n)$ with $n=\gamma_q$,
$\gamma_\qb$, and $\gamma_g$, respectively.

\begin{table*}[t]
\caption{Degeneracy of multiparton states for the configurations and irreps
  shown, assuming a single level for each type of constituent with
  degeneracies $\gamma_q$, $\gamma_\qb$, and $\gamma_g$. `Total'' uses all
  color configurations. The second ``Total'' column produces the same
  expressions as the first one in terms of Young tableaux dimensions (see
  text).  ``Irred'', uses just the irreducible color configurations. The last
  column displays the bag model minimal sum of $\omega$'s required for the
  configuration (see Sec.~\ref{sec:3}).}
\label{tab:3}
\begin{tabular}{cccccccc}
Irrep & Configuration & \multicolumn{2}{c}{Total}  & \multicolumn{2}{c}{Irred} & min $\sum_\alpha \omega_\alpha$
\cr
\hline 
${\bf 1}$ & $ q \qb$ 
& $\gamma_q \gamma_\qb$ & $[1]_q [1]_\qb$ 
& 0 & 0 
& 4.086
\cr
 & $g^2$ 
& $\frac{1}{2} \gamma_g (\gamma_g + 1)$ & $[2]_g$ 
& 0 & 0
& 5.488
\cr
& $\qb^3$ & 
$ \frac{1}{6} \gamma_\qb (\gamma_\qb + 1) (\gamma_\qb + 2)$ & $[3]_\qb$
& 0 & 0
& 6.129
\cr
& $q^3$ 
& $\frac{1}{6} \gamma_q (\gamma_q + 1) (\gamma_q + 2)$ & $[3]_q$
& 0 & 0 
& 6.129
\cr
& $ q \qb g $ 
& $ \gamma_q \gamma_\qb \gamma_g $ & $[1]_q [1]_\qb [1]_g $
&  0 & 0
& 6.830
\cr
& $g^3$ &  $\frac{1}{3} \gamma_g (\gamma_g^2 + 2)$ & $([3]+[1^3])_g$
& 0 & 0
& 8.232
\cr
\hline 
${\bf 3}$ &
$\qb$ & $ \gamma_\qb$ & $[1]_\qb$
& $\gamma_\qb$ & $[1]_\qb$
& 2.043
\cr
& $q^2$ 
& $\frac{1}{2} \gamma_q (\gamma_q + 1)$ & $[2]_q$
& $\frac{1}{2}  \gamma_q (\gamma_q + 1)$ & $[2]_q$
& 4.086
\cr
&
$\qb g $ 
& $ \gamma_\qb \gamma_g $ & $ [1]_\qb [1]_g $
& $ \gamma_\qb \gamma_g $ & $ [1]_\qb [1]_g $
& 4.787
\cr
&$ q \qb^2$ 
& $ \gamma_q \gamma_\qb^2$ & $ [1]_q ([2]+[1^2])_\qb $
& $ 0 $ & 0
& 6.129
\cr
& $ q^2 g $ 
& $ \gamma_q^2 \gamma_g $ & $([2]+[1^2])_q [1]_g$ 
& $ \gamma_q^2 \gamma_g $ & $([2]+[1^2])_q [1]_g$ 
& 6.830
\cr
& $ \qb g^2 $ 
& $\frac{1}{2} \gamma_\qb \gamma_g (3 \gamma_g + 1) $ & $[1]_\qb (2 \, [2]+[1^2])_g$
& $ \gamma_\qb \gamma_g^2 $ & $[1]_\qb ([2] + [1^2])_g $
& 7.531
\cr
& $q^3\qb$ 
& $\frac{1}{2}\gamma_q^2(\gamma_q+1)\gamma_\qb $ & $ ([3]+[2,1])_q [1]_\qb $
& $0$ & $0$
& 8.172
\cr
& $\qb^4$ 
& $\frac{1}{8}\gamma_\qb (\gamma_\qb^2-1) (\gamma_\qb+2)$ & $ [3,1]_\qb $
& $0$ &  $0$
& 8.172
\cr
& $q \qb^2 g$ 
& $ 2 \gamma_q \gamma_\qb^2 \gamma_g $ & $2 [1]_q ([2]+[1^2])_\qb [1]_g$
& $0$ & $0$
& 8.873
\cr
& $q^2 g^2$ 
& $\frac{1}{2} \gamma_q \gamma_g (3 \gamma_q \gamma_g + 1)$ 
& $ (2\,[2]_q + [1^2]_q) [2]_g 
$
& $\frac{1}{4} \gamma_q \gamma_g (5 \gamma_q \gamma_g - \gamma_q - \gamma_g + 1)$ 
& $ ([2]_q + [1^2]_q ) [2]_g 
$
& 9.574
\cr
& & 
& $ 
+ ([2]_q + 2\,[1^2]_q) [1^2]_g $
&
& $ 
+ ( [2]_q + 2\,[1^2]_q ) [1^2]_g $
\cr
& $\qb g^3$ 
& $\frac{1}{3} \gamma_\qb \gamma_g ( 5 \gamma_g^2 + 1 ) $ 
& $ [1]_\qb (2\,[3] + 3\, [2,1] + 2\, [1^3])_g$
&
$\frac{1}{6} \gamma_\qb \gamma_g ( 5 \gamma_g + 2 ) ( \gamma_g - 1 ) $ 
& $ [1]_\qb ( 2\, [2,1] + [1^3])_g$
& 10.275
\cr
\hline 
${\bf 8}$ &
$g$ 
& $\gamma_g$ & $ [1]_g $
& $\gamma_g$ & $ [1]_g $
& 2.744
\cr
&
$q \qb $ 
&  $\gamma_q \gamma_\qb $ & $ [1]_q [1]_\qb $
& $\gamma_q \gamma_\qb$ & $ [1]_q [1]_\qb $
& 4.086
\cr
& $g^2$ 
& $\gamma_g^2$ & $ ([2]+[1^2])_g$
& $\gamma_g^2$ & $ ([2]+[1^2])_g$
& 5.488 
\cr
& $ \qb^3 $ 
& $ \frac{1}{3} \gamma_\qb ( \gamma_\qb^2 - 1) $ & $[2,1]_\qb $
& $ \frac{1}{3} \gamma_\qb ( \gamma_\qb^2 - 1) $ & $[2,1]_\qb $
& 6.129
\cr
& $q^3$ 
& $ \frac{1}{3} \gamma_q ( \gamma_q^2 - 1 ) $ & $[2,1]_q $
& $ \frac{1}{3} \gamma_q ( \gamma_q^2 - 1 ) $ & $[2,1]_q $
& 6.129
\cr
& $ q \qb g $ 
& $ 3 \gamma_q \gamma_\qb \gamma_g $ & $ 3\,[1]_q [1]_\qb [1]_g $
& $ 2 \gamma_q \gamma_\qb \gamma_g $ & $ 2\,[1]_q [1]_\qb [1]_g $
& 6.830
\cr
& $g^3$ 
& $\frac{1}{3} \gamma_g ( 4 \gamma_g^2 - 1 ) $ & $([3] + 3\,[2,1] + [1^3])_g$
& $\frac{1}{6}  \gamma_g ( 5 \gamma_g + 2 )  ( \gamma_g - 1 ) $
& $(2\,[2,1] + [1^3])_g$
& 8.232
\cr
\hline 
\end{tabular}\\
\end{table*}

\subsection{Minimal coupling of the Polyakov loop}
\label{sec:3.E}

Here we turn to the approach based on treating the Polyakov loop as a random
variable, and its probability distribution function.

Let $Z(T)$ denote the total partition function, understood as the sum
over all configurations (irreducible or not) and in any color
irrep. Likewise, let $Z_\mu(T)$ be the sum over all states in the irrep
$\bar{\mu}$, counting each irrep once. Therefore,
\begin{equation}
Z(T) = \Tr( e^{-\beta H} ) = \sum_\mu n_\mu Z_\mu
,
\qquad
Z_\mu(T) = \frac{1}{n_\mu} \Tr( e^{-\beta H} P_{\bar{\mu}} )
.
\label{eq:4.1}
\end{equation}
Here, $H$ is the total Hamiltonian, $\Tr$ refers to the full Hilbert
space spanned by multiparticle states of quarks, antiquarks, and
gluons. $P_{\bar{\mu}}$ projects to multiparticle states in the irrep
$\bar{\mu}$.  Being partition functions in various spaces, the
functions $Z(T)$, and $Z_\mu(T)$ are all real and positive and
coincide for $\mu$ and $\bar{\mu}$ due to $C$-symmetry.

In order to obtain $Z_\mu$, let us introduce its conjugate function
\begin{equation}
Z(\Omega,T) = \Tr ( e^{-\beta H} U(\Omega) )
,
\end{equation}
where $U(\Omega)$ represents the $\SU(3)$ rotation $\Omega$ on the multiparton
states.  ($U(\Omega)$ and $H$ commute due to color symmetry.)  Use of the
orthonormality relations of the characters in \Eq{2.1} produces the relations
\begin{eqnarray}
Z_\mu(T) &=& \int d\Omega \, \chi_\mu(\Omega) Z(\Omega,T)
,
\nonumber \\ 
Z(\Omega,T) &=&
\sum_\mu \chi_\mu^*(\Omega) Z_\mu(T)
,
\nonumber \\ 
Z(T) &=& Z(1,T) .
\label{eq:4.4}
\end{eqnarray}
Once again, the function $Z(\Omega,T)$ represents the (unnormalized)
probability density of the random variable $\Omega$ on the manifold of the
group $\SU(3)$. Therefore, this function is not only real (this would follow
from $C$-invariance) but also non negative definite. In addition
$Z(\Omega,0)=1$ since at zero temperature only the vacuum state remains.

To avoid any confusion, let us remark that $Z_\mu(T)$, and $Z(\Omega,T)$ play
similar roles as $Z_{{\rm QCD},\mu}(T)$ and $Z_{\rm QCD}(\Omega,T)$,
respectively, however, $Z(T)$ includes all irreps and so it does not match
$Z_{\rm QCD}(T)$. The latter function contains just the singlet states and so
it matches $Z_{\bf 1}(T)$ of the independent particle model discussed here.

As is well-known, for a purely one-body Hamiltonian $H$, the sum
over multiparticle states reduces to single particle
sums~\cite{huang1987statistical}:
\begin{equation}
\log \Tr_{\rm Fock} e^{-\beta H} = -\zeta \, \Tr_{\mbox{\scriptsize \rm o-p}} 
\log(1 - \zeta e^{-\beta H})
.
\end{equation}
where $\Tr_{\mbox{\scriptsize \rm o-p}} $ is the trace over the one-particle
subspace and $\zeta$ indicates the statistics of the particle, $\zeta= \pm 1$
for bosons and fermions respectively. The coupling to the Polyakov loop 
is done by using 
\begin{equation}
\log U(\Omega) = -\beta A_0 Q
,
\end{equation}
$\beta A_0$ are the Lie group parameters (actually, in the present
context $\Omega$ is temperature independent, and so $A_0$ temperature
dependent) and the charge operator $Q$ represents the color group
generators. $Q$ is of one-body type. Generally this corresponds to
formally consider a minimal coupling, $ h \to h + A_0 $.  Thus, we
have more generally,
\begin{equation}
\log \Tr_{\rm Fock} (e^{-\beta H} U(\Omega)) = 
-\zeta \, \Tr_{\mbox{\scriptsize \rm o-p}} 
\log(1 - \zeta e^{-\beta H} U(\Omega))
.
\end{equation}

In our case, there are three types of particles, $q$, $\qb$ and $g$, each one
giving a factor in $Z$:
\begin{equation}
Z(\Omega) = Z_q Z_\qb Z_g
.
\label{eq:4.13}
\end{equation}
These partition functions are given by
\begin{eqnarray}
\log Z_q(\Omega,T)
&=&
\sum_\alpha\!{}^q  \, \gamma_\alpha \,\chi_{\bf 3} \!\left(
\log (1 + \Omega \, e^{-\beta\epsilon_\alpha} )
\right)
,
\nonumber \\
\log Z_\qb(\Omega,T)
&=&
\sum_\alpha\!{}^\qb  \, \gamma_\alpha \,\chi_{\bf \bar{3}} \!\left(
\log (1 + \Omega \, e^{-\beta \epsilon_\alpha } )
\right)
,
\nonumber \\
\log Z_g(\Omega,T)
&=&
- \sum_\alpha\!{}^g \, \gamma_\alpha \,\chi_{\bf 8} \!\left(
\log (1 - \Omega \,e^{-\beta\epsilon_\alpha} )
\right)
.
\label{eq:4.14}
\end{eqnarray}
In each case the sum over $\alpha$ runs on the corresponding set of
single-particle spin-flavor levels (rather than states) with degeneracy
$\gamma_\alpha$. The characters are taken on the group algebra, that is,
$\chi_\mu(\sum_n c_n \Omega^n)=\sum_n c_n\chi_\mu(\Omega^n)$. Alternatively
one can work with traces and matrices. For quarks the character is just the
trace in the fundamental representation and $\Omega$ can be identified with
the unitary $3\times 3$ matrix itself. For antiquarks the same trace with
$\Omega^\dagger$ appears. For gluons the trace is taken in the adjoint
representation and $\Omega$ is represented by an $8 \times 8$ unitary
matrix $\Omega_A$. $C$-invariance implies
\begin{equation}
Z_\qb = (Z_q)^*
,\qquad
Z_g = (Z_g)^*
.
\end{equation}

In order to proceed, it is computationally more convenient to reduce the
number of variables. To this end, we introduce the auxiliary
$\Omega$-independent functions
\begin{eqnarray}
z_q(T) 
&=& \tr \, e^{-\beta h_q } = 
\sum_\alpha\!{}^q \, \gamma_\alpha \, e^{-\beta \epsilon_\alpha }
,
\nonumber \\
z_g(T) 
&=& \tr \, e^{-\beta h_g } = 
\sum_\alpha\!{}^g \, \gamma_\alpha \, e^{-\beta \epsilon_\alpha }
. \label{eq:zqzg}
\end{eqnarray}
These are essentially single-particle partition functions which need to be
computed only once for each value of the temperatures. In terms of these,
\begin{eqnarray}
\log Z_q(\Omega,T)
&=&
\sum_{n=1}^\infty \frac{(-1)^{n+1}}{n} \,\chi_{\bf 3}(\Omega^n)
\, z_q(T/n)
,
\nonumber \\
\log Z_g(\Omega,T)
&=&
\sum_{n=1}^\infty \frac{1}{n} \,\chi_{\bf 8}(\Omega^n) \, z_g(T/n)
.
\label{eq:4.17}
\end{eqnarray}
The antifundamental and adjoint characters can be reduced to the fundamental
one by using
\begin{equation}
\chi_{\bf \bar{3}} (\Omega) = \chi_{\bf 3}^*( \Omega)
,
\quad
\chi_{\bf 8} (\Omega) = \chi_{\bf 3}( \Omega) \chi_{\bf 3}^*( \Omega) - 1 
.
\label{eq:4.18}
\end{equation}

All the functions of $\Omega$ that we are using are class functions, and this
implies that they depend only on the eigenvalues of $\Omega$ in the
fundamental representation, which we denote $\omega_i$, $i=1,2,3$. These
complex eigenvalues fulfill the constraints $|\omega_i|=1$ and
$\omega_1\omega_2\omega_3=1$. The characters can be written as
\begin{equation}
\chi_{\bf 3}(\Omega) = \sum_{i=1}^3\omega_i
, \qquad
\chi_{\bf 8}(\Omega) = \sum_{i,j=1}^3\omega_i \omega_j^* - 1
.
\end{equation}
So we only need to compute once the following functions  for each value of
$\omega\in\U(1)$, and of the temperatures:
\begin{eqnarray}
\log \hat{Z}_q(\omega,T)
&=&
\sum_{n=1}^\infty \frac{(-1)^{n+1}}{n} \,\omega^n
\, z_q(T/n)
,
\nonumber \\
\log \hat{Z}_g(\omega,T)
&=&
\sum_{n=1}^\infty \frac{1}{n} \,\omega^n \, z_g(T/n)
.
\label{eq:4.19}
\end{eqnarray}

Finally, using these functions, we can write~\footnote{Note that the characters
  in \Eq{4.14} can be computed in closed form using the method described below
  in the next section (see Eqs. (\ref{eq:3.27}) and (\ref{eq:4.43})), prior to
  summing over levels. Such procedure would be less convenient here because it
  requires to take the sum anew for each value of $\Omega$ and $T$.}
\begin{eqnarray}
\log Z_q(\Omega,T)
&=&
\sum_{i=1}^3 \log \hat{Z}_q(\omega_i,T)
,
\label{eq:4.20a}
\\
\log Z_g(\Omega,T)
&=&
\sum_{i,j=1}^3 
\log \hat{Z}_g(\omega_i\omega_j^*,T)
- \log \hat{Z}_g(1,T)
.
\nonumber 
\end{eqnarray}

\subsection{Group integration}
\label{sec:3.F}

The functions of $\Omega$ are actually functions of the pair
$(\omega_1,\omega_2)$, so they can be expressed as periodic functions of
$\phi_1$ and $\phi_2$, with $\omega_i= e^{i\phi_i}$. The set of
eigenvalues is covered once by taking $(\phi_1,\phi_2)\in
[-\pi,\pi]\times[-\pi,\pi]$. Moreover, for class functions
$\chi(\Omega)$, as those considered here, the $\SU(3)$ group integration
can be written as
\begin{eqnarray}
&& \int d\Omega \, \chi(\Omega) =  \int_{-\pi}^\pi \frac{d\phi_1}{2\pi}
\frac{d\phi_2}{2\pi} \frac{1}{6} 
\nonumber \\ && \times
|\omega_1-\omega_2|^2|\omega_1-\omega_3|^2|\omega_2-\omega_3|^2 
\chi(\omega_1,\omega_2)
.
\label{eq:4.26}
\end{eqnarray}
The analysis of a class function $\chi(\Omega)$ in terms of characters,
cf. \Eq{4.4}, exposes the weight of each irrep $\mu$. The plots of class
functions of $\Omega$ can be done using $(\phi_1,\phi_2)$, however, this
introduces a distortion since the natural manifold is the plane
$\phi_1+\phi_2+\phi_3=0$ in ${\mathbb R}^3$, with orthonormal coordinates there. This
is equivalent to using the Gell-Mann matrices $\lambda_3$ and $\lambda_8$ to
express $-i\log \Omega$ for diagonal $\Omega={\rm
  diag}(\omega_1,\omega_2,\omega_3)$. More explicitly, we introduce the new
coordinates $\varphi_3$, $\varphi_8$
\begin{equation}
-i\log \Omega = {\rm diag}(\phi_1,\phi_2,\phi_3) = 
\varphi_3\frac{\lambda_3}{\sqrt{2}}
+
\varphi_8\frac{\lambda_8}{\sqrt{2}}
\end{equation}
In terms of these coordinates the $D_3$ symmetry of such class function
$\chi(\Omega)$ is more immediate: symmetry under exchange of $\omega_1$ and
$\omega_2$ implies reflection under the $\varphi_8$ axis, whereas exchange of
$\omega_2$ and $\omega_3$ corresponds to a rotation by an angle $2\pi/3$.  If
$\chi(\Omega)$ contains only autoconjugated irreps, the symmetry is extended
to rotations of angle $\pi$ and so to $D_6$. Besides, in these coordinates,
the periodicity takes the form
\begin{eqnarray}
\chi(\varphi_3,\varphi_8) &=& 
\chi(\varphi_3+\sqrt{2}\pi ,\varphi_8+\sqrt{6}\pi) 
\nonumber \\
&=& \chi(\varphi_3-\sqrt{2}\pi ,\varphi_8+\sqrt{6}\pi)
.
\end{eqnarray}
The set of eigenvalues is covered once by taking $(\varphi_3,\varphi_8)\in
[-\sqrt{2}\pi,\sqrt{2}\pi]\times [-\sqrt{3/2}\pi,\sqrt{3/2}\pi]$, and the
measure becomes
\begin{eqnarray}
&& \int d\Omega \, \chi(\Omega) = 
\int_{-\sqrt{2}\pi}^{\sqrt{2}\pi} \frac{d\varphi_3}{2\pi}
\int_{-\sqrt{3/2}\pi}^{\sqrt{3/2}\pi} \frac{d\varphi_8}{2\pi}
\frac{1}{6\sqrt{3}} \nonumber \\ &&
\times
|\omega_1-\omega_2|^2|\omega_1-\omega_3|^2|\omega_2-\omega_3|^2 
\chi(\omega_1,\omega_2)
.
\end{eqnarray}

\subsection{Expansions in the number of constituents}
\label{sec:3.G}

We note that the counting of degeneracies of multiparticle states, as done in
Sec. \ref{sec:3.D}, can be recovered by integration over the Polyakov loop
variable $\Omega$, by inserting the proper characters.  Specifically, let
$Z_{f,\mu}$ denote the polynomials in $\gamma_q$, $\gamma_\qb$ and $\gamma_g$
shown in the ``Total'' column of Table \ref{tab:3}. They are obtained by
computing the partition function of a fictitious theory with a single level
for each of the three species. The corresponding $Z_f(\Omega)$ serves as
generating function of those polynomials:
\begin{eqnarray}
Z_f(\Omega)
&=& \sum_\mu \chi_\mu^*(\Omega) Z_{f,\mu}
\label{eq:3.26}\\
\log Z_f(\Omega)
&=&
 \gamma_q \, \chi_{\bf 3}(\log(1 + \Omega \, q ))
+ \gamma_\qb  \, \chi_{\bf \bar{3}}(\log(1 + \Omega \, \qb)) \nonumber \\ 
&& - \gamma_g  \, \chi_{\bf 8}(\log(1 - \Omega \, g))
.
\label{eq:3.27}
\end{eqnarray}
Here the symbols $q$, $\qb$ and $g$ denote the corresponding ($T$-dependent)
single-particle Boltzmann weights. We will use them to tag the various
constituents present in the multiparticle states, namely, by expanding
$Z_f(\Omega)$ in powers of $q$, $\qb$ and $g$.

To obtain the expansion of the $Z_{f,\mu}$ in the number of constituents one
can proceed by expanding $Z_f(\Omega)$ in powers of $q$, $\qb$ and $g$,
inserting the required factor $\chi_\mu(\Omega)$, using $\Omega ={ \rm
  diag}(\omega_1,\omega_2,\omega_3)$, and finally integrating over $\Omega$,
using e.g. \Eq{4.26}. That integration can be conveniently done by residues.

Alternatively, $Z_f(\Omega)$ can be obtained directly in the form in
\Eq{3.26}. Following \cite{Sasaki:2012bi}, we first evaluate the characters in
closed form, and this produces
\begin{eqnarray}
\chi_{\bf 3}(\log(1 + \Omega \, q ) )
&=&
\log( \chi_{\bf 1} + \chi_{\bf 3} q + \chi_{\bf {\bar 3}} q^2 + \chi_{\bf 1} q^3)
,
\nonumber \\
\chi_{\bf {\bar 3}}(\log(1 + \Omega \, \bar{q} ) )
&=&
\log( \chi_{\bf 1} + \chi_{\bf \bar{3}} \bar{q} + \chi_{\bf 3} \bar{q}^2 
+ \chi_{\bf 1} \bar{q}^3)
,
\nonumber \\
\chi_{\bf 8}(\log(1 - \Omega \, g ) )
&=&
\log( 
\chi_{\bf 1} (1 + g^8)
- \chi_{\bf 8} ( g + g^7 ) \nonumber \\ && 
+ ( \chi_{\bf 8} + \chi_{\bf 10} + \chi_{\bf \bar{10}} ) (g^2+g^6)
\nonumber \\ &&
- ( \chi_{\bf 1} + \chi_{\bf 8} + \chi_{\bf 10} + \chi_{\bf \bar{10}} +
\chi_{\bf 27} )  (g^3+g^5) \nonumber \\ &&
+  2 ( \chi_{\bf 8} + \chi_{\bf 27} )  g^4
)
,
\label{eq:4.43}
\end{eqnarray}
where in the r.h.s. of these equations we have used the notation
$\chi_n \equiv \chi_n(\Omega)$. These closed forms are specific for
the logarithmic function since traces of log of polynomials give log
of polynomials. The coefficients are class functions of $\Omega$ and
they can be expressed systematically in terms of the characters. It
also follows that $Z_{f}(\Omega)$ is a rational function of the group
characters, therefore $Z_{f,\mu}$ can be computed in closed form for
any given value of $\mu$ and arbitrary but concrete values of the
degeneracies, which are necessarily integer numbers.

After expansion of $Z_f(\Omega)$ in powers of $q$, $\qb$ and $g$, the products
of characters are reduced by using the known $\SU(3)$ Clebsch-Gordan series
(see Appendix \ref{sec:B.3}) as well as the character properties
\begin{equation}
\chi_{\mu\oplus\nu} (\Omega) = \chi_\mu( \Omega) + \chi_\nu( \Omega) 
,
\quad
\chi_{\mu\otimes\nu} (\Omega) = \chi_\mu( \Omega) \chi_\nu( \Omega) 
.
\end{equation}
This method directly produces the character expansion of
$Z_f(\Omega)$. Explicitly, up to two constituents:
\begin{widetext}
\begin{eqnarray}
Z_f(\Omega) &=&
\chi_{\bf 1}
+
\gamma_q q \chi_{\bf 3} 
+ \gamma_\qb \qb \chi_{\bf\bar{3}}
+ \gamma_g g \chi_{\bf 8}
\nonumber \\ &&
+
 \left(
  \gamma_q \gamma_\qb q \qb 
+ \frac{1}{2} \left(\gamma_g^2 + \gamma_g\right) g^2 
\right) \chi_{\bf 1} 
+ \left( 
 \frac{1}{2} \left(\gamma_\qb^2 + \gamma_\qb\right) \qb^2
+ \gamma_q \gamma_g q g  
\right) \chi_{\bf 3} 
+ \left( 
  \frac{1}{2} \left(\gamma_q^2 + \gamma_q\right) q^2 
+ \gamma_\qb \gamma_g \qb g 
\right)  \chi_{\bf\bar{3}}
\nonumber \\ &&
+ \left( 
 \frac{1}{2} \left(\gamma_q^2 - \gamma_q\right) q^2
+ \gamma_\qb \gamma_g \qb g 
\right) \chi_{\bf 6} 
+  \left( 
 \frac{1}{2} \left(\gamma_\qb^2 - \gamma_\qb\right) \qb ^2
+ \gamma_q \gamma_g q g
\right) \chi_{\bf\bar{6}}
+ \left( 
 \gamma_q \gamma_\qb q \qb 
+ \gamma_g^2 g^2 
\right) \chi_{\bf 8}
\nonumber \\ &&
+ \frac{1}{2} \left(\gamma_g^2 - \gamma_g\right) g^2 \chi_{\bf 10}
+ \frac{1}{2} \left(\gamma_g^2 - \gamma_g\right) g^2 \chi_{\bf\bar{10}}
+  \gamma_q \gamma_g q g \chi_{\bf 15}  
+  \gamma_\qb \gamma_g \qb g \chi_{\bf\bar{15}}
+ \frac{1}{2} \left(\gamma_g^2 + \gamma_g\right) g^2 \chi_{\bf 27}
+ O(c^3)
\label{eq:4.45}
.
\end{eqnarray}
\end{widetext}
(Here $O(c^3)$ denotes terms with three or more constituents.) As it should,
the formula just obtained reproduces the results labeled as ``Total'' in Table
\ref{tab:3}. Using instead the results labeled as ``Irred'' in that table
(plus those in Table \ref{tab:1} for irreps different from ${\bf 1}$, ${\bf
  3}$, ${\bf\bar{3}}$, and ${\bf 8}$), we can also write the corresponding
result for the irreducible contributions $L_f(\Omega)$. However, to the order
shown in \Eq{4.45} the two functions do not differ except in the singlet part
(which reduces to just $\chi_1$ in $L_f(\Omega)$).

Recalling that the reducible color configurations contain non-empty subsets of
constituents forming a singlet by themselves (i.e., forming a dynamical
hadron), it follows that all the configurations (in $Z_f$) can be generated
from the irreducible ones (in $L_f$) by adding all possible new
singlets. Since the singlets are in $Z_{f,\bf 1}$, this suggests the relation
\begin{equation}
Z_f(\Omega) \approx L_f (\Omega)\, Z_{f,\bf 1}
,
\label{eq:4.44}
\end{equation}
or equivalently, using the notation in Sec. \ref{sec:2.A},
\begin{equation}
L_f (\Omega) \approx \tilde{L}_f(\Omega) := \frac{Z_f(\Omega)}{Z_{f,\bf 1}}
.
\label{eq:4.44a}
\end{equation}
As it turns out, the conjecture in \Eq{4.44} is exact for up to three
constituents but it fails for four or more constituents. For instance,
considering up to pentaquark configurations~\footnote{The name refers to the
  fact that adding a heavy quark, these are heavy baryon pentaquark
  configurations.} and neglecting gluonic terms, one finds
\begin{widetext}
\begin{eqnarray}
\tilde{L}_{f,{\bf 3}} = \frac{Z_{f,{\bf 3}}}{Z_{f,{\bf 1}}}
&=&
\frac{
[1]_\qb \,\qb + [2]_q \, q^2 + [1]_q ([2]+[1^2])_\qb \, q \qb^2
+ ([3]+[2,1])_q[1]_\qb \, q^3 \qb + [3,1]_\qb \, \qb^4 + O(c^5) + O(g)
}{
1
+ [1]_q [1]_\qb q \, \qb 
+ [3]_q q^3 + [3]_\qb \, \qb^3
+ O(c^4) + O(g)
}
\nonumber \\
&=&
[1]_\qb \,\qb + [2]_q \, q^2 
- [3]_q[1]_\qb \, q^3 \qb - [4]_\qb \, \qb^4 + O(c^5) + O(g)
.
\label{eq:3.32}
\end{eqnarray}
\end{widetext}
The negative weights in the pentaquark contributions $q^3 \qb$ and $\qb^4$
indicate that dividing by $Z_{f,{\bf 1}}$ tends to oversubtract the reducible
terms from the total. As follows from Table \ref{tab:3}, the result obtained
by including only color irreducible configurations is, instead,
\begin{equation}
L_{f,{\bf 3}} = [1]_\qb \,\qb + [2]_q \, q^2 + O(c^5)+O(g)
.
\end{equation}

The function $L_f(\Omega)$ exists (is well defined) and likely it is possible
to write it in closed form as a generating function for the irreducible terms,
but we have not found such an expression. As a less satisfactory alternative,
in concrete models, what we do is to take the full $Z_{f,\mu}$ and
$\tilde{L}_{f,\mu}$ as upper and lower bounds, respectively, or estimates, of
the true $L_{f,\mu}$ in the calculation to all orders in the number of
constituents.

\section{Estimates based on the Polyakov constituent quark model}
\label{sec:cqm}

In order to obtain an overall estimate of the spectrum of the heavy hadrons in
the sum rule \Eq{2.7} we will adopt the widely used PNJL
model~\cite{Meisinger:1995ih,Fukushima:2003fw,Megias:2004hj%
  ,Megias:2006bn,Ratti:2005jh,Sasaki:2006ww,Ciminale:2007sr%
  ,Contrera:2007wu,Schaefer:2007pw,Costa:2008dp,Mao:2009aq%
  ,Sakai:2010rp,Radzhabov:2010dd,Zhang:2010kn}. Here the Polyakov loop will be
treated {\em explicitly} as a collective quantum and local variable, which is
integrated over the group manifold. This makes calculations rather
straightforward, since the integration projects configurations in the presence
of a heavy and irreducible color source which are globally color singlets. In
the next section we will show that this is fully equivalent to using group
characters, as discussed in the next sections, and this is particular
advantageous in cases where the Polyakov loop is not introduced {\em
  explicitly} such as the bag model.

\subsection{The Polyakov constituent quark model}
\label{sec:cqm.a}

In this Section we consider a model in the spirit of describing
$Z_{\rm QCD}$ using free constituent dynamical quarks (and possibly
gluons) with a Polyakov variable at each point of the space,
$\Omega(\vx)$. The dynamics is determined by that of the Polyakov loop
variable. This is similar to the most usual implementation of the PNJL
model except that we avoid taking a mean-field approximation, and
$\Omega(\vx)$ is kept as a quantum and local degree of freedom. We
start out from the formulation motivated in previous
works~\cite{Megias:2004hj,Megias:2006bn,Megias:2005qf,Megias:2006df},
but neglecting details not essential for our argument~\footnote{For
  simplicity, we consider mass-degenerate quarks. Also we leave
  implicit the detailed model of the quark dynamics originating
  constituent masses or the chiral transition, for instance, a
  Nambu-Jona--Lasinio model.}. The partition function is given by
\begin{equation}
Z_{\rm PCM} = \int \prod_{\vx}d\Omega(\vx) \, e^{-S_{\rm PCM}(\Omega,T)} \,,
\label{eq:3}
\end{equation}
where $d\Omega(\vx)$ is the invariant $\SU(N_c)$ measure at each point. In the
simplest version the action contains a contribution from dynamical quarks
(and antiquarks)
and another from the Polyakov loop, and explicit dynamical gluons
are not included,
\begin{equation}
S_{\rm PCM} (\Omega,T) =  S_P (\Omega,T) + S_q (\Omega,T)
.
\label{eq:Spq}
\end{equation}
(Note that $S$ depends {\em functionally} on the full Polyakov loop
configuration ${\Omega(\vx)}$ on $\R^3$, and not on a variable $\Omega$ as,
e.g., in \Eq{2.9}.)

The action $S_P(\Omega,T)$ would follow from gluodynamics, and in
particular it is responsible for spontaneous breaking of center
symmetry \cite{'tHooft:1979uj} above the transition temperature
\cite{Polyakov:1978vu,Susskind:1979up}. However, we only need to model
some low temperature properties of $S_P(\Omega,T)$. We first consider
the correlation of Polyakov loop variables at the same point.  We
assume that, at low temperatures, $S_P (\Omega)$ is close to zero and
consequently, the distribution of $\Omega$ (in the absence of the
quark term $S_q$) locally coincides with the Haar measure. In this
approximation $\Omega$ is a completely random variable with equal
probability to take any group value.  This would manifest in a very
small expectation value of the Polyakov loop in the adjoint
representation, even if this is not an order parameter of center
symmetry in gluodynamics. Such small expectation value is actually
observed \cite{Dumitru:2003hp,Gupta:2007ax}. It is noteworthy that a
mean-field treatment, as in the PNJL version, does not naturally tend to
suppress $\langle \chi_{\bf 8}(\Omega)\rangle$. This fact makes the usual mean-field version of the
PNJL model unsuitable to describe the expectation value of the
Polyakov loop in higher irreps.

The part of the action depending on the quarks follows from the fermion
determinant and reads
\begin{eqnarray}
S_q (\Omega,T) &=& - 2 N_f  \int \frac{d^3
  x d^3 p}{(2\pi)^3} \bigg[ 
 \Tr_F \big( \log \big( 1+\Omega ({\vx}) \, e^{-\beta E_p}\big) \big) \nonumber \\ 
&& + \Tr_F \big( \log \big( 1+\Omega^\dagger ({\vx}) \,e^{-\beta E_p}\big)
\big) \bigg] \nonumber \\ 
&=& - 2 N_f  \int \frac{d^3
  x d^3 p}{(2\pi)^3} \bigg[ 
 \chi_{\bf 3}\big( \log \big( 1+\Omega ({\vx}) \, e^{-\beta E_p}\big) \big) \nonumber \\ 
&& + \chi_{\bf \bar{3}} \big( \log \big( 1+\Omega ({\vx}) \,e^{-\beta E_p}\big)
\big) \bigg]
 .
\label{eq:quark-action}
\end{eqnarray}
Here $E_p = \sqrt{\vp^2+M_q^2}$ is the energy of the quarks. In chiral quark
models one takes $M_q$ to be the constituent quark mass, i.e., a non-vanishing
quantity at zero current quark mass. The Polyakov loop corresponds to a
chemical potential in color space. The color traces in \Eq{quark-action} can
be written explicitly in terms of characters of $\Omega$, using \Eq{4.43}.

When the full action $S_P+S_q$ is used, the distribution of $\Omega(\vx)$
departs from the Haar measure as a color source can be screened by quarks and
antiquarks near it. Each quark brings a penalty $e^{-\beta E_p}$ and so, at
lower temperatures, the effect is smaller and also involves fewer
constituents.

\subsection{Confined domains approximation}
\label{subsec:confined_domains}

As noted above, the usual mean-field implementation of the PNJL model is not
suited to describe the Polyakov loop in higher representations. So we will
adopt a different approach here.

A color source $\mu$ at $\vx_0$ is screened by quarks or antiquarks at points
$\vx$ (carrying Polyakov loop variables $\Omega(\vx)$). Therefore, in
principle, for $n$ constituents, $S_P(\Omega,T)$ needs to be modeled to
describe the correlations $\langle \chi_\mu(\Omega(\vx_0))
\chi_{\mu_1}(\Omega(\vx_1))\cdots \chi_{\mu_n}(\Omega(\vx_n)) \rangle$ as a
function of the temperature. It is not hard to make a model valid for the
confined phase and $n=1$, namely
\begin{equation}
\langle \chi_\mu(\Omega (\vx_0)) \chi_{\bar{\mu}}(\Omega (\vx_1))
  \rangle_{S_P} = e^{-\beta \sigma_\mu |\vx_0-\vx_1| } 
\,,
\label{eq:corr-func}
\end{equation}
where $\sigma_\mu$ is the string tension between color charges in the
irrep $\mu$~\footnote{Isolated color charges in irreps with trivial
  triality can be screened by gluons. In those cases the $\sigma_\mu
  r$ form of the potential refers to distances not so large that
  string breaking is favored. Likewise, at short distances, 
one has attractive Coulomb terms $\sim -\alpha_s/r$,
which would lead to a unbounded from
below free energy.
However, as discussed in
\cite{RuizArriola:2012wd}, quantization effects are expected to
dominate over the classical behavior regularizing the divergences. As a
result
these Coulomb terms will be neglected.
}. It can be verified that the correct
normalization from the Haar measure,
$\langle\chi_\mu\chi_{\bar\mu}\rangle_{\SU(N_c)}=1$, is reproduced at
the coincidence limit. As argued in \cite{RuizArriola:2012wd}, this
term combines with the kinetic energy $E_p$, to produce a
source-antiquark Hamiltonian (for $\mu={\bf 3}$) with confining
potential $\sigma r$.  Quantization of such model would produce the
energy levels $\Delta$ to be used in \Eq{2.7}. Here we remain at a
semiclassical level, retaining $\vx$ and $\vp$ simultaneously.

The extension of \Eq{corr-func} to a larger number of constituents remains a
challenging problem. The correct counting of color states is guaranteed but it
is not trivial to fulfill other requirements such as cluster decomposition and
existence of a thermodynamic limit, or effective Lorentz invariance
restoration.  Another issue, which would help as guidance to construct the
model, is that of the degree of consistency with the hadron resonance gas
picture. The hadron resonance gas picture has been shown to be consistent with
strong coupling QCD at leading orders \cite{Langelage:2010yn}, but this
picture might fail at a higher orders. This can happen if a conflict arises
between statistics of identical particles for quarks and for hadrons, with
four or more constituents. Within the Polyakov constituent model we find that
the correlations of the local Polyakov loop variables can be chosen as to
reproduce the form of a hadron gas if a single meson or a single baryon is
involved, but obstructions are likely to arise in more general situations.
We will analyze this important issue elsewhere.

In order to bypass these difficulties, we observe that, according to
\Eq{corr-func}, Polyakov loop variables are uncorrelated if they are
sufficiently separated and tend to full correlation when they are
close to each other. (This is not in contradiction with our previous
assumption that $\Omega (\vx)$ is a completely random variable in the
absence of quarks: two very close variables would move together, but
still at random on $\SU(N_c)$.)  In view of this, we assume a simple
model \cite{Megias:2004hj,Megias:2006bn} in which the space is divided
in (confinement) domains. Polyakov loop variables in the same domain
take identical values and are distributed according to the Haar
measure of $\SU(N_c)$ (in the absence of dynamical quarks or
gluons). Variables in different domains are fully uncorrelated. From
\Eq{corr-func} it follows that the typical size of the domain,
$V_\sigma$, depends on the temperature, being larger as the
temperature increases.  \Eq{corr-func} suggests $V_\sigma$ scaling as
$T^3/\sigma^3$ (taking for $\sigma$ that of the fundamental irrep),
but we keep this quantity as a parameter.

Under the previous assumption, a color source in one domain cannot be screened
by quarks in a different domain, each domain becomes independent and contains
a single Polyakov loop variable. This allows us to immediately write down the
function similar to $Z_{\rm QCD}(\Omega,T)$ for the constituent quark model,
\begin{equation}
Z_{\rm PCM}(\Omega,T) = 
e^{ - \beta V_\sigma (\mathcal{L}_q+\mathcal{L}_g)}
.
\label{eq:ZOqg}
\end{equation}
$\mathcal{L}_q$ is the Lagrangian density corresponding to $S_q$ (i.e., $S_q$
removing $\int d^3x$ and adding a factor $T$). This Lagrangian would depend on
$x$ (a point in the domain) only through $\Omega$, which is now an external
parameter. $S_P$ is no longer present since its non trivial effect in forming
the domains has already been used. Thus $Z_{\rm PCM}(\Omega,T)$ represents the
unnormalized probability density distribution of the Polyakov loop variable
(relative to the Haar measure) in the Polyakov constituent model.

In \eq{ZOqg} we have also included a dynamical gluon Lagrangian (not
present in the previous version of \Eq{Spq}) similar to that of the
quarks~\cite{Meisinger:2001cq,Meisinger:2003id},
\begin{eqnarray}
{\cal L}_g (x) &=& 2 T \int \frac{d^3 p}{(2\pi)^3} 
 \Tr_A \big( \log \big( 1 - \Omega ({\vx}) \, e^{-\beta w_p}\big) \big) \nonumber \\ 
&=& 2 T \int \frac{d^3 p}{(2\pi)^3} 
 \chi_{\bf 8}\big( \log \big( 1 - \Omega ({\vx}) \, e^{-\beta w_p}\big) \big) 
.
\label{eq:gluon-action}
\end{eqnarray}
where $\Tr_A$ is the trace in the Adjoint representation and we have used the
identity $\Tr_A F(\Omega) = \chi_{\bf 8} [F(\Omega)]$.  Here
$w_p=\sqrt{\vp^2+M_g^2}$, where $M_g$ represents a constituent gluon mass. (We
still assume two polarizations for massive gluons although other degeneracies
could be implemented as well.)  The former version, without gluons, is
recovered by taking the infinitely heavy gluon limit. Likewise, the
corresponding model for gluodynamics is recovered by setting $N_f=0$ or the
infinitely heavy quark limit. The adjoint trace of the logarithm can be
expressed in terms of the characters of $\Omega$ \cite{Sasaki:2012bi} using
\Eq{4.43} (see also \cite{Ruggieri:2012ny}).

\subsection{Expansions in the number of constituents}
\label{sec:cqm.b}

At low temperatures quark and gluon states have a Boltzmann suppression and
hence we can reorganize the low temperature expansion as an expansion in the
number of constituents. According to the model specified by the quark and
gluon actions given by \Eq{quark-action} and \Eq{gluon-action} we must expand
in powers of $\Omega$.

The formal similarities between the Polyakov constituent model in
\Eq{ZOqg} and the independent parton model in Eqs. (\ref{eq:4.13}) and
(\ref{eq:4.14}) are obvious. Therefore, the machinery developed in
Sec. \ref{sec:3.E} applies here in a very direct way. In this section
we carry out the calculation using the group integration method.

To be more specific, let us introduce the function
\begin{equation}
J(M,T) := \int\frac{d^3p}{(2\pi)^3} e^{- \beta \sqrt{M^2+\vp^2}}
=
\frac{T M^2}{2\pi^2} K_2(\beta M)
,
\end{equation}
$K_2(x)$ being the modified Bessel function. Its low temperature behavior is
given by
\begin{equation}
J(M,T) \approx \left(\frac{M T}{2\pi}\right)^{3/2} e^{-\beta M}
\quad  (T \ll M)
,
\end{equation}
whereas in the massless limit
\begin{equation}
J(0,T) = \frac{T^3}{\pi^2}
.
\label{eq:4.9}
\end{equation}

It follows that the role played by $z_g$ of Sec.~\ref{sec:3.E} would
correspond to $2 V_\sigma J(M_g,T)$ here, whereas $z_q$ would correspond to $2
N_f V_\sigma J(M_q,T)$. Under these identifications, the effective degeneracy
of states is $\gamma_\alpha = 2 V_\sigma J(M_g,T)/e^{-M_g/T}$ and
$\gamma_\alpha = 2 N_f V_\sigma J(M_q,T)/e^{-M_q/T }$ for gluons and quarks
respectively, cf. Eq~(\ref{eq:zqzg}). However, this result for the
degeneracies corresponds to a semiclassical estimate, as a continuum of states
is assumed. Such description is not expected to be reliable for the lowest
states. In fact it leads to effective degeneracies in the range
$0<\gamma_\alpha <1$ which produce negative values for the Polyakov loop at
very low $T$ in some representations, e.g., $\tilde{L}_{\bf 6}$. In order to
correct this problem, it is standard in the semiclassical method to amend the
degeneracies in the form $\gamma_\alpha \to \gamma_\alpha + c$, for suitable
$c$, usually $c=1$ or $2$. Under this prescription, the identification of the
$z_q$ and~$z_g$ functions within the Polyakov constituent quark model
becomes~\footnote{It is important to note here that any replacements $T\to
  T/n$, as done e.g. in \Eq{4.17}, apply to the explicit $T$ in $J(M,T)$ and
  are not to be taken on a possible temperature dependence in $V_\sigma$.}
\begin{eqnarray}
z_q(T)= z_\qb(T) &=& 2 N_f V_\sigma J(M_q,T) + c \, e^{-M_q/T} \,, \nonumber \\
z_g(T) &=& 2 V_\sigma J(M_g,T) + c \, e^{-M_g/T} \,. \label{eq:zqzgPCM}
\end{eqnarray}
These are the expressions to be applied in \Eq{4.17}.
We will consider from now on the value $c=1$. This value is sufficient to
solve the problem for most representations considered.

\begin{figure}[t]
\begin{center}
\epsfig{figure=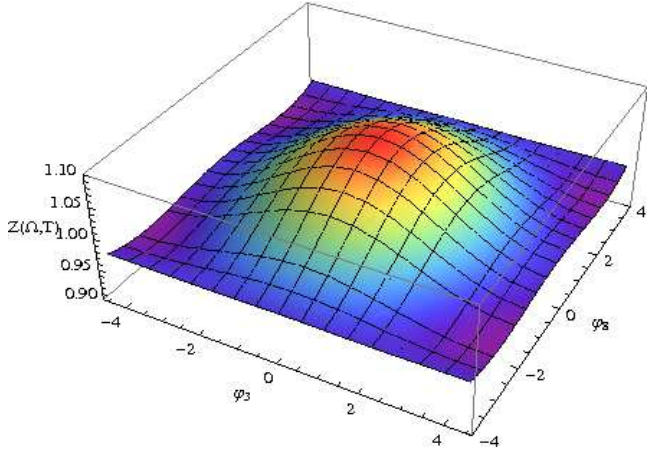,height=60mm,width=85mm}
\end{center}
\caption{The function $Z(\Omega,T)$ on the plane
  $(\varphi_3,\varphi_8)$, for $T = 70\MeV$. We have
  used $N_f=2$, considered as constituent quark and gluon masses
  $M_q=300\MeV$ and $M_g= 664\MeV$ respectively, and the volume
  rule $V_\sigma = 8\pi T^3/\sigma^3$ with $\sigma = (425\MeV)^2$.}
\label{fig:ZPCM70}
\end{figure}

\begin{figure}[t]
\begin{center}
\epsfig{figure=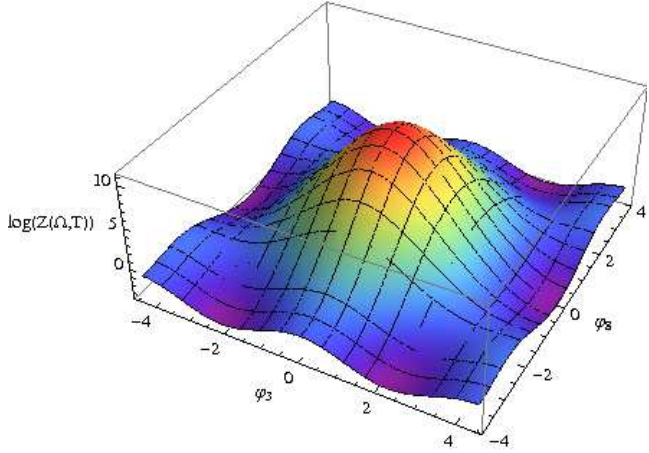,height=60mm,width=85mm}
\end{center}
\caption{The function $\log Z(\Omega,T)$ on the plane
  $(\varphi_3,\varphi_8)$, for $N_f=2$ and $T = 300\MeV$. See Fig.~\ref{fig:ZPCM70} for other details.}
\label{fig:ZPCM300}
\end{figure}

Making use of these prescriptions, $Z_{\rm PCM}(\Omega,T)$ can be computed to
all orders in the number of constituents.  In Figs.~\ref{fig:ZPCM70} and
\ref{fig:ZPCM300} we consider two different temperatures, ~$T=70\MeV$ and
~$T=300\MeV$, corresponding to the confined and the deconfined phases. At low
temperatures the distribution tends to be uniformly distributed (note in
Fig.~\ref{fig:ZPCM70} the small range in the $z$-axis), while at high
temperatures it tends to concentrate at $\Omega=1$. (Note that
Fig.~\ref{fig:ZPCM300} displays the logarithm of $Z$.)
\vspace{10mm}

\begin{figure}[t]
\begin{center}
\epsfig{figure=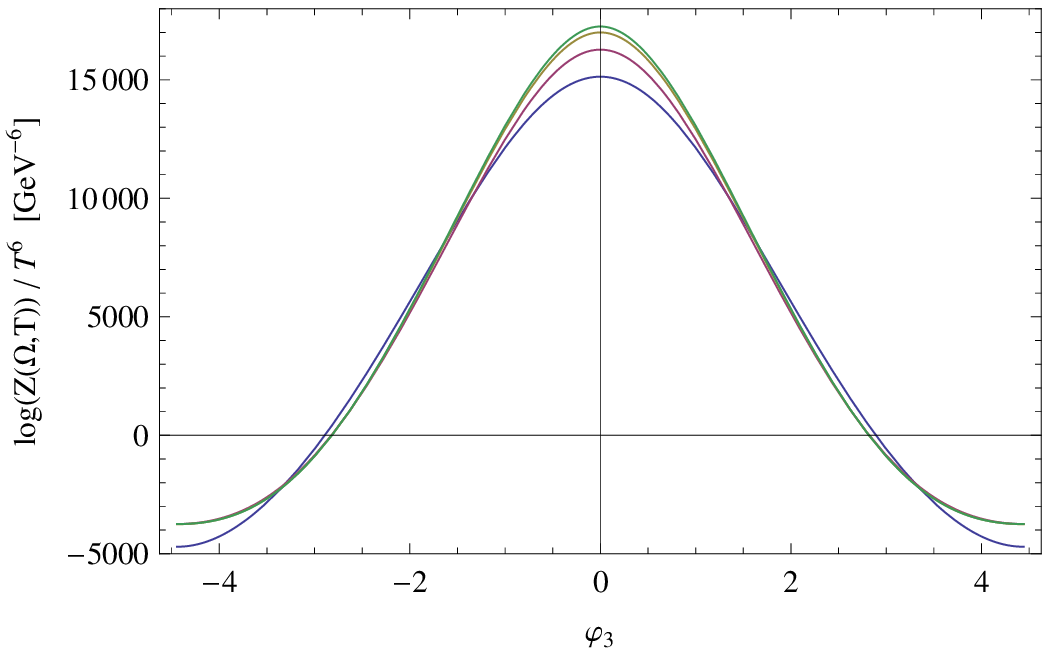,height=60mm,width=85mm}
\end{center}
\caption{Plots of $(\log Z(\Omega,T))/T^6$ as a
  function of $\varphi_3$ for $\varphi_8=0$, for $N_f=2$ and several
  temperatures: from bottom to top $T = 0.3, \, 1, \, 2$ and $5 \, \GeV$.}
\label{fig:logZT6}
\end{figure}

The behavior of $Z(\Omega,T)$ at high temperatures follows from
Fig.~\ref{fig:logZT6} where the ratio $\log Z(\Omega,T)$ over $T^6$ is
displayed. The existence of a limiting profile implies that all the functions
$Z_\mu(T)$ grow exponentially as $\exp(\kappa T^6)$.  As discussed below
for the bag model, a state-independent cavity of volume $V$ yields a behavior
$\exp(\kappa V T^3)$. The power $T^6$ is a consequence of the
modeling $V_\sigma \sim T^3$ and mimics, in the constituent model, the
Hagedorn behavior displayed by the MIT bag.

In what follows we present analytic results for the Polyakov loop
based on expansions in the number of constituents. These are obtained
following the method already explained for \Eq{4.45}. We consider the
two types of estimates for $\langle \chi_\mu(\Omega)\rangle$ discussed
at the end of Sec. \ref{sec:2.A}, namely,
\begin{equation}
\tilde{L}_\mu(T) = \frac{Z_\mu(T)}{Z_{\bf 1}(T)}
,
\qquad
L_\mu(T) = Z_{\mu,{\rm irred}}(T)
.
\end{equation}
In the first expression the partition functions contain all types of color
configurations (reducible and irreducible). In the second one only irreducible
configurations are retained.

For the singlet partition function (i.e., the partition function in the
absence of any colored source) one obtains
\begin{widetext}
\begin{eqnarray}
Z_{\bf 1} &=&
1+\frac{1}{2}  \left(G_1^2+G_2+2 Q_1 \bar{Q}_1\right)
\nonumber\\&&
+\frac{1}{6}  \left(
2 G_1^3 + 4 G_3
+ 6 G_1 Q_1 \bar{Q}_1 + Q_1^3 + 3 Q_1 Q_2
+ 2 Q_3 + \bar{Q}_1^3 + 3 \bar{Q}_1 \bar{Q}_2
+ 2 \bar{Q}_3\right)
\nonumber\\&&
+ \frac{1}{6}  \big(
2 G_1^4 + 3 G_2^2 - 2 G_3 G_1 + 3 G_4
+ 3 (3G_1^2  + G_2 ) Q_1 \bar{Q}_1
+ 2 G_1 ( Q_1^3 - Q_3 + \bar{Q}_1^3 - \bar{Q}_3 )
\nonumber\\&&
+ 3 Q_1^2 \bar{Q}_1^2 + 3 Q_2 \bar{Q}_2 \big)
+ O(c^5)
.
\label{eq:Z1}
\end{eqnarray}
\end{widetext}
Here we have defined
\begin{eqnarray}
Q_n(T) &=& 2 N_f V_\sigma J(M_q,T/n) + c \, e^{-n M_q/T} \,, \\
G_n(T) &=& 2 V_\sigma J(M_g,T/n) + c \, e^{-n M_g/T} \,,
\end{eqnarray}
for quarks and gluons respectively, and we choose $c=1$. The
quantity $\bar{Q}_n$ is numerically identical to $Q_n$ but we
distinguish both quantities in order to display the content of each
term in the constituents: each factor $Q_n$, $\bar{Q}_n$ or $G_n$
count as $n$ quarks, antiquarks or gluons, respectively. So for
instance, a term $G_1G_2 Q_3 \bar{Q}_1^2$ has a content
$g^3q^3\bar{q}^2$.

If $Q_n$, $\bar{Q}_n$ and $G_n$ are replaced, respectively, by $\gamma_q q^n$,
$\gamma_\qb \qb^n$, and $\gamma_g g^n$, in $Z_{\bf 1}$, the formulas for the
singlet labeled as ``Total'' in Table \ref{tab:3} are reproduced. A similar
statement holds for any irrep. 

For the irreps of lowest order the expansions produce
\begin{widetext}
\begin{eqnarray}
\tilde{L}_{\bf 3}  &=&
\bar{Q}_1 
\nonumber\\&&
+\frac{1}{2}  \left(2 G_1
   \bar{Q}_1+Q_1^2+Q_2\right)
\nonumber\\&&
+ G_1 \left(G_1 \bar{Q}_1 + Q_1^2\right)
\nonumber\\&&
+\frac{1}{24}  \big(
4 (5 G_1^3 - 3 G_1 G_2 -2  G_3) \bar{Q}_1
+ 6 G_1^2 \left(5 Q_1^2 -Q_2 \right) 
- 6 G_2 ( Q_1^2 - Q_2 ) 
\nonumber\\&&
- 4 ( Q_1^3 + 3 Q_1 Q_2 + 2 Q_3 ) \bar{Q}_1 
- \bar{Q}_1^4 
- 6 \bar{Q}_2 \bar{Q}_1^2 - 8 \bar{Q}_1 \bar{Q}_3
- 3 \bar{Q}_2^2-6 \bar{Q}_4 \big)
+O(c^5)
,
\label{eq:Lt3}
\end{eqnarray}
\begin{eqnarray}
\tilde{L}_{\bf 6} &=&
\frac{1}{2}  \left(2 G_1
   Q_1+\bar{Q}_1^2-\bar{Q}_2\right)
\nonumber\\&&
+ \frac{1}{2}  \left(
2 G_1 \bar{Q}_1^2 + ( 3 G_1^2 - G_2 ) Q_1
+ \left(Q_1^2+Q_2\right) \bar{Q}_1
\right)
\nonumber\\&&
+ \frac{1}{12} \big(
6 ( 3 G_1^3 - G_1 G_2) Q_1
+ 12 G_1 Q_1^2 \bar{Q}_1
+ 18 G_1^2 \bar{Q}_1^2 - 6 G_2 \bar{Q}_2
+ Q_1^4 - 4 Q_1 Q_3 + 3 Q_2^2 \big) 
+ O(c^5)%
,
\label{eq:Lt6}
\end{eqnarray}
\begin{eqnarray}
\tilde{L}_{\bf 8} &=&
G_1 
\nonumber\\&&
+ \left(G_1^2+Q_1
   \bar{Q}_1\right)
\nonumber\\&&
+ \frac{1}{6}  \left(
5 G_1^3 - 3 G_1 G_2 - 2 G_3
+ 12 G_1 Q_1 \bar{Q}_1
+ 2 Q_1^3 - 2 Q_3 + 2 \bar{Q}_1^3 -2 \bar{Q}_3
 \right)
\nonumber\\&&
+ \frac{1}{6}  \big(
3 G_1^2 ( G_1^2 - G_2)
+ 3 (5 G_1^2 - G_2) Q_1 \bar{Q}_1
\nonumber\\&&
+ G_1 ( 5 Q_1^3  - 3 Q_1 Q_2 - 2 Q_3
+ 5 \bar{Q}_1^3 - 3 \bar{Q}_1 \bar{Q}_2 - 2 \bar{Q}_3 ) 
  \big)
+ O(c^5)
,
\label{eq:Lt8}
\end{eqnarray}
\begin{eqnarray}
\tilde{L}_{\bf 10} &=&
\frac{1}{2} \left(G_1^2-G_2\right) 
\nonumber\\&&
+\frac{1}{6} 
   \left(
4 G_1^3 + 2 G_3
+ 6 G_1 Q_1 \bar{Q}_1 
+ \bar{Q}_1^3-3 \bar{Q}_1 \bar{Q}_2 + 2 \bar{Q}_3
\right)
\nonumber\\&&
+\frac{1}{12} \big(
7 G_1^4 - 3 G_2^2 - 4 G_1 G_3
+ 24 G_1^2 Q_1 \bar{Q}_1  
+ G_1 \left( 4 Q_1^3 - 4 Q_3 + 6 \bar{Q}_1^3 - 6 \bar{Q}_1 \bar{Q}_2
 \right)
\nonumber\\&&
+ 3 \left( Q_1^2 + Q_2 \right) \left( \bar{Q}_1^2 - \bar{Q}_2 \right)
    \big)
+ O(c^5)
,
\label{eq:Lt10}
\end{eqnarray}
\begin{eqnarray}
\tilde{L}_{\bf 15^\prime} &=&
\frac{1}{2} \left( G_1^2 - G_2 \right) \bar{Q}_1 
\nonumber\\&&
+\frac{1}{24} \big(
 12 G_1 Q_1 \left(\bar{Q}_1^2-\bar{Q}_2\right)
+ 12 G_1^2 Q_1^2 - 12 G_2 Q_2 + 24 G_1^3 \bar{Q}_1
\nonumber\\&&
+ \bar{Q}_1^4 - 6 \bar{Q}_1^2 \bar{Q}_2
+ 8 \bar{Q}_1 \bar{Q}_3 + 3 \bar{Q}_2^2  - 6 \bar{Q}_4 \big)
+ O(c^5)
,
\label{eq:Lt15p}
\end{eqnarray}
\begin{eqnarray}
\tilde{L}_{\bf 15} &=&
G_1 \bar{Q}_1 
\nonumber\\&&
+ \frac{1}{2} \left(4 G_1^2 \bar{Q}_1 + 2 G_1 Q_1^2 
+ Q_1 ( \bar{Q}_1^2-\bar{Q}_2 ) \right)
\nonumber\\&&
+\frac{1}{24} \big( 
12 (5 G_1^3 - G_1 G_2 ) \bar{Q}_1
+ 6 G_1^2 ( 9 Q_1^2 - Q_2 ) - 6 G_2 ( Q_1^2 - Q_2 )
\nonumber\\&&
+ 12 G_1 Q_1 ( 3 \bar{Q}_1^2 - \bar{Q}_2 )
+ 8 ( Q_1^3 - Q_3 ) \bar{Q}_1 + 3 \bar{Q}_1^4 
- 6 \bar{Q}_2 \bar{Q}_1^2 - 3 \bar{Q}_2^2 + 6 \bar{Q}_4 \big)
+ O(c^5)
,
\label{eq:Lt15}
\end{eqnarray}
\begin{eqnarray}
\tilde{L}_{\bf 24} &=&
\frac{1}{2} G_1 \left( 2 G_1 \bar{Q}_1 + Q_1^2 - Q_2 \right)
\nonumber\\&&
+\frac{1}{6}  \big( 6 G_1 Q_1 \bar{Q}_1^2 + 12 G_1^3 \bar{Q}_1
+ 3 G_1^2 \left( 3 Q_1^2 - Q_2\right)
\nonumber\\&&
+ \left( Q_1^3 - 3 Q_1 Q_2 + 2 Q_3 \right) \bar{Q}_1
\big)
+ O(c^5)
,
\label{eq:Lt24}
\end{eqnarray}
\begin{eqnarray}
\tilde{L}_{\bf 27} &=&
\frac{1}{2} \left(G_1^2+G_2\right) 
\nonumber\\&&
+ \left( G_1 Q_1 \bar{Q}_1+G_1^3 \right)
\nonumber\\&&
+ \frac{1}{8} \big(
9 G_1^4 + 3 G_2^2 -2 G_4 
- 2 G_1^2 G_2 + 24 G_1^2 Q_1 \bar{Q}_1
\nonumber\\&&
+ 4 G_1 \left( Q_1^3 - Q_1 Q_2 + \bar{Q}_1^3 - \bar{Q}_1 \bar{Q}_2 \right)
+ 2 \left( Q_1^2 - Q_2 \right) \left(\bar{Q}_1^2-\bar{Q}_2\right) \big)
+ O(c^5)
,
\label{eq:Lt27}
\end{eqnarray}
\begin{eqnarray}
\tilde{L}_{\bf 35} &=&
\frac{1}{3} \left(G_1^3-G_3\right) 
\nonumber\\&&
+ \frac{1}{8} ( 5 G_1^4 + 2 G_1^2 G_2 - G_2^2 + 2 G_4 )
+ G_1^2 Q_1 \bar{Q}_1 
+ \frac{1}{6} G_1 \left( Q_1^3 - 3 Q_1 Q_2 + 2 Q_3 \right)
+ O(c^5)
,
\label{eq:Lt35}
\end{eqnarray}

\begin{eqnarray}
\tilde{L}_{\bf 42} &=&
\frac{1}{2} \left(G_1^2+G_2\right) \bar{Q}_1 
\nonumber\\&&
+\frac{1}{4} \left( 2 G_1 \left(G_2 \bar{Q}_1 
+ Q_1 \left(\bar{Q}_1^2 - \bar{Q}_2 \right) \right)
+ 6 G_1^3 \bar{Q}_1 + G_1^2 \left(3 Q_1^2-Q_2\right)
+ G_2 \left( Q_1^2 + Q_2 \right) \right)
+ O(c^5)
,
\label{eq:Lt42}
\end{eqnarray}
\begin{eqnarray}
\tilde{L}_{\bf 64} &=&
\frac{1}{6} \left(G_1^3+3 G_2 G_1+2 G_3\right) 
+\frac{1}{2} \left(G_1^2+G_2\right) 
   \left(G_1^2+Q_1 \bar{Q}_1\right)+O(c^5)
.
\label{eq:Lt64}
\end{eqnarray}
\end{widetext}

For these same irreps, we have worked out the counting of irreducible color
configurations for up to three constituents (see Table \ref{tab:1}), and up to
four constituents for the fundamental representation (see Table
\ref{tab:2}). As already noted, up to three constituents the two estimates
coincide and the same property holds for any single-particle Hamiltonian, i.e.,
\begin{equation}
L_\mu(T) = \tilde{L}_\mu(T) + O(c^4)
. \label{eq:LLtilde}
\end{equation}
The two estimates $L_\mu$ and $\tilde{L}_\mu$ differ by terms of $O(c^5)$ for
irreps beyond ${\bf 27}$, that require at least three constituents to be
screened.

For the fundamental representation, up to four constituents, we find
\begin{eqnarray}
L_{\bf 3}  &=&
\bar{Q}_1 
\nonumber\\&&
+ \frac{1}{2}  \left(2 G_1 \bar{Q}_1+Q_1^2+Q_2 \right)
\nonumber\\&&
+ G_1 \left(G_1 \bar{Q}_1 + Q_1^2\right)
\nonumber\\&&
+ \frac{1}{4} (5 G_1^2 Q_1^2 - G_2 Q_1^2 - G_1^2 Q_2 + G_2 Q_2) 
\nonumber\\&&
+ \frac{1}{6} (5 G_1^3 - 3 G_1 G_2 - 2 G_3) \bar{Q}_1
+ O(c^5)
,
\label{eq:L3}
\end{eqnarray}
which differs from $\tilde{L}_{\bf 3}$ at $O(c^4)$. $L_{\bf 3}$ reproduces the
result quoted in Table \ref{tab:3}.

\begin{figure}[t]
\begin{center}
\epsfig{figure=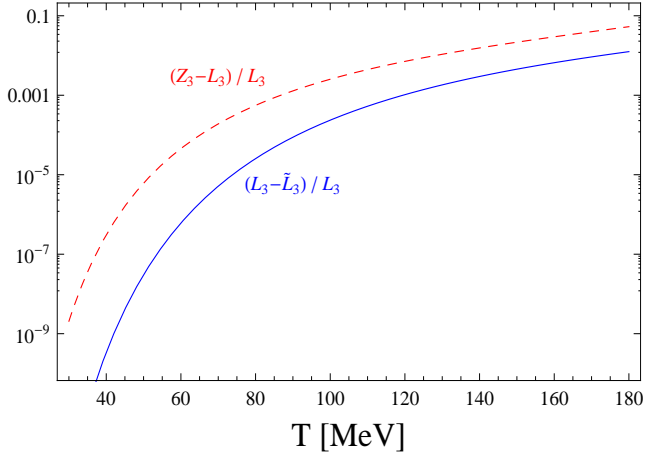,height=60mm,width=85mm}
\end{center}
\caption{Difference between $Z_{{\rm PCM},{\bf 3}}$ and $L_{{\rm
      PCM},{\bf 3}}$ on the one hand, and between $L_{{\rm PCM},{\bf
      3}}$ and $\tilde{L}_{{\rm PCM},{\bf 3}}$ on the other
  (normalized to $L_{{\rm PCM},{\bf 3}}$) as a function of $T$ (in
  $\MeV$), cf.~Eqs.~(\ref{eq:Lt3})-(\ref{eq:L3}). In this plot we have
  included up to four constituents for the solid blue line, and up
  to three constituents for the dashed red line. See
  Fig.~\ref{fig:ZPCM70} for other details.}
\label{fig:L3L3t}
\end{figure}

\begin{figure}[t]
\begin{center}
\epsfig{figure=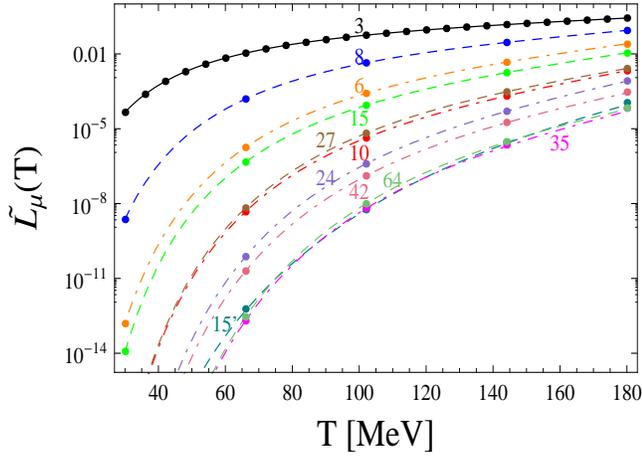,height=60mm,width=85mm}
\end{center}
\caption{$\tilde{L}_{{\rm PCM},\mu}$ as a function of $T$ (in $\MeV$),
  for several irreps. From top to bottom $\mu= {\bf 3}$, ${\bf 8}$,
  ${\bf 6}$, ${\bf 15}$, ${\bf 27}$, ${\bf 10}$, ${\bf 24}$, ${\bf
    42}$, ${\bf 64}$, ${\bf 15'}$, and ${\bf 35}$. Lines correspond to
  the result using the analytical formulas up to four constituents,
  cf. Eqs.~(\ref{eq:Lt3})-(\ref{eq:Lt64}), while points are the
  numerical result to all orders. We include only a few points for
  irreps other than ${\bf 3}$, for the clarity and appearance of the figure. See
  Fig.~\ref{fig:ZPCM70} for other details.}
\label{fig:Lpcm1}
\end{figure}

We plot in Fig.~\ref{fig:L3L3t} the Polyakov loop in the fundamental
representation, computed within the two estimates, cf.~Eqs.~(\ref{eq:Lt3}) and
(\ref{eq:L3}). The difference increases with temperature, but it is rather
small $\lesssim 1\%$ already for temperatures close to the phase
transition. In this figure it is shown also the difference between $Z_{\bf 3}$ and
$L_{\bf 3}$, for which the values are noticeably larger. In Fig.~\ref{fig:Lpcm1} we
display the Polyakov loop in several representations~$\tilde{L}_\mu$. We have
considered in these plots a phenomenological value for the constituent gluon
mass, $M_g = 664\MeV$, which leads to an exponential suppression at low
temperature and it is consistent with strong coupling models of
gluodynamics~\cite{Meisinger:2003id}.

\begin{figure}[t]
\begin{center}
\epsfig{figure=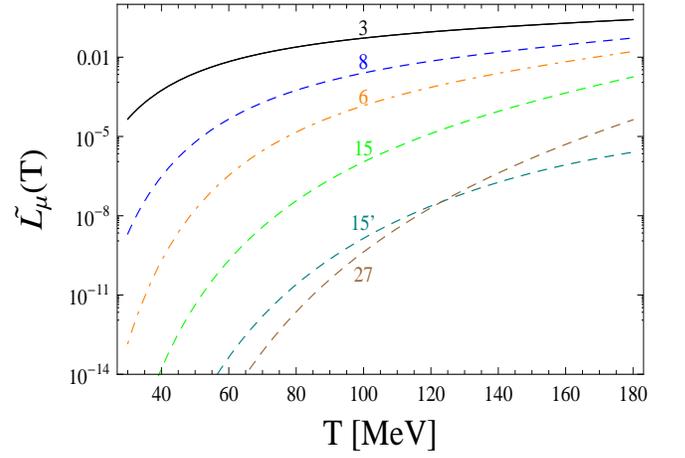,height=60mm,width=85mm}
\end{center}
\caption{As Fig.~\ref{fig:Lpcm1} but including only quarks and antiquarks. Irreps ${\bf 35}$, ${\bf 42}$ and ${\bf 64}$ lead to values for the Polyakov loop identically zero. See text for discussion on other irreps.}
\label{fig:Lpcm1q}
\end{figure}

\begin{figure}[t]
\begin{center}
\epsfig{figure=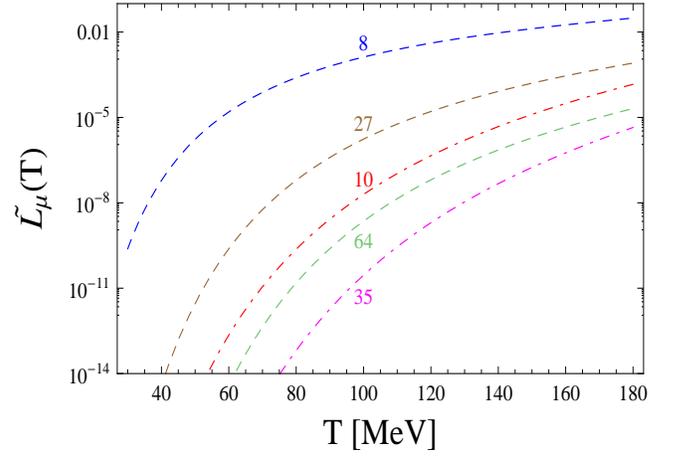,height=60mm,width=85mm}
\end{center}
\caption{As Fig.~\ref{fig:Lpcm1} but including only gluons. Irreps ${\bf 3}$, ${\bf 6}$, ${\bf 15}$, ${\bf 15^\prime}$, ${\bf 24}$ and ${\bf 42}$ lead to values for the Polyakov loop identically zero.}
\label{fig:Lpcm1g}
\end{figure}

We show in Figs.~\ref{fig:Lpcm1q} and \ref{fig:Lpcm1g} the same information as
in Fig.~\ref{fig:Lpcm1}, but including only quarks and antiquarks in the first
case, and only gluons in the latter. Only those irreps which lead to positive
results are depicted. Exceptionally, the irreps ${\bf 10}$ and ${\bf 24}$ lead
to negative results in the quarks and antiquarks contribution for $T < 123
\MeV$ and $T < 171 \MeV$ respectively, indicating that the prescription of
Eq.~(\ref{eq:zqzgPCM}) with $c=1$ is still insufficient in those cases. Note
however that this does not happen in the gluonic terms, and the combined quark
plus gluonic contribution is positive for all temperatures,
c.f. Fig.~\ref{fig:Lpcm1}.

\begin{figure}[t]
\begin{center}
\epsfig{figure=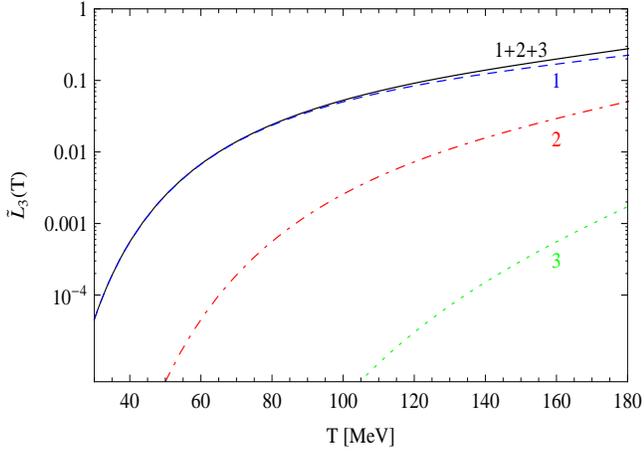,height=60mm,width=85mm}
\end{center}
\caption{$\tilde{L}_{{\rm PCM},{\bf 3}}$ as a function of $T$ (in $\MeV$). Labels 1, 2 and 3 correspond to configurations with one, two and three constituents, respectively. $1+2+3$ corresponds to the sum of all the configurations up to three constituents.}
\label{fig:L3123}
\end{figure}

\begin{figure}[t]
\begin{center}
\epsfig{figure=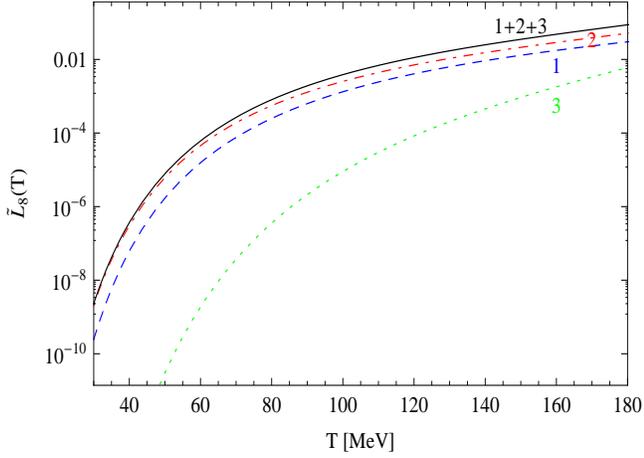,height=60mm,width=85mm}
\end{center}
\caption{As in Fig.~\ref{fig:L3123} for the adjoint representation.}
\label{fig:L8123}
\end{figure}

\begin{figure}[t]
\begin{center}
\epsfig{figure=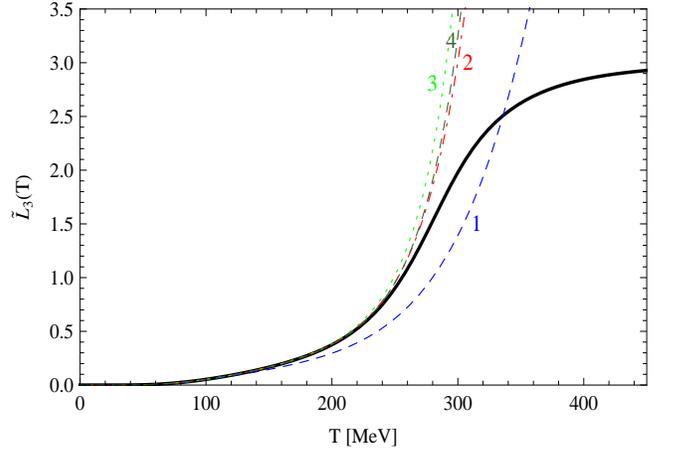,height=60mm,width=85mm}
\end{center}
\caption{$\tilde{L}_{{\rm PCM},{\bf 3}}$ as a function of $T$ (in $\MeV$). Dashed lines represent the result using the analytical expansion up to one, two, three and four constituents, as indicated in the labels. The solid line correspond to the numerical result to all orders. See fig.~\ref{fig:ZPCM70} for other details.}
\label{fig:Lpcm3}
\end{figure}

\begin{figure}[t]
\begin{center}
\epsfig{figure=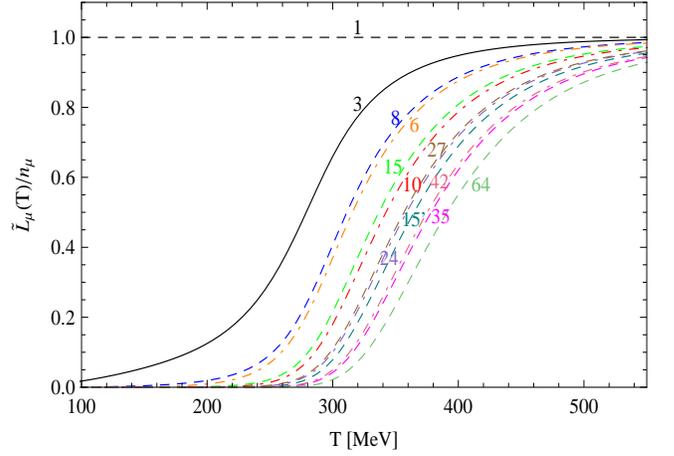,height=60mm,width=85mm}
\end{center}
\caption{$\tilde{L}_{{\rm PCM},\mu}$ (normalized to $n_\mu$) as a function of $T$ (in $\MeV$), for several irreps. From top to bottom,  $\mu=  {\bf 1}$, ${\bf 3}$, ${\bf 8}$, ${\bf 6}$, ${\bf 15}$, ${\bf 10}$, ${\bf 27}$, ${\bf 24}$, ${\bf 15^\prime}$, ${\bf 42}$, ${\bf 35}$, and ${\bf 64}$. The curves are obtained from the numerical result to all orders in the expansion in the number of constituents. See fig.~\ref{fig:ZPCM70} for other details.}
\label{fig:Lmu}
\end{figure}

It is shown in Figs.~\ref{fig:L3123} and \ref{fig:L8123} the separate
contribution of configurations with one, two and three constituents in
$\tilde{L}_{\bf 3}$ and $\tilde{L}_{\bf 8}$. The convergence of the results is
manifest. Fig.~\ref{fig:Lpcm3} displays the behavior of $\tilde{L}_{\bf 3}$
for a wider range of temperatures. The numerical result including all orders
in the expansion in the number of constituents tend to the value $n_{\bf 3}=3$
at high temperatures. One can see in the figure that the analytical formulas,
valid at low temperatures, break down at $T \approx 250\MeV$ when up to four
constituents are included. In order to compare all the representations, we
plot in Fig.~\ref{fig:Lmu} the result of $\tilde{L}_{\mu}/n_\mu$ including all
orders in the expansion in the number of constituents. The value of the
Polyakov loop in the representation $\mu$ tends to $n_\mu$ at high
temperature. Note however that higher representations suffer a stronger
``Polyakov cooling''~\cite{Megias:2004hj}.  A consequence of that is that the
inflexion temperature is shifted towards higher values for representations
with increasing dimensionality. The upward shift of the curves has to do with
the higher masses of the relevant states in higher representations.

\section{Estimates based on the bag model: selected configurations}
\label{sec:3}

In order to obtain an overall estimate of the spectrum of the heavy
hadrons in the sum rule \Eq{2.7} we will use a simplified version of
the MIT bag model \cite{Chodos:1974je,Hasenfratz:1977dt}. We expect
that this approach will provide a picture of the Polyakov loop in
different representations, and in particular, of the scaling rules at
low temperatures. This would be an alternative to the Casimir scaling
assumption, which is justified for temperatures above the crossover to
the deconfining regime, where perturbation theory eventually applies.

Specifically, to obtain the spectrum of $\Delta$ (\Eq{2.7}), we consider
states with zero, one or more quarks, antiquarks and gluons, occupying the
allowed modes in the spherical cavity, and
\begin{eqnarray}
\Delta &=& \min_R\left(
\frac{1}{R}(\sum_\alpha n_\alpha \omega_\alpha - Z_0) + \frac{4\pi}{3}R^3 B \right)
+ \sum_\alpha n_\alpha m_\alpha \nonumber \\
&=&
T_B (\sum_\alpha n_\alpha \omega_\alpha - Z_0)^{3/4} + \sum_\alpha n_\alpha m_\alpha \, 
.
\label{eq:3.1}
\end{eqnarray}
Here $R$ is the bag radius, $Z_0$ a dimensionless parameter for the zero point
energy, $B$ is the bag constant representing the QCD vacuum energy density,
$\omega$ the modes in the spherical cavity and $m$ a mass term for the
constituents (quarks, antiquarks, gluons) which we introduce
additively. $n_\alpha$ is the occupation number of the spin-flavor state $\alpha$.

In the MIT bag, the radius is not fixed but selected in each hadron state by
equating internal and external pressures. This produces the energy scale
\begin{equation}
T_B = \frac{4}{3}(4\pi B)^{1/4}
.
\end{equation}
Note that no center-of-mass corrections are required in $\Delta$, due to the
infinite mass of the source.

In practice we take $Z_0=0$, and adopt $B=(166\MeV)^4$. This
corresponds to $T_B= 416.7\MeV$.  We include quarks with flavors $u$,
$d$ and $s$, and take $m=0$ for $u$ and $d$ quarks, and also for
gluons. For $s$ quarks we take $109\MeV$, although the limiting cases
$m_s=0$ (three chiral flavors) and $m_s=\infty$ (two flavors), are
also occasionally considered. In those limiting cases $B$, or
equivalently $T_B$, is the only scale, so the calculation can be used
directly for other values of $B$.

The modes in the spherical cavity are classified by the radial quantum number
$n=1,2,\ldots$, the angular momentum $j=1/2,3/2,\ldots$ for quarks and
antiquarks and $j=1,2,\ldots$ for gluons, the third component of angular
momentum $m_j$, and the parity label $t= \pm 1$. This is taken such that the
parity is $t (-1)^{j-1/2}$ for quarks, $t (-1)^{j+1/2}$ for antiquarks, and $t
(-1)^j$ for gluons. The allowed values of $\omega$ are
determined by the relations \cite{Bhaduri:1988gc,Barnes:1977hg}:
\begin{eqnarray}
0 &=& j_{j+1/2}(\omega)-j_{j-1/2}(\omega)
 \quad ~~~(t=+1, \quad q~~\mbox{or}~~\bar{q})
,
\nonumber \\
0 &=& j_{j+1/2}(\omega)+j_{j-1/2}(\omega)
\quad ~~~( t=-1, \quad q~~\mbox{or}~~\bar{q})
,
\nonumber \\
0 &=& j_j(\omega)
\qquad\qquad\qquad\qquad 
~~( t=+1, \quad \mbox{gluon})
,
\nonumber \\
0 &=& \frac{j}{j+1}j_{j+1}(\omega) - j_{j-1}(\omega)
\quad (t=-1, \quad \mbox{gluon} )
.
\label{eq:3.10}
\end{eqnarray}
($\omega>0$, and $j_\ell$ is the $\ell$-th spherical Bessel function.) For a
given type of particle, the $\omega$'s depend on the quantum numbers
$(n,j,t)$.  The lowest values of $\omega$ (i.e., $n=1$ and $j=1/2$ or $j=1$)
for the four types of states in \Eq{3.10} are $2.043$, $3.812$, $4.493$, and
$2.744$, respectively.

A MIT bag multiquark investigation, including the stability properties, was
carried out in Ref.~\cite{Zouzou:1993ri}. The non-additivity of the quark
model provides a further attraction for increasing number of quarks.

In general each bag state in the sum \Eq{2.7} is a multiparticle state
composed of quarks, antiquarks and gluons, $q^{n_q}\qb^{n_\qb}g^{n_g}$,
occupying certain bag levels, with color state coupled to the irrep
$\bar{\mu}$. Note that, although the MIT bag is not strictly an independent
particle system, the construction of Secs. \ref{sec:3.B}, \ref{sec:3.C},
and \ref{sec:3.D} apply equally well in this case.

Numerical results for this section (MIT bag model and few constituents) are
displayed in Appendix~\ref{app:C} together with the lowest bag states.

\section{Estimates based on the bag model: all configurations}
\label{sec:4}

In this section we address the bag model calculation trying to include all
possible sets of constituents. The goal is to sum up all possible fillings of
irreducible type of the bag levels with constituents coupled to a given color
irrep $\bar{\mu}$. Although in principle this is an improvement over the
previous section, two problems appear: in this approach it is not practical to
obtain separately the contributions of different sets of constituents, and
more importantly, it is not obvious how to isolate the irreducible color
configurations, i.e., the confining ones. So in this approach we have to
content ourselves by obtaining some estimates.

In addition, in the MIT bag, the number of multiparticle states increases so
quickly with the energy that the partition function diverges beyond a certain
Hagedorn temperature \cite{Hagedorn:1965st}, as discussed e.g. in
\cite{Kapusta:1981ay}. The Hagedorn temperature, $T_H$, is lowered (and so
more restrictive) as the number of species in the bag increases. We discuss
this effect below.

\subsection{The fixed-radius bag system}
\label{sec:4.A}

If $H$ were the Hamiltonian of non-interacting particles, it would be
immediate to express $Z$ in terms of sums over single particle states. The MIT
bag introduces an attractive interaction through the rearrangement of the bag
radius $R$, but its Hamiltonian has a particularly simple form which can be
exploited by transforming it into an independent particle model looking
form. To do so, we introduce an auxiliary bag system with a cavity of fixed
radius $R$.  The numerical value of $R$ is not relevant since this quantity is
eliminated in the final result for the MIT bag. In addition, we introduce the
following two non-interacting Hamiltonians (which obviously commute with each
other)
\begin{equation}
H^\prime_k = \frac{1}{R}\sum_{\alpha,c} \omega_\alpha a_{\alpha,c}^\dagger a_{\alpha,c}
,
\qquad
H_m =  \sum_{\alpha,c} m_\alpha  a_{\alpha,c}^\dagger a_{\alpha,c}
.
\label{eq:3.1a}
\end{equation}
Here $\alpha$ and $c$ indicate the spin-flavor and color labels and
$a_{\alpha,c}$ is the corresponding annihilation operator. To distinguish them
from the MIT bag quantities, in what follows we use a prime to indicate
quantities related to this auxiliary {\em fixed-radius bag} system.

In terms of these auxiliary one-body Hamiltonians, the Hamiltonian of the MIT
bag model (i.e., with relaxation of $R$ to its most stable value) that
reproduces \Eq{3.1}, takes the form
\begin{equation}
H = T_B (R H^\prime_k - Z_0)^{3/4} + H_m
.
\label{eq:3.1b}
\end{equation}

To proceed further let $f_{\frac{3}{4}}(t)$ denote the inverse Laplace
transform of $e^{-s^{3/4}}$ (see Appendix~\ref{app:B} for details). This
allows us to write the relation
\begin{equation}
Z(\Omega,T) = 
\int_0^\infty d\beta^\prime
\bar{f}(\beta^\prime,\beta)
Z^\prime(\Omega,T^\prime,T) \,,
\label{eq:4.7}
\end{equation}
where $T^\prime=1/\beta^\prime$,
\begin{equation}
\bar{f}(\beta^\prime,\beta) = 
\frac{e^{\beta^\prime Z_0/R} }{R (\beta T_B)^{4/3}}
f_{\frac{3}{4}}\left(\frac{\beta^\prime}{R (\beta T_B)^{4/3}}\right) \,,
\end{equation}
and
\begin{equation}
Z^\prime(\Omega,T^\prime,T) = \Tr(e^{-\beta^\prime H^\prime_k - \beta H_m} 
\,U(\Omega)
).
\end{equation}

$Z^\prime(\Omega,T^\prime,T)$ describes the partition function of a
non-interacting system (with two temperatures), namely, that of a bag of fixed
radius $R$. This systems does not have a Hagedorn temperature. Of course, this
partition function can also be analyzed in terms of $\SU(3)$ irreps (similar
to \Eq{4.4}) to give $Z^\prime_\mu(T^\prime,T)$, and it follows that
\begin{equation}
Z_\mu(T) = 
\int_0^\infty d\beta^\prime 
\bar{f}(\beta^\prime,\beta)
Z^\prime_\mu(T^\prime,T)
.
\label{eq:4.7b}
\end{equation}

The virtue of $H^\prime_k$ and $H_m$ is that they are one-body operators,
leading to a non-interacting system (more precisely, interacting just with an
external potential) and this is not spoiled by the presence of $U(\Omega)$
(which is just equivalent to minimal coupling).  Thus, using the expressions in
terms of $T$ and $T^\prime$ we may profit from the results of section
\ref{sec:3.E} in a rather straightforward manner.  In particular, the
partition function of the fixed-radius system fulfills the factorization
property as in \Eq{4.13},
\begin{equation}
Z^\prime = Z^\prime_q Z^\prime_\qb Z^\prime_g \, , 
\end{equation}
where these partition functions are given now (compare with \Eq{4.14}) by
\begin{eqnarray}
\log Z^\prime_q(\Omega,T^\prime,T)
&=&
\sum_\alpha\!{}^q  \, \gamma_\alpha \,\chi_{\bf 3} \!\left(
\log (1 + \Omega \, e^{-\beta^\prime \omega_\alpha/R - \beta m_\alpha } )
\right)
,
\nonumber \\
\log Z^\prime_\qb(\Omega,T^\prime,T)
&=&
\sum_\alpha\!{}^\qb  \, \gamma_\alpha \,\chi_{\bf \bar{3}} \!\left(
\log (1 + \Omega \, e^{-\beta^\prime \omega_\alpha/R - \beta m_\alpha } )
\right)
,
\nonumber \\
\log Z^\prime_g(\Omega,T^\prime,T)
&=&
- \sum_\alpha\!{}^g \, \gamma_\alpha \,\chi_{\bf 8} \!\left(
\log (1 - \Omega \,e^{-\beta^\prime \omega_\alpha/R  - \beta m_\alpha } )
\right) \, . \nonumber \\ 
\end{eqnarray}
Likewise, the following symmetry properties hold 
\begin{equation}
Z^\prime_\qb = (Z^\prime_q)^*
,\qquad
Z^\prime_g = (Z^\prime_g)^*
.
\end{equation}
Similarly,  we can introduce the auxiliary $\Omega$-independent functions
\begin{eqnarray}
z^\prime_q(T^\prime,T) 
&=&
\sum_\alpha\!{}^q \, \gamma_\alpha \, e^{-\beta^\prime \omega_\alpha/R - \beta m_\alpha }
,
\nonumber \\
z^\prime_g(T^\prime,T) 
&=&
\sum_\alpha\!{}^g \, \gamma_\alpha \, e^{-\beta^\prime \omega_\alpha/R - \beta m_\alpha }
,
\end{eqnarray}
and the relation corresponding to \Eq{4.17} becomes
\begin{eqnarray}
\log Z^\prime_q(\Omega,T^\prime,T)
&=&
\sum_{n=1}^\infty \frac{(-1)^{n+1}}{n} \,\chi_{\bf 3}(\Omega^n)
\, z^\prime_q(T^\prime/n,T/n)
,
\nonumber \\
\log Z^\prime_g(\Omega,T^\prime,T)
&=&
\sum_{n=1}^\infty \frac{1}{n} \,\chi_{\bf 8}(\Omega^n) \, z^\prime_g(T^\prime/n,T/n)
.
\end{eqnarray}

The antifundamental and adjoint characters can be reduced to the fundamental
one using \Eq{4.18}. The analogous of \Eq{4.19} is then 
\begin{eqnarray}
\log \hat{Z}^\prime_q(\omega,T^\prime,T)
&=&
\sum_{n=1}^\infty \frac{(-1)^{n+1}}{n} \,\omega^n
\, z^\prime_q(T^\prime/n,T/n)
,
\nonumber \\
\log \hat{Z}^\prime_g(\omega,T^\prime,T)
&=&
\sum_{n=1}^\infty \frac{1}{n} \,\omega^n \, z^\prime_g(T^\prime/n,T/n)
.
\label{eq:4.19a}
\end{eqnarray}
Finally, 
\begin{eqnarray}
\log Z^\prime_q(\Omega,T^\prime,T)
&=&
\sum_{i=1}^3 \log \hat{Z}^\prime_q(\omega_i,T^\prime,T)
,
\nonumber \\
\log Z^\prime_g(\Omega,T^\prime,T)
&=&
\sum_{i,j=1}^3 
\log \hat{Z}^\prime_g(\omega_i\omega_j^*,T^\prime,T) 
- \log \hat{Z}^\prime_g(1,T^\prime,T) \, .\nonumber \\ 
\label{eq:4.20}
\end{eqnarray}

Before going any further, let us give more details on the functions
$z^\prime_q$ and $z^\prime_g$. If we assume $N_f$ massless flavors, and
$N_f^\prime$ degenerated flavors with mass $m_s$, as well as massless gluons,
we can write
\begin{eqnarray}
z^\prime_q(T^\prime,T) 
&=&
(N_f+N^\prime_f e^{-\beta m_s})\sum_\alpha\!{}^q \, \gamma_\alpha \, e^{-\beta^\prime \omega_\alpha/R }
,
\nonumber \\
z^\prime_g(T^\prime,T) 
&=&
\sum_\alpha\!{}^g \, \gamma_\alpha \, e^{-\beta^\prime \omega_\alpha/R }
.
\label{eq:4.21}
\end{eqnarray}
Here $\sum_\alpha^q$ no longer includes flavor. For both, quarks and gluons,
$\alpha=(n,j,t)$ and $\gamma_\alpha=2j+1$.  So numerically, we need to compute
functions of a single variable $T^\prime$.

The sums over levels are not finite. To deal with this problem, say for
gluons, we separate the level sums as
\begin{equation}
z_g^\prime(T^\prime) 
=
(\sum_{\omega_\alpha<\omega_{\rm max}}^{}
+ 
\sum_{\omega_\alpha \ge \omega_{\rm max}}^{}
)
\, \gamma_\alpha \, e^{-\beta^\prime \omega_\alpha/R }
:=
z^\prime_<(T^\prime) + z^\prime_>(T^\prime)
.
\end{equation}
The finite sum $z^\prime_<(T^\prime)$ is done numerically. On the other hand,
$z^\prime_>(T^\prime)$ is approximated using the asymptotic form of the
spectrum. As is very well-known from the black-body problem, the cumulative
number of single-particle states in a cavity of volume $V$ is
$N_1(E)=VE^3/(3\pi^2)$ plus terms that are subdominant for large $E$. We have
included a spin degeneracy factor of $2$ valid for both quarks and gluons. For
our spherical bag of radius $R$, this gives for the cumulative number of
states and the corresponding density of single-particle levels in the large
$E$ region
\begin{equation}
N_1(E) \approx \frac{4}{9\pi} R^3 E^3, \quad
\rho_1(E) \approx \frac{4}{3\pi} R^3 E^2
.
\label{eq:4.24}
\end{equation}
This allows us to write
\begin{eqnarray}
z^\prime_>(T^\prime) &=& \int_{\omega_{\rm max}/R}^{+\infty} dE \rho_1(E) \,
e^{-\beta^\prime E}
\\ 
&\approx&
\frac{4}{3\pi} R^3 T^\prime \left[
\left( \frac{\omega_{\rm max}}{R} + T^\prime \right)^2 + T^\prime{}^2
\right] e^{-\beta^\prime \omega_{\rm max}/R}
.
\nonumber
\end{eqnarray}
This asymptotic expression of $z^\prime_>(T^\prime)$ holds both for quarks and
gluons. In our calculation we have taken $\omega_{\rm max} = 100$, although a
much smaller value would probably be sufficient. We have verified that
$z^\prime_{q,g}(T^\prime,T)$ are not sensitive to this particular choice of
$\omega_{\rm max}$.

Although certainly possible, it is technically more difficult to deal with
functions of several variables, so in what follows we consider just $N_f$
massless quark flavors and $N^\prime_f=0$ in \Eq{4.21}. The choices $N_f=2$
($N_f=3$) would correspond to taking the limit of a heavy (light) strange
quark. $N_f=0$ corresponds to gluodynamics. As a consequence, the dependence
on $T$ disappears in the various primed functions. In addition, $R$ or $T_B$
become the only scale, for fixed-radius bag and MIT bag systems, respectively.

\begin{figure}[t]
\begin{center}
\epsfig{figure=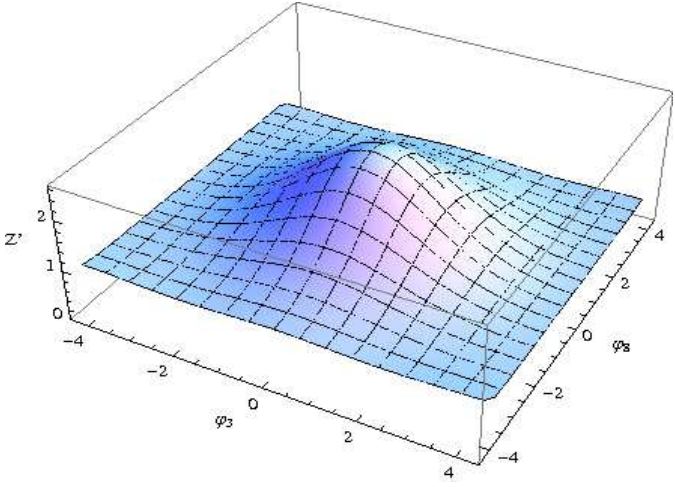,height=60mm,width=85mm}
\end{center}
\caption{The function $Z^\prime(\Omega,T^\prime)$ on the plane
  $(\varphi_3,\varphi_8)$, for $N_f=3$ and $R T^\prime=1/2$.}
\label{fig:9nb1}
\end{figure}

\begin{figure}[t]
\begin{center}
\epsfig{figure=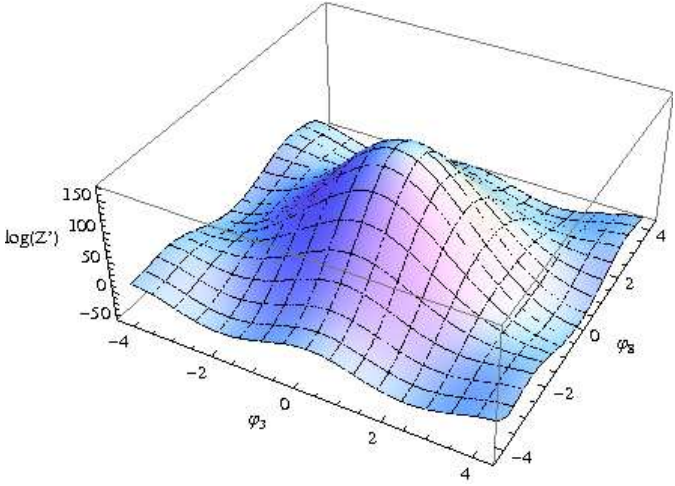,height=60mm,width=85mm}
\end{center}
\caption{The function $\log Z^\prime(\Omega,T^\prime)$ on the plane
  $(\varphi_3,\varphi_8)$, for $N_f=3$ and $R T^\prime=2$.}
\label{fig:9nb2}
\end{figure}

In the fixed bag-radius problem the partition function converges for all
temperatures $T^\prime$, nevertheless, this function increases rather quickly
with the temperature. In Fig. \ref{fig:9nb1} we display
$Z^\prime(\Omega,T^\prime)$ as a function of $\Omega$ (in the plane
$(\varphi_3,\varphi_8)$) for $N_f=3$ and $R T^\prime = 0.5$.  As shown in
\Eq{4.4}, the central value $Z^\prime(\Omega=1,T^\prime)$ is just the full
partition function, adding all irreps. In Fig. \ref{fig:9nb2} we display the
logarithm of the same function, this time for $R T^\prime = 2$. The direct plot
of the partition function would look similar to a two-dimensional Dirac delta
distribution centered at $\Omega=1$ in the plane $(\varphi_3,\varphi_8)$, with
a certain normalization. This suggests that in the large-$T^\prime$ limit,
we would have
\begin{eqnarray}
\frac{Z^\prime(\Omega,T^\prime)}{ \int d\Omega^\prime 
Z^\prime(\Omega^\prime,T^\prime)}&=&
\frac{1}{ Z^\prime_{\bf 1}(T^\prime) }
\sum_\mu \chi^*_\mu(\Omega) 
Z^\prime_\mu(T^\prime) \nonumber \\ 
&\underset{T^\prime\to\infty}{\sim}&
\delta(\Omega,1) = \sum_\mu n_\mu \chi^*_\mu(\Omega)
.
\end{eqnarray}
Therefore, in the large temperature limit
\begin{equation}
\frac{ Z^\prime_\mu(T^\prime) }{ Z^\prime_{\bf 1}(T^\prime) }
\sim n_\mu
.
\label{eq:4.31}
\end{equation}
\begin{figure}[t]
\begin{center}
\epsfig{figure=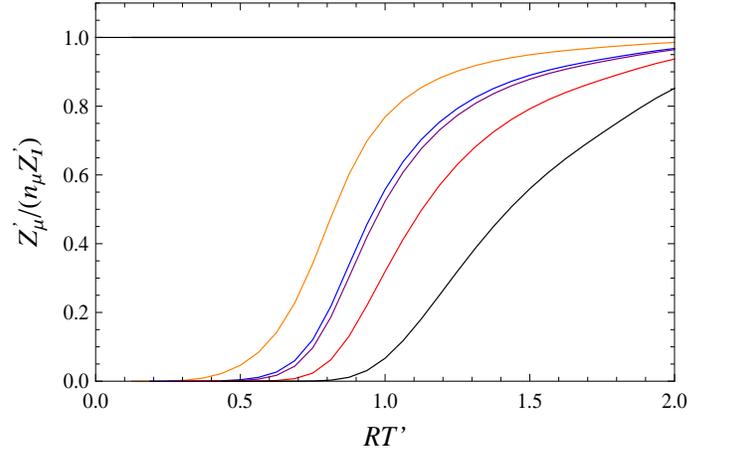,height=60mm,width=90mm}
\end{center}
\caption{Plot of $Z^\prime_\mu(T^\prime)/(n_\mu Z^\prime_{\bf 1}(T^\prime))$
  as a function of $R T^\prime$ for several irreps. From top to bottom, $\mu=
  {\bf 1}$, ${\bf 3}$, ${\bf 8}$, ${\bf 6}$, ${\bf 10}$, and ${\bf 64}$.  Note
  that the ratio $Z^\prime_\mu/Z^\prime_{\bf 1}$ is just the estimate
  $\tilde{L}^\prime_\mu$ introduced below, in \Eq{4.49}.  }
\label{fig:9nb4}
\end{figure}
This conjecture is indeed supported by the numerical results shown in
Fig. \ref{fig:9nb4}. The interpretation is that of an equipartition of
populations between all available modes, in the large temperature limit.
\footnote{The discussion always refers to a bag with quarks and antiquarks, in
  addition to gluons. In a bag with just gluons, i.e. $N_f=0$, the functions
  $Z^\prime_\mu(T^\prime,T)$ and $Z_\mu(T)$ would vanish identically for
  triality non-trivial irreps and in this case the equipartition would take
  place on the triality preserving irreps. In the fixed-radius bag there is no
  phase transition and so no spontaneous breaking of the center
  symmetry. Hadron resonance gas estimates, as considered in this work, refer
  always to the confined phase.}

\begin{figure}[t]
\begin{center}
\epsfig{figure=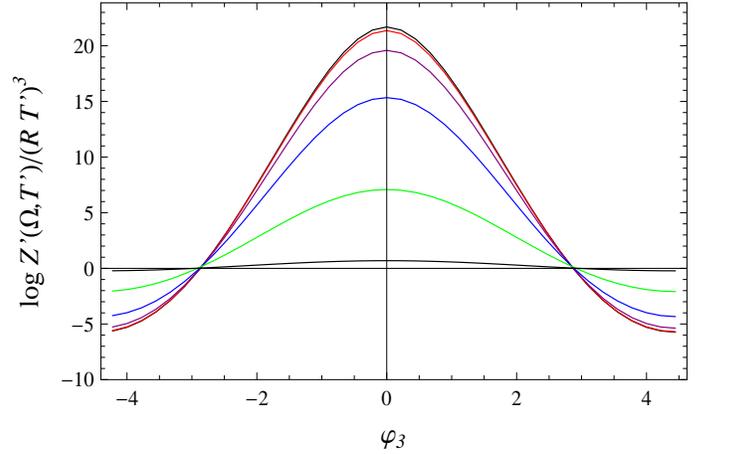,height=60mm,width=90mm}
\end{center}
\caption{Plots of $(\log Z^\prime(\Omega,T^\prime))/(R T^\prime)^3$ as a
  function of $\varphi_3$ for $\varphi_8=0$, for $N_f=3$ and several
  temperatures: from bottom to top $R T^\prime = 0.25,0.5,1,2,5,10$.  The
  theoretical large-$T^\prime$ limit at the maximum is $19(\pi/3)^3 =21.82$.
Remarkably the value of $\varphi_3$ for which
$Z^\prime(\Omega,T^\prime)=1$ turns out to be very similar (although not exactly
equal) for all temperatures, namely, $\varphi_3=2.9$. For gluons and for
quarks a similar effect takes place at $\varphi_3=2.2$ and $2.99$, respectively.
}
\label{fig:9nb3}
\end{figure}
The function $Z^\prime(\Omega,T^\prime)$ increases very rapidly with the
temperature and this directly causes the existence of a Hagedorn
temperature for $Z(\Omega,T)$.  In order to investigate whether the same
critical temperature exists for the functions $Z_\mu(T)$ we need to see the
large temperature behavior of the $Z^\prime_\mu(T^\prime)$. To this end, we
display in Fig. \ref{fig:9nb3} the ratio $\log Z^\prime(\Omega,T^\prime)$ over
$(R T^\prime)^3$, on the $\varphi_3$ axis (the plot on the $\varphi_8$
axis is qualitatively similar), for several temperatures. The plot suggests
that the ratio tends to a fixed limiting profile $P(\Omega)$
\begin{equation}
P(\Omega)  = \lim_{T^\prime\to \infty} \frac{\log Z^\prime(\Omega,T^\prime)}{(R
  T^\prime)^3}
.
\end{equation}
This function has its absolute maximum at $\Omega=1$. (Obvious modifications
apply when $N_f=0$, due to center symmetry.) The existence and shape of the
limiting function $P(\Omega)$ implies that the height of
$Z^\prime(\Omega,T^\prime)$ increases exponentially as $\exp(\kappa (R
T^\prime)^3)$ (with $\kappa=P(1)$) while its width in the $\Omega$ plane
decreases as a power, $1/(R T^\prime)^{3/2}$.  Therefore the integral
$Z^\prime_{\bf 1}(T^\prime)$, as well as all other $Z^\prime_\mu(T^\prime)$,
grow exponentially at a common rate $\exp(\kappa (R T^\prime)^3)$. As a
consequence the Hagedorn temperature is common to all irreps. The argument
also shows that the total number of {\em active} irreps at a temperature
$T^\prime$ grows at a rate $(R T^\prime)^3$.

\subsection{Hagedorn temperature}

The existence of a critical temperature, $T_H$, in the MIT bag was noted very
early \cite{Chodos:1974je}. In order to determine the value of $T_H$, we first
obtain the asymptotic form of $Z^\prime(T^\prime)$ at large temperatures. This
can be done by using the asymptotic form of the single-particle level density
$\rho_1(E)$ in \Eq{4.24}. This yields $z^\prime_g(T^\prime)\sim (8/3\pi)(R
T^\prime)^3$ for gluons, with an extra factor $N_f$ for quarks. These
expressions can be inserted in \Eq{4.19a} (here it enters the Riemann $\zeta$
function) and then in \Eq{4.20} for $\Omega=1$.  Adding the contributions from
quarks, antiquarks and gluons, this finally produces
\begin{eqnarray}
&&
\log Z^\prime(T^\prime)
\sim
\kappa \, R^3 T^\prime{}^3
,
\nonumber \\ 
&&
\kappa = 
\frac{\pi^3}{135}
(7 N_c N_f + 4(N_c^2-1) ) 
,
\label{eq:4.33}
\end{eqnarray}
where $N_c=3$ is the number of colors. For $N_f=3$, this value of $\kappa$
reproduces the large-$T^\prime$ limit at the maximum in
Fig. \ref{fig:9nb3}. Of course, the asymptotic form in \Eq{4.33} is nothing
else than the well-known result for the partition function of a gas of free
massless particles in a volume $V=(4\pi/3)R^3$, and we have just checked the
consistency of our formulas.

The parameter $\kappa$ controls the Hagedorn temperature. Let
$\rho^\prime(E^\prime)$ denote the multiparticle density of energy levels in
the fixed-radius bag system,
\begin{equation}
Z^\prime(T^\prime) = \int_0^\infty dE^\prime \, \rho^\prime(E^\prime) \, 
e^{-\beta^\prime E^\prime }
.
\end{equation}
The behavior in the asymptotic regime of large energies and temperatures can
be obtained by using a saddle point approximation in the integral together
with \Eq{4.33}. This gives, for the cumulative number of states,
\begin{equation}
N^\prime(E^\prime) \sim e^{\frac{4}{3}(3\kappa)^{1/4} (R E^\prime)^{3/4}}
.
\end{equation}

The MIT bag and fixed-radius bag Hamiltonians are functionally related,
namely, $H=T_B (R H^\prime_k)^{3/4}$. The same functional relation follows for
the eigenvalues, $E = T_B (R E^\prime)^{3/4}$, while degeneracies are
unchanged. It follows that the cumulative number of states transforms as a
scalar quantity, $N(E)= N^\prime(E^\prime)$. This immediately implies
\begin{equation}
N(E) \sim e^{E/T_H}
,
\qquad
T_H = \frac{3}{4}\frac{T_B}{(3\kappa)^{1/4}}
=
\left(\frac{4\pi B}{3\kappa}\right)^{1/4}
.
\end{equation}
This coincides with standard result \cite{Kapusta:1981ay}.  Here we see that
the effective attraction between constituents, leading to rearrangement of the
radius of the bag, produces a slide in the multi-particle spectrum. Higher
states are brought down in such a way that $N^\prime(E^\prime)$ goes into
$N(E)$.

Similar relations hold when $\Omega$ is present. As for the partition
function, one can introduce the cumulative number of states and density of
levels,
\begin{eqnarray}
N(\Omega,E) &=& \Tr(\Theta(E-H)U(\Omega))
,
\nonumber \\
\rho(\Omega,E) &=& \Tr(\delta(E-H)U(\Omega))
,
\end{eqnarray}
and analogously for the primed system. Likewise, the scalar transformation
property still holds in the presence of $\Omega$,
\begin{equation}
N(\Omega,E) = N^\prime(\Omega,E^\prime)
.
\end{equation}

As discussed previously, the $Z^\prime_\mu(T^\prime)$ all grow exponentially
with a common rate $e^{\kappa(R T^\prime)^3}$, and this implies a common
limiting temperature equal to $T_H$ for all the $Z_\mu(T)$. The situation for
$Z(\Omega,T)$ is slightly different.  $Z^\prime(\Omega,T^\prime)$ grows at an
$\Omega$-dependent rate $e^{ P(\Omega) (R T^\prime)^3}$, so in this case we
expect also an $\Omega$-dependent critical temperature
\begin{equation}
T_c(\Omega) = (\kappa/P(\Omega))^{1/4} T_H
\ge T_H
.
\end{equation}
This critical temperature coincides with $T_H$ when $\Omega$ is the identity
element and increases as $\Omega$ departs from it.

In order to obtain the partition function $Z(\Omega,T)$ itself, we have to
apply \Eq{4.7}. The small-$t$ asymptotic form of $f_{\frac{3}{4}}(t)$, \Eq{A.1},
implies the following large-$T^\prime$ behavior
\begin{equation}
\bar{f}(\beta^\prime,\beta) \sim 
\frac{1}{R}\left(\frac{T}{T_B}\right)^{4/3}
e^{-\frac{1}{3} (R T^\prime)^3 \left(\frac{3}{4}\frac{T_B}{T}\right)^4 }
.
\end{equation}
From the asymptotic form of $Z^\prime(\Omega,T^\prime)$ it follows that the
$\beta^\prime$ integral in \Eq{4.7} diverges at small $\beta^\prime$ when
$T \ge T_c(\Omega)$, as expected.

\begin{figure}[t]
\begin{center}
\epsfig{figure=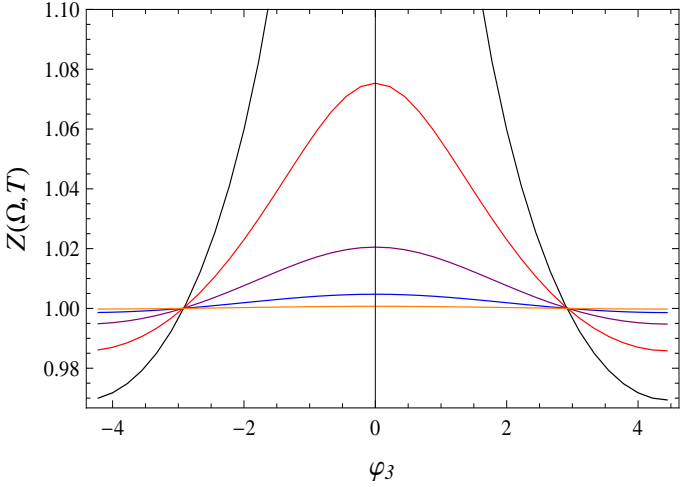,height=65mm,width=90mm}
\end{center}
\caption{The function $Z(\Omega,T)$ on the $\varphi_3$ axis, for $N_f=3$ and
  for several temperatures. From bottom to top, $T/T_B = 0.15625$, $0.1875$,
  $0.21875$, $0.25$, and $0.28125$. The last temperature is above $T_H$ and $Z$
  diverges in a neighborhood of $\varphi_3=0$.}
\label{fig:9nb5}
\end{figure}
For $N_f=3$ (our preferred case) $T_H = 0.2637\, T_B$, so we cannot consider
temperatures larger that this in the calculation of $Z_\mu(T)$. The form of
$Z(\Omega,T)$ at $T=0.25 T_B$, for $N_f=3$ is displayed in
Fig. \ref{fig:9nb5}.

Before ending this section, we briefly mention the large temperature
behavior of the quark constituent model. The asymptotic forms discussed for
the fixed-radius bag also apply for the quark constituent model by changing
$V=(4\pi/3)R^3$ to $V=V_\sigma$. E.g.  $N_1(E) \sim V_\sigma E^3/(3\pi^2)$ at
large energies (and this is consistent with the form of $J(0,T)$ in
\Eq{4.9}). Consequently, at large temperatures
\begin{equation}
\log Z_{\rm PCM}(1,T) 
\sim
\frac{3}{4\pi} \kappa V_\sigma T^3
,
\end{equation}
where $\kappa$ is that of \Eq{4.33}. The full dependence on the temperature
requires to specify how $V_\sigma$ depends on $T$. An increasing function
would imply a steeper dependence of $Z_{\rm PCM}$ on $T$, as compared to the
fixed-radius cavity, and so effectively to attraction between constituents.

In general it should be expected that the constituent model behaves as the
fixed-radius bag with volume $V_\sigma$. This is specially so for $T$ larger
than $R^{-1}$ and $M$, since then the discrete spectrum of the cavity can
be approximated by a continuous semiclassical one and the mass (not present in
the bag) can also be disregarded.

\subsection{Irreducible color configurations}
\label{sec:4.C}

All the previous considerations refer to the partition functions
$Z_\mu(T)$. These functions include all types of configurations, reducible and
irreducible, as defined in Sec. \ref{sec:3.C}. We can introduce the
corresponding irreducible version $L_\mu(T)$, which only includes the
irreducible color configurations. $L_\mu(T)$ is the bag estimate of the
Polyakov loop in the irrep $\mu$, introduced in \Eq{2.7a}. Using similar
constructions to those in \Eq{4.1} and \Eq{4.4}, one can define $L(T)$ (adding
up all $\SU(3)$ irreps) and $L(\Omega,T)$. (Note that those relations are all
linear.) Furthermore, we can also consider the fixed-radius bag versions,
$L^\prime_\mu(T^\prime,T)$, $L^\prime(T^\prime,T)$, and
$L^\prime(\Omega,T^\prime,T)$. They will fulfill the corresponding versions of
Eqs.~(\ref{eq:4.7}) and (\ref{eq:4.7b}):
\begin{eqnarray}
L(\Omega,T) &=& 
\int_0^\infty d\beta^\prime 
\bar{f}(\beta^\prime,\beta)
L^\prime(\Omega,T^\prime,T)
,
\nonumber \\
L_\mu(T) &=& 
\int_0^\infty d\beta^\prime 
\bar{f}(\beta^\prime,\beta)
L^\prime_\mu(T^\prime,T)
.
\label{eq:4.41}
\end{eqnarray}

The isolation of the irreducible configurations to obtain $L_\mu(T)$ or
$L^\prime_\mu(T^\prime,T)$ can be done for each given finite set of
constituents. Unfortunately, we have not devised a method to do so when all
such sets are taken together, i.e., in the form of a generating functional
producing just the irreducible color configurations. In view of that, we take
$Z$ (including all color configurations, irreducible or not) and $\tilde{L}$
(the ratio $Z/Z_{\bf 1}$) as upper and lower estimates of $L$.

The estimate of irreducible terms by the ratio $Z/Z_{\bf 1}$ makes sense (and
it is exact to three constituents) only in the {\em non interacting case}, and
so in the fixed-radius bag problem. So we define
\begin{equation}
\tilde{L}^\prime_\mu(T^\prime,T) :=
\frac{Z^\prime_\mu(T^\prime,T)}{Z^\prime_{\bf 1}(T^\prime,T)} ,
\label{eq:4.49}
\end{equation}
and use it as a lower bound or estimate on the true
$L^\prime_\mu(T^\prime,T)$:
\begin{equation}
\tilde{L}^\prime_\mu(T^\prime,T)
\underset{\sim}{<}
L^\prime_\mu(T^\prime,T)
\underset{\sim}{<}
Z^\prime_\mu(T^\prime,T)
.
\end{equation}
From $\tilde{L}^\prime_\mu(T^\prime,T)$ we then obtain a lower estimate 
$\tilde{L}_\mu(T)$ in the MIT bag by applying \Eq{4.41}.

It is noteworthy that the lower estimates become exact in the low temperature
limit: in this limit only the states with fewest constituents have a chance to
get populated.

We do not know how $L_\mu(T)$ behaves as the temperature increases. However,
using \Eq{4.41}, setting $Z_0=N_f^\prime=0$ in the bag parameters as before,
it follows that
\begin{equation}
L_\mu(T) = \int_0^\infty d\beta^\prime x f_{\frac{3}{4}}(x\beta^\prime) 
L^\prime_\mu(T^\prime)
,
\qquad
x= \left(\frac{T}{T_B}\right)^{4/3}
.
\end{equation}
As $T \to \infty$, $x f_{\frac{3}{4}}(x\beta^\prime)$ becomes a normalized
Dirac delta at $\beta^\prime=0$, and therefore
\begin{equation}
\lim_{T\to \infty} L_\mu(T)
=
\lim_{T^\prime \to \infty} L^\prime_\mu(T^\prime)
.
\label{eq:4.51}
\end{equation}

As discussed above, the upper estimate $Z^\prime_\mu(T^\prime,T)$ increases
exponentially with the temperature, as $e^{\kappa (R T^\prime)^3}$, and
produces a divergence in the upper estimate of $L_\mu(T)$ for temperatures
beyond $T_H$.

On the other hand, from \Eq{4.31} and Fig. \ref{fig:9nb4}, the lower estimate
$\tilde{L}^\prime_\mu(T^\prime,T)$ tends to a constant value $n_\mu$.
Therefore, \Eq{4.51} implies for the MIT bag
\begin{equation}
\lim_{T\to \infty} \tilde{L}_\mu(T)
= n_\mu
.
\end{equation}
This result is remarkable because it indeed coincides with the exact
QCD result: for temperatures above the crossover, in the perturbative
regime, the Polyakov loop tends to the identity element and so
$L_{{\rm QCD},\mu} \to n_\mu$~\footnote{Appropriate modifications are
  to be noted in the case of gluodynamics, to account for the
  additional center symmetry.}. Nevertheless, there is no sound reason
to believe that $\tilde{L}_\mu(T)$ should be a valid estimate of
$L_{{\rm QCD},\mu}(T)$ at all temperatures. Certainly the true
$L_\mu(T)$ (irreducible configurations in the MIT bag) increases
without bound with $T$: even the simplest configuration $Q\bar{q}$
(heavy meson in $L_{\bf 3}$) contains an ever increasing set of single
particle states that are activated as $T$ grows. What is less clear is
whether $L_\mu(T)$ has a Hagedorn temperature, as $Z_\mu(T)$, since to
settle that point requires to estimate the growth of the number of
irreducible color configurations as compared to the total ones. As
already noted, estimates based on the hadron resonance model are not
reliable at temperatures above the QCD crossover.

\begin{figure}[t]
\begin{center}
\epsfig{figure=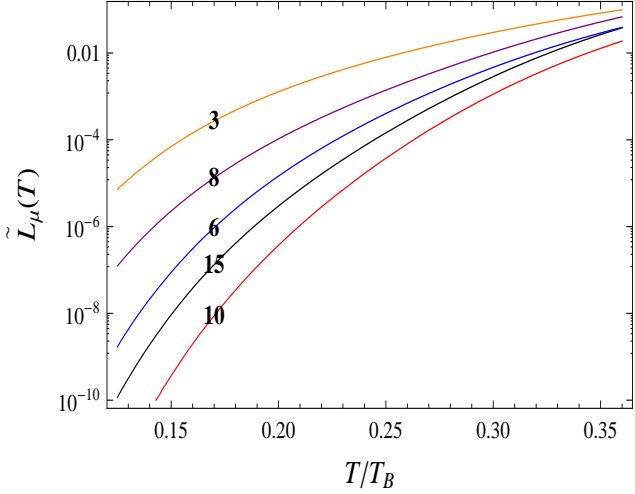,height=65mm,width=85mm}
\end{center}
\caption{$\tilde{L}_\mu(T)$ as a function of $T/T_B$, for several irreps. From
  top to bottom $\mu= {\bf 3}$, ${\bf 8}$, ${\bf 6}$, ${\bf 15}$, ${\bf 10}$.} 
\label{fig:9nb7}
\end{figure}

\begin{figure}[t]
\begin{center}
\epsfig{figure=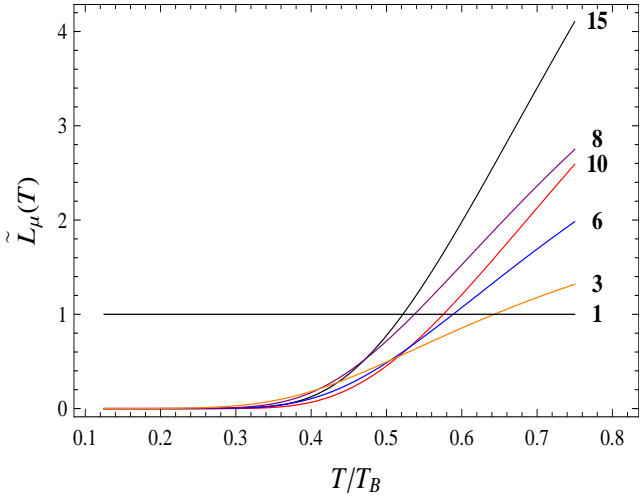,height=65mm,width=85mm}
\end{center}
\caption{As Fig. \ref{fig:9nb7} (in non logarithmic scale) for a wider range
  of temperatures.  From bottom to top $\mu= {\bf 1}$, ${\bf 3}$, ${\bf 6}$,
  ${\bf 10}$, ${\bf 8}$, ${\bf 15}$. The asymptotic value is $n_\mu$.}
\label{fig:9nb6}
\end{figure}

Finally, in Figs.~\ref{fig:9nb7} and \ref{fig:9nb6} we display the
behavior of $\tilde{L}_\mu(T)$ for several irreps below the QCD
crossover.

\section{Polyakov loop scaling in single-particle models at low temperatures}
\label{sec:cqm.c}

At temperatures well below the crossover, the Boltzmann weight factor
becomes determinant in selecting the dominant mechanisms contributing
to the expectation value of the Polyakov loop in each irrep. In
principle, this is very different from a Casimir scaling behavior,
based on perturbative estimates, although in both cases higher irreps
tend to be suppressed.

Sources carrying higher representations require a larger number of
constituents for its screening, and this number gives the leading term in the
low temperature scaling. From Table \ref{tab:1}, in gluodynamics this counting
gives (up to proportionality constants)
\begin{equation}
L_{\bf 10} \sim L_{\bf 27} \sim L_{\bf 8}^2,
\qquad
L_{\bf 35} \sim L_{\bf 64} \sim L_{\bf 8}^3
.
\label{eq:5.27}
\end{equation}
On the other hand, in QCD and assuming gluonic modes to be heavy, one
obtains
\begin{equation}
L_{\bf 6} \sim L_{\bf 8} \sim L_{\bf 3}^2,
\qquad
L_{\bf 10} \sim L_{\bf 15} \sim L_{\bf 3}^3,
\qquad
L_{\bf 15^\prime} \sim L_{\bf 24} \sim L_{\bf 3}^4
.
\label{eq:5.28}
\end{equation}

The previous scalings assume that a purely single-particle description is
appropriate. In the MIT bag the energy is not strictly additive and deviations
take place. Specifically, in gluodynamics the lightest mode is $1^+$ with
$\omega_g=2.744$. Due to the symmetry of the gluon wavefunction, the
degeneracies of $\mu={\bf 8}$, ${\bf 10}$, ${\bf 27}$, ${\bf 35}$, and ${\bf
  64}$ are, respectively, $3$, $3$, $6$, $8$, and $10$ for the corresponding
lightest states. Therefore, at low temperatures the model produces the
following relations:
\begin{eqnarray}
&&
\log(\frac{1}{3}L_{\bf 8}) \sim 2^{-3/4}\log(\frac{1}{3}L_{\bf 10})
\sim 3^{-3/4}\log(\frac{1}{8}L_{\bf 35})
,\nonumber \\
&&
L_{\bf 27} \sim 2 L_{\bf 10}
,\qquad
L_{\bf 64} \sim \frac{4}{5} L_{\bf 35}
. 
\end{eqnarray}

In QCD, the lightest quark mode is $\frac{1}{2}^+$, with $\omega_q=2.043$. For
$\mu = {\bf 3}$ and ${\bf 6}$, one and two quark configurations, respectively,
are dominant, while the one gluon configurations still dominates in the ${\bf
  8}$, so
\begin{eqnarray}
\log(\frac{1}{2N_f}L_{\bf 3})
&\sim& 2^{-3/4} \log(\frac{1}{N_f(2N_f-1)}L_{\bf 6})
\nonumber \\
&\sim& (\omega_q/\omega_g)^{3/4} \log(\frac{1}{3}L_{\bf 8})
.
\end{eqnarray}

Coming back to single-particle models, it is clear from their derivation, that
all the expansions presented in Sec. \ref{sec:cqm.b} for $Z_\mu(T)$,
$\tilde{L}_\mu(T)$ or $L_\mu(T)$, (Eqs. (\ref{eq:Z1}), (\ref{eq:Lt3}) to
(\ref{eq:Lt64}), and (\ref{eq:L3})) hold not only for the Polyakov constituent
quark model, but also for any model with a {\em one-body Hamiltonian}, where
the total energy is additive. This simply requires to define $Q_n$,
$\bar{Q}_n$ and $G_n$ as
\begin{equation}
Q_n(T) = \bar{Q}_n(T) =\sum_\alpha\!{}^q \gamma_\alpha e^{-n \beta \varepsilon_\alpha}
,
\quad
G_n(T) = \sum_\alpha\!{}^g \gamma_\alpha e^{-n \beta \varepsilon_\alpha}
,
\end{equation}
where $C$-symmetry is assumed.

Incidentally, using these expressions for $Q_n$ and $G_n$, it is easy
to establish the validity of the following inequalities
\begin{equation}
G_1^2 \ge G_2
,\qquad
5 G_1^3 \ge 3 G_1 G_2 + 2 G_3
,
\end{equation}
for any spectrum. They are sufficient to show that the $O(c^4)$ term of
$L_{\bf 3}$ in \Eq{L3} is never negative. In fact, by construction, each
configuration $q^l \qb^n g^m$ has a non negative weight in $L_\mu(T)$ for any
irrep, provided $Q_n$, $\bar{Q}_n$ and $G_n$ derive from a one-body
Hamiltonian~\footnote{This property needs not strictly apply for the Polyakov
  constituent quark model because there the effective degeneracy is controlled
  by $V_\sigma$ and the required inequalities are not guaranteed if the
  $\gamma_\alpha$ are allowed to be non integer.}. As we have noted this is not
the case for the terms in $\tilde{L}_\mu$ (see \Eq{3.32}).

The interesting observation is that, in a strict single-particle model, the
functions $Q_n(T)$, $\bar{Q}_n(T)$ and $G_n(T)$ for different values of $n$
are not independent, instead they are related by
\begin{equation} 
Q_n(T) = Q_1(T/n)
,\quad
G_n(T) = G_1(T/n)
.
\end{equation}
This means that, in such models, all the $\langle\chi_\mu(\Omega)\rangle(T)$
depend on just two independent functions of $T$, namely, $Q_1(T)$ and
$G_1(T)$. In turn, this implies that the two functions $\langle\chi_{\bf
  3}(\Omega)\rangle(T)$ and $\langle\chi_{\bf 8}(\Omega)\rangle(T)$ fully fix
the expectation values of the Polyakov loop in all other irreps. This holds
regardless of the concrete form of the spectrum and degeneracies, as long as
the Hamiltonian acts additively for quarks, antiquarks and gluons. In
gluodynamics, there is a single independent function, $G_1(T)$ which is fixed
by $\langle\chi_{\bf 8}(\Omega)\rangle(T)$. It should be noted that in an
effective model like the Polyakov constituent quark model there is a further
function of $T$, namely, $V_\sigma(T)$, since the substitution $T\to T/n$ is
not applied in $V_\sigma$.

More quantitatively, we can express $L_{\mu}(T)$ using the auxiliary variables
$L_{{\bf 3},n} = L_{\bf 3}(T/n)$, and $L_{{\bf 8},n} = L_{\bf 8}(T/n)$. In our
counting $L_{{\bf 3},n}$ and $L_{{\bf 8},n}$ are of $O(c^n)$. In this way one
obtains relations which are much more detailed than those in
Eqs. (\ref{eq:5.27}) and (\ref{eq:5.28}):
\begin{widetext}
\begin{eqnarray}
L_{\bf 6} &=& 
\frac{1}{2}  \left( 2 L_{\bf 3} L_{\bf 8}
+ L_{\bf 3}^2 - L_{{\bf 3},2}\right) 
- \frac{1}{2}  \left( 2 L_{\bf 3}^3
+ L_{\bf 3} L_{\bf 8}^2 + L_{\bf 3} L_{{\bf 8},2} 
+ L_{\bf 3}^2 L_{\bf 8} + L_{{\bf 3},2} L_{\bf 8} \right)
+ O(c^4)
,
\nonumber \\
L_{\bf 10} &=& 
\frac{1}{2}  \left(L_{\bf 8}^2-L_{{\bf 8},2}\right)+\frac{1}{6}
    \left(L_{\bf 3}^3-3 L_{{\bf 3},2} L_{\bf 3}+2
   L_{{\bf 3},3}-2 L_{\bf 8}^3+2 L_{{\bf 8},3}\right)
+ O(c^4)
,
\nonumber \\
L_{\bf 15} &=&
 L_{\bf 3} L_{\bf 8}
- \frac{1}{2} \left(
L_{\bf 3}^3 
+ L_{\bf 3} L_{{\bf 3},2} - L_{\bf 3}^2 L_{\bf 8}
+ L_{{\bf 3},2} L_{\bf 8} \right)
+ O(c^4)
,
\nonumber \\
L_{\bf 27} &=&
\frac{1}{2} \left( L_{\bf 8}^2 + L_{{\bf 8},2} \right)
+ O(c^4)
.
\end{eqnarray}
\end{widetext}
In these expressions $L_{\bf 3}=L_{{\bf 3},1}$ and $L_{\bf 8}=L_{{\bf 8},1}$,
and also $L_{\bf \bar{3}} = L_{\bf 3}$. Results for other irreps are rather
immediate to the order presented here and they are omitted.  The same formulas
hold in gluodynamics, setting $L_{\bf 3}$ to zero. On the other hand, if
gluons are neglected, $L_{\bf 8}$ is no longer an independent function and
this function starts at $O(c^2)$:
\begin{eqnarray}
L_{\bf 6} &=&
\frac{1}{2} \left(L_{\bf 3}^2 - L_{{\bf 3},2} \right)
+ O(c^4)
,
\nonumber \\
L_{\bf 8} &=&
L_{\bf 3}^2 
- \frac{1}{3} \left( L_{\bf 3}^3 + 3 L_{{\bf 3},2} L_{\bf 3} 
+ 2 L_{{\bf 3},3} \right)
+ O(c^4)
.
\end{eqnarray}

\begin{figure*}[t]
\begin{center}
\epsfig{figure=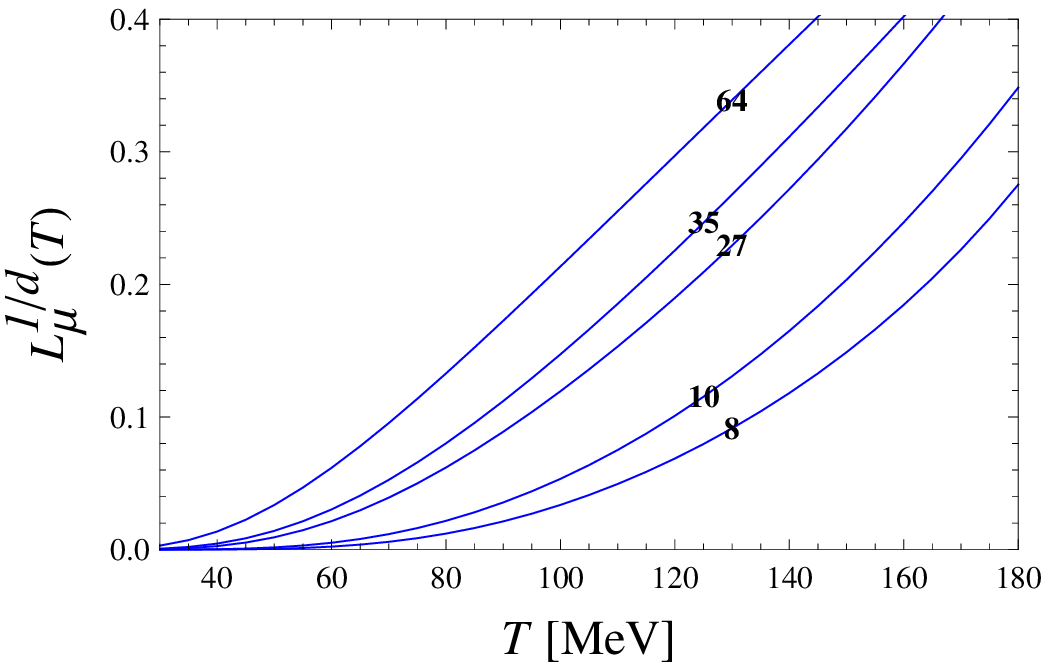,height=60mm,width=85mm}
\epsfig{figure=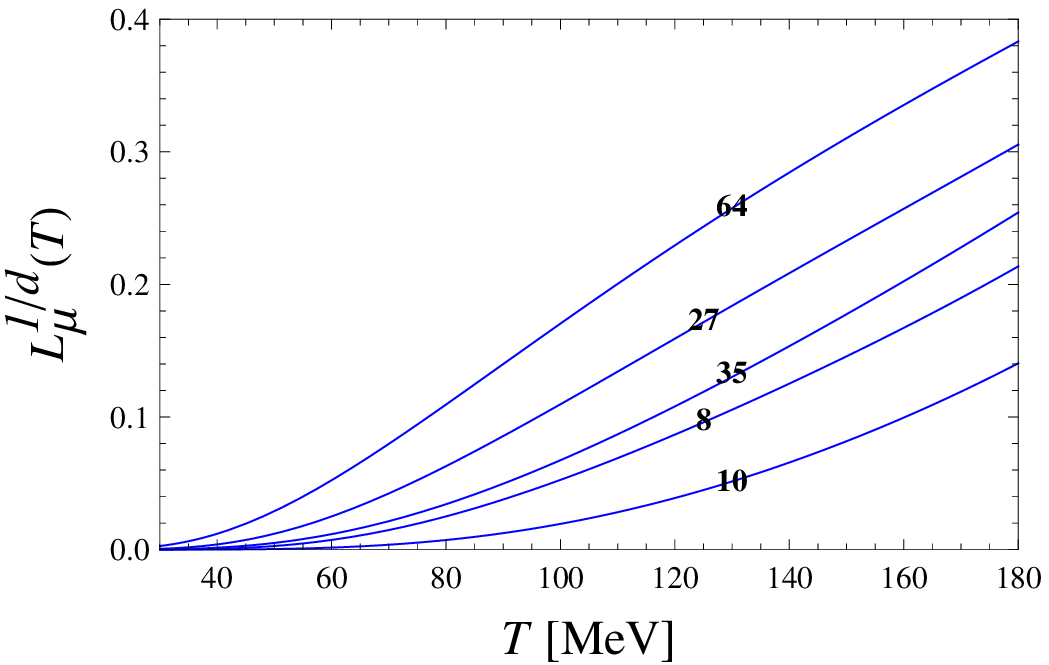,height=60mm,width=85mm}
\end{center}
\caption{ $L_\mu^{1/d_\mu}(T)$ for the MIT bag model (left panel) and the
  constituent quark model (right panel). Here $d_\mu=C_2(\mu)/C_2({\bf 3})$,
  is the ratio of the two body Casimir operator for the irrep $\mu={\bf 8},
  {\bf 10}, {\bf 27}, {\bf 35}, {\bf 64}$, relative
  to the fundamental one. We consider gluodynamics and $N_c=3$. }
\label{fig:casimir}
\end{figure*}

In a recent paper \cite{Mykkanen:2012ri} a thorough study establishes the
Casimir scaling properties of the Polyakov loop in pure gluodynamics and large
$N_c$, above the phase transition, and at sufficiently large temperatures.
That means that $L_\mu^{1/d_\mu}(T)$ approaches a $\mu$-independent value at
high temperatures. Here $d_\mu=C_2(\mu)/C_2({\bf 3})$, is the ratio of the two
body Casimir operator for the irrep $\mu$ relative to the fundamental one. One
can see from their calculation that sizable deviations from this scaling
behavior take place for lower temperatures. It is interesting to see whether
the models studied in the present work comply to this Casimir scaling
departures. As we have repeatedly stressed, the low temperature behavior is
dominated by the lightest energy gap associated with the screening of heavy
sources, and as such, implement {\em different} scaling rules. For the case
considered in \cite{Mykkanen:2012ri}, and combining our results in
Fig. \ref{fig:casimir}, we generally agree with the monotonous offset of the
Casimir scaling below the phase transition as the dimension of the
representation increases, for $\mu={\bf 8}$, ${\bf 10}$, ${\bf 27}$, ${\bf
  35}$, and ${\bf 64}$ (albeit with inversions in ${\bf 8}$ and ${\bf
  10}$, and in ${\bf 27}$ and ${\bf 35}$ in the constituent quark model).  In
this regard, it would be interesting to pursue Casimir scaling violations to
temperatures below $0.75\,T_c$, where the data of Ref. \cite{Mykkanen:2012ri}
stop. It is remarkable that, even though Polyakov loops are exponentially
suppressed at low temperatures, as established here, lattice calculations may
provide a clear signal on the light of Casimir scaling.

\section{Conclusions}
\label{sec:conclusions}

In the present paper we have extended a hadronic sum rule for the Polyakov
loop developed previously for the fundamental representation to higher
representations of the $\SU(3) $ gauge group. Quite generally we find that for
any irreducible $\SU(3)$ representation $\mu$ there is a finite energy gap
$\Delta_\mu$ which corresponds to the lightest state having an infinitely
heavy source transforming according to the irrep $\mu$ screened by a
conjugated dynamical charge. This provides the Boltzmann factor controlling
the low temperature exponential suppression of the Polyakov loop. While these
gaps could in principle be computed on the lattice, they can be compared to
existing hadronic physical states to some approximation. Regardless of whether
or not these states can be extracted from suitable experiments, they provide a
basis for which the completeness of the hadronic spectrum can be tested.

One can alternatively formulate the problem using the Polyakov loop
itself as a quantum and local dynamical variable and hence define the
corresponding partition function as an integral over $\Omega$ which is
proportional to the probability of having a given value of $\Omega$ at
a given temperature. We have found a low temperature character
expansion which involves group representations with higher
dimensionality.

Within an independent particle model picture the hadronic sum rule generates
scaling properties at low temperatures which offer an alternative to Casimir
scaling and could possibly be tested on the lattice.  One good feature of
these models is the correct counting of multiparton states. We have found that
by analyzing the models within a low temperature expansion an equivalent
multipartonic decomposition of the Fock space arises in terms of the number of
constituents. The clustering of these states into hadrons turns out to be more
tricky, as there is no completely unambiguous way of defining the hadronic
partition function. This has to do with the emergence of globally color
singlet states which are reducible, i.e.  clusters that i) are decomposable
into subclusters which are themselves color singlets, ii) can be dissociated
with an energy much smaller than the mass of the subclusters, and iii) within
the approximation that these states interact through residual spin-flavor
forces, they factorize in the partition function. From this point of view, a
six quark state such as the deuteron or the weakly bound $X(3872)$ made as a
large $c\bar c \bar n n $ and a small $c \bar c$ component would factorize in
the partition function sum. On the other hand, some states in this category
create sharp resonances which in the standard HRG model are counted on their
own, as suggested by the quantum virial expansion
analysis~\cite{Dashen:1974yy,Venugopalan:1992hy}. By using as an example the
MIT bag model, which is free from center of mass corrections due to heavy
source, we find very visible consequences of considering all states or just
irreducible ones. Therefore, we expect that future lattice calculations may
shed some light on this intriguing issue which already shows up in a low
temperature expansion.

Saturating the sum rule even to moderate temperatures requires a large
number of excited states. This is due to the lack of mass gaps in the
spectrum besides the finite and large lowest excitation gaps,
$\Delta_\mu$, generating by screening static color charges in a given
irreducible representation, $\mu$. Thus, even for heavy quarks,
relativistic corrections become important for high excitations.  Our
model calculations implemented relativistic features.

As we have stressed in this paper it is also crucial to maintain at
any rate the quantum and local character of the Polyakov loop, as
simple-minded mean-field approximations applied to the Polyakov loop
in e.g. a PNJL approach washes out all information on higher
representations.

As we mentioned in the Introduction, the definition of a hadron as a true or
effective elementary degree of freedom from a thermodynamic point of view is
subtle. In this regard it is remarkable that lattice calculations for the
trace anomaly are reproduced in the bulk and within uncertainties by the
Hadron Resonance Gas model taking PDG
states~\cite{Karsch:2003zq,Borsanyi:2010cj,Huovinen:2009yb,Borsanyi:2013bia}.
The partition function actually counts the total number of states {\em at
  large} temperatures. In a previous paper~\cite{Megias:2012kb} we profited
from the results found in the RQM~\cite{Godfrey:1985xj,Capstick:1986bm}. The
extension of these quark model calculations needed for the saturation of our
generalized sum rule given by Eq.~(\ref{eq:2.7}) beyond the fundamental
representation remains to be done (for a sketch see e.g.
\cite{RuizArriola:2012wd}). Finally, calculations for tetraquark and
pentaquark excited states require assumptions about the color structure of
interactions which actually correspond to a specific choice on the interacting
forces from the model building point of view. However, the very existence of
a hadronic limit imposes non trivial constraints on the microscopic quark and
gluon interactions. We hope the present paper will stimulate further
developments in setting up a Polyakov spectroscopy exploiting the partition
function character of the Polyakov loop in different representations.

\begin{acknowledgments}
We thank M. Panero and O. Kaczmarek for correspondence.
This work has been supported by Plan Nacional de Altas Energ\'{\i}as
(FPA2011-25948), DGI (FIS2011-24149), Junta de Andaluc\'{\i}a grant
FQM-225, Spanish MINECO's Consolider-Ingenio 2010 Programme CPAN
(CSD2007-00042) and Centro de Excelencia Severo Ochoa Programme grant
SEV-2012-0234. The research of E.M. is supported by the Juan
de la Cierva Program of the Spanish MINECO.
\end{acknowledgments}

\appendix

\section{Counting of irreducible color configurations}
\label{app:A}

In this section we describe the counting of reducible and irreducible color
configurations. Results up to three constituents are given in Table
\ref{tab:1}, and up to four constituents for the fundamental representation in
Table \ref{tab:2}.

The total multiplicity (adding the number of irreducible plus reducible
configurations) can be computed straightforwardly when all spin-flavor states
are different. Namely, it suffices to compute the Clebsch-Gordan series of the
product of $n_q$ ${\bf 3}$'s, $n_\qb$ ${\bf \bar{3}}$'s, and $n_g$ ${\bf
  8}$'s, and count how many times each irrep appears there. For instance
\begin{equation}
{\bf 8}\otimes {\bf 8} = 
{\bf 1} \oplus {\bf 8} \oplus {\bf 8} \oplus {\bf 10} \oplus {\bf \bar{10}}
 \oplus {\bf 27}
,
\end{equation}
implies that the configuration $(0,0,1^2)$ (two gluons in different
spin-flavor states) can ``screen'' a singlet source (or rather form a singlet)
in one way, an adjoint source in two ways, etc. Unfortunately, we do not have
such a simple counting when some spin-flavor labels are repeated, nor to
separate reducible from irreducible configurations.

We recall that a color configuration is reducible when a non-empty subset of
the constituents forms a singlet. This is different from reducibility under
the group (minimal subspace). For instance, in the example above the two
gluons screening an adjoint source span a space ${\bf 8} \times {\bf 8}$,
which is therefore reducible under the group. However, the color
configurations are themselves irreducible since a non-empty subset of
constituents forming a singlet would make use of the two gluons, and the
adjoint source would not be screened (or equivalently, screening the source
requires at least one gluon, and the other gluon cannot form a singlet by
itself).

In general, we have a certain number of constituents, $q$, $\qb$, $g$ in
various spin-flavor states, $\alpha$, $\beta$, etc, and color states $i$, $j$,
etc. To each such constituent we assign an operator (that can be regarded as a
creation operator), $q_\alpha^i$, $\bar{q}_{\alpha,i}$, or $g_{\alpha,j}^i$.
The operators $q$ and $\qb$ are fermionic and $g$ is bosonic.  The matrix
$g_{\alpha,j}^i$ (with respect to $ij$) is traceless. Let
$\alpha_1,\ldots,\alpha_{n_q}$ be the spin-flavor states of the $n_q$ quarks in
the configuration, and similarly $\beta_1,\ldots,\beta_{n_\qb}$ for
antiquarks, and $\gamma_1,\ldots,\gamma_{n_g}$ for gluons. This defines a
vector space
\begin{equation}
{\cal H} = {\rm lin}\big\{
q_{\alpha_1}^{i_1}\cdots q_{\alpha_{n_q}}^{i_{n_q}}
\qb_{\beta_1, j_1}\cdots \qb_{\beta_{n_\qb}, j_{n_\qb}}^{}
g_{\gamma_1,l_1}^{k_1}\cdots g_{\gamma_{n_g},l_{n_g}}^{k_{n_g}}
\big\}\,,
\end{equation}
where the spin-flavor labels are fixed and the color labels run from 1 to
3. ${\cal H}$ carries a representation of $\SU(3)$ which in
general will be reducible
\begin{equation}
{\cal H} = 
\bigoplus_\mu \left(
\bigoplus_{k=1}^{t_\mu}{\cal H}_\mu^{(k)}\right)
.
\end{equation}
Each ${\cal H}_\mu^{(k)}$ carries an irrep $\bar\mu$. (Recall that the source
is in the irrep $\mu$ and therefore the dynamical particles are in $\bar\mu$.)
The multiplicity $t_\mu$ is the total shown in Table \ref{tab:1}
for each irrep $\mu$ and for each configuration of constituents $(
\{\alpha_1\ldots\alpha_{n_q}\},
\{\beta_1\ldots\beta_{n_\qb}\},\{\gamma_1\ldots\gamma_{n_g}\} )$.

Instead of working with the $n_\mu$-dimensional spaces ${\cal H}_\mu^{(k)}$,
it is preferable to work with one-dimensional spaces by coupling the states of
${\cal H}_\mu^{(k)}$ to a source in the irrep $\mu$ to form a singlet. In
$\SU(3)$ an irrep of the type $(n,m)=[n+m,m]$ can be represented by a color
tensor $\Omega^{i_1,\ldots,i_n}_{j_1,\ldots,j_m}$ completely symmetric in both
upper and in lower indices and traceless
(i.e. $\Omega^{k,i_2,\ldots,i_n}_{k,j_2,\ldots,j_m}=0$)
\cite{Coleman:1988bk}. The components of
$\Omega^{i_1,\ldots,i_n}_{j_1,\ldots,j_m}$ are c-numbers. The abovementioned
two-gluon configurations $(0,0,1^2)$ screening the adjoint source in two
ways correspond to the two combinations
\begin{equation}
\Omega^i_j g_{\alpha,k}^j g_{\beta,i}^k
,
\quad
\Omega^i_j g_{\alpha,i}^k g_{\beta,k}^j
.
\label{eq:A.9}
\end{equation}
These operators are linearly independent and span a two-dimensional space. By
changing $\Omega$ (or by applying group rotations with fixed $\Omega$) each
operator would generate an 8-dimensional irreducible space. For $(0,0,2)$,
i.e. $\alpha=\beta$, the states are no longer linearly independent and the
dimension of the spanned space becomes one (the antisymmetric combination
disappears).

For an arbitrary irrep $\mu$, because of irreducibility, any choice of
$\Omega^{i_1,\ldots,i_n}_{j_1,\ldots,j_m}$ gives equivalent results. The
simplest choice is to take only one component different from zero (plus those
needed to fulfill the constraints of symmetry and tracelessness).

All operators can be generated by using the appropriate source $\Omega$, the
$q$, $\qb$, $g$ operators, and contracting all indices. Besides the Kronecker
delta, the only invariant tensor that can be used is the $\epsilon_{ijk}$ (or
$\epsilon^{ijk}$). Not all operators so obtained will be linearly
independent. As noted above, the total dimension of the space is known
beforehand when spin-flavor states are all different, so in this case it is
easy to check when a sufficient number of independent operators have been
accounted for. To go to the case in which some of the spin-flavor states are
equal, it is sufficient to take these same operators and identify spin-flavor
labels there, and then count the new number of linearly independent operators.

The basis of operators obtained in this way will contain both reducible and
irreducible color configurations in general. To separate and then count the
two types of configurations the prescription is to identify first the subspace
of reducible ones, the ``irreducible'' subspace being the orthogonal
complement.

It is sufficient to consider the case when all spin-flavor states are
different. For a given set $(n_q,n_\qb,n_g)$, the construction of the {\em
  reducible} color configurations of type $\mu$ is recursive. First one
considers all $\mu$-configurations with a smaller number of constituents and
then completes the set by adding all possible non-empty {\em singlet}
configurations so that the total number of constituents $(n_q,n_\qb,n_g)$ is
attained. This procedure produces the subspace of reducible configurations.

Not all reducible states so obtained will be linearly independent. For
instance, for pentaquark configurations of the type $(0,1^4,0)$ screening a
fundamental source,
\begin{equation}
(\Omega^i \qb_{\alpha,i}) (\qb_{\beta,j} \qb_{\gamma,k} \qb_{\delta,l} \epsilon^{jkl})
,
\end{equation}
one could expect to form four states by screening the source with either
$\alpha$, $\beta$, $\gamma$ or $\delta$, however, just three of them are
linearly independent. In the present case the Clebsch-Gordan series is
\begin{equation}
{\bf \bar{3}} \otimes {\bf \bar{3}} \otimes {\bf \bar{3}} \otimes {\bf
  \bar{3}} = {\bf \bar{3}} \oplus {\bf \bar{3}} \oplus {\bf \bar{3}} \oplus
{\bf 6} \oplus {\bf 6} \oplus {\bf \bar{15} }\oplus {\bf \bar{15} } \oplus
{\bf \bar{15} } \oplus {\bf \bar{15^\prime} } ,
\end{equation}
therefore the three reducible configurations saturate the series and there are
no irreducible states (i.e., $(0+3)$ in Table \ref{tab:2}).

For another example, consider the configuration $(0,1,1^2)$ with
Clebsch-Gordan series
\begin{equation}
{\bf \bar{3}} \otimes {\bf 8} \otimes {\bf 8} = 
{\bf \bar{3}} \oplus {\bf \bar{3}} \oplus {\bf \bar{3}} \oplus
{\bf 6} \oplus  \cdots
,
\end{equation}
so there are three ways to screen a fundamental source:
\begin{equation}
\Omega^i \qb_{\alpha,i} g_{\beta,j}^k g_{\gamma,k}^j,\quad
\Omega^i \qb_{\alpha,j} g_{\beta,i}^k g_{\gamma,k}^j,\quad
\Omega^i \qb_{\alpha,j} g_{\beta,k}^j g_{\gamma,j}^k,
\end{equation}
the first one is reducible and the two last irreducible and they all are
linearly independent.

Finally, for a less trivial case consider $(0,1,1^3)$ screening a
fundamental source. The Clebsch-Gordan series produces ten ${\bf \bar{3}}$.
The operators one can write down are of the form
\begin{equation}
\Omega^i \qb_{\alpha,i} g_{\beta,j}^k g_{\gamma,k}^l g_{\delta,l}^j,
\quad
\Omega^i \qb_{\alpha,j} g_{\beta,i}^j g_{\gamma,k}^l g_{\delta,l}^k,
\quad
\Omega^i \qb_{\alpha,j} g_{\beta,i}^k g_{\gamma,k}^l g_{\delta,l}^j,
\end{equation}
plus those obtained by permutation of the gluonic labels. Operators of the two
first types are reducible and span a space of dimension five (two from the
first type and three more from the second one). The third type produces six
states but only five linearly independent ones, that is, one combination (the
symmetric one) is actually reducible, and the other five irreducible.

If there are repeated spin-flavor labels, one can apply the above procedure
from scratch or instead use previous results obtained for the case of
different labels. In the latter case one should be aware that after setting,
say $\beta=\alpha$, in the basis obtained for $\beta\not=\alpha$ some states
that were previously irreducible can become reducible (that is, linear
combination of manifestly reducible ones). So the correct procedure is again
to first identify the reducible space, and then compute its orthogonal
complement.

\section{Further computational details}
\label{app:B}

\subsection{Truncation of infinite sums}

To deal with the infinite sums over $n$ in
$Z_{q,g}^\prime(\omega,T^\prime,T)$, \Eq{4.19a}, we simply cut off the sums
when the $z^\prime$'s become smaller than $\varepsilon=10^{-10}$.

\subsection{The function $f_{\frac{3}{4}}(t)$}

The function $f_{\frac{3}{4}}(t)$ is defined by the relation
\begin{equation}
\int_0^\infty f_{\frac{3}{4}}(t) e^{-s t} \, dt  = e^{-s^{3/4}}
.
\label{eq:A.2}
\end{equation}
$f_{\frac{3}{4}}(t)$ is a positive normalized unimodal function which vanishes
(with all its derivatives) at $t=0$ and $t=\infty$.

Use of the Bromwich inversion formula allows to express $f_{\frac{3}{4}}(t)$ as
\begin{equation}
f_{\frac{3}{4}}(t)
=
\int_0^\infty \frac{dx}{\pi}
\cos(tx-\cos(\pi/8)x^{3/4})e^{-\sin(\pi/8)x^{3/4}}
.
\end{equation}
The integrand is oscillating and we find that making the change of variables
$x=y^{4n/3}$ with $n=1$ or $2$ improves the convergence for large and small $x$. We
use the integral in the range $0.2\le t \le 6$. For larger $t$ a fit is used
of the form
\begin{equation}
f_{\frac{3}{4}}(t) \sim 
  0.4663\frac{1}{t^2}
- 0.4180\frac{1}{t^3} 
+ 0.8403\frac{1}{t^4}
\qquad
(t>6)
.
\end{equation}
A saddle point approximation in \Eq{A.2} allows to obtain the small $t$
behavior of the former function, namely,
\begin{equation}
f_{\frac{3}{4}}(t) \underset{t\to 0}{\sim} e^{-\frac{1}{4}\left(\frac{3}{4 \, t}\right)^3} \,.
\label{eq:A.1}
\end{equation}
On this basis, for small $t$ we use the fit
\begin{widetext}
\begin{equation}
f_{\frac{3}{4}}(t)
\sim
\exp\left(
-\frac{1}{4}\left(\frac{3}{4 \, t}\right)^3
(1 - 5.401 \, t^2 - 10.35 \, t^3 + 24.07 \, t^4)
\right) \qquad
(t < 0.2)
.
\end{equation}
\end{widetext}

\subsection{Clebsch-Gordan series}
\label{sec:B.3}

In order to obtain the expansion in \Eq{3.26} from \Eq{3.27} one can use
the expressions in \Eq{4.43} and expand in powers of $q$, $\qb$ and $g$.
The required Clebsch-Gordan series can be obtained by using back and forth the
following two sets of equations:
\begin{eqnarray}
\chi_{\bf 1} &=& 1 
\nonumber \\
\chi_{{\bf 6}} &=& \chi_{\bf 3}^2 - \chi_{\bf 3}^* 
\nonumber \\
\chi_{{\bf 8}} &=& \chi_{\bf 3} \chi_{\bf 3}^* - 1
\nonumber \\
\chi_{{\bf 10}} &=& -2 \chi_{\bf 3} \chi_{\bf 3}^* + \chi_{\bf 3}^3 + 1
\nonumber \\
\chi_{{\bf 15}} &=& \chi_{\bf 3}^2 \chi_{\bf 3}^* - \chi_{\bf 3} - \chi_{\bf 3}^{*2}
\nonumber \\
\chi_{\bf 15^\prime} &=& -3 \chi_{\bf 3}^2 \chi_{\bf 3}^* + \chi_{\bf 3}^4
+ 2 \chi_{\bf 3} + \chi_{\bf 3}^{*2}
\nonumber \\
\chi_{{\bf 24}} &=& \chi_{\bf 3} \chi_{\bf 3}^{*3} - 2 \chi_{\bf 3}^2 \chi_{\bf 3}^*
+ 2 \chi_{\bf 3}-\chi_{\bf 3}^{*2}
\nonumber \\
\chi_{{\bf 27}} &=& \chi_{\bf 3}^2 \chi_{\bf 3}^{*2} - \chi_{\bf 3}^3 
- \chi_{\bf 3}^{*3}
\nonumber \\
\chi_{{\bf 35}} &=& \chi_{\bf 3}^4 \chi_{\bf 3}^* - 3 \chi_{\bf 3}^2 \chi_{\bf 3}^{*2}
+ 4 \chi_{\bf 3} \chi_{\bf 3}^* - \chi_{\bf 3}^3 + \chi_{\bf 3}^{*3} - 1
\nonumber \\
\chi_{{\bf 42}} &=& \chi_{\bf 3}^3 \chi_{\bf 3}^{*2} + \chi_{\bf 3}^2 \chi_{\bf 3}^*
- 2 \chi_{\bf 3} \chi_{\bf 3}^{*3} - \chi_{\bf 3}^4 - \chi_{\bf 3} + 2 \chi_{\bf 3}^{*2} 
\nonumber \\
\chi_{{\bf 64}} &=& -2 \chi_{\bf 3}^4 \chi_{\bf 3}^* 
+ \chi_{\bf 3}^3 \chi_{\bf 3}^{*3} + 3 \chi_{\bf 3}^2 \chi_{\bf 3}^{*2}
- 2 \chi_{\bf 3} \chi_{\bf 3}^{*4} - 5 \chi_{\bf 3} \chi_{\bf 3}^*
\nonumber \\ &&
+ 2 \chi_{\bf 3}^3 + 2 \chi_{\bf 3}^{*3} + 1
\label{eq:B6}
\end{eqnarray}

\begin{eqnarray}
1 &=& \chi_{\bf 1}
\nonumber \\
\chi_{\bf 3}^2 &=& \chi_{\bf 3}^*+\chi_{\bf 6}
\nonumber \\
\chi_{\bf 3}^3 &=& \chi_{\bf 1} + 2 \chi_{{\bf 8}} + \chi_{\bf 10}
\nonumber \\
\chi_{\bf 3}^4 &=& 3 \chi_{\bf 3} + 2 \chi_{\bf 6}^* + 3 \chi_{\bf 15}
+ \chi_{\bf 15^\prime}
\nonumber \\
\chi_{\bf 3} \chi_{\bf 3}^* &=& \chi_{\bf 1} + \chi_{{\bf 8}}
\nonumber \\
\chi_{\bf 3}^2 \chi_{\bf 3}^* &=& 2 \chi_{\bf 3}+\chi_{\bf 6}^* + \chi_{{\bf 15}}
\nonumber \\
\chi_{\bf 3}^3 \chi_{\bf 3}^* &=& 3 \chi_{\bf 3}^* + 3 \chi_{\bf 6}
+ 2 \chi_{\bf 15}^*+\chi_{\bf 24}^*
\label{eq:B7} \\
\chi_{\bf 3}^4 \chi_{\bf 3}^* &=& 3 \chi_{\bf 1} + 8 \chi_{\bf 8}
+ 4 \chi_{\bf 10} + 2 \chi_{\bf 10}^* + 3 \chi_{{\bf 27}} + \chi_{\bf 35}
\nonumber \\
\chi_{\bf 3}^2 \chi_{\bf 3}^{*2} &=& 2 \chi_{\bf 1} + 4 \chi_{\bf 8} + \chi_{\bf 10}
+ \chi_{\bf 10}^* + \chi_{\bf 27}
\nonumber \\
\chi_{\bf 3}^3 \chi_{\bf 3}^{*2} &=& 6 \chi_{\bf 3} + 5 \chi_{\bf 6}^*
+ 6 \chi_{\bf 15} + \chi_{\bf 15^\prime} + 2 \chi_{\bf 24} + \chi_{\bf 42}
\nonumber \\
\chi_{\bf 3}^3 \chi_{\bf 3}^{*3} &=& 6 \chi_{\bf 1} + 17 \chi_{\bf 8}
+ 7 \chi_{\bf 10} + 7 \chi_{\bf 10}^* + 9 \chi_{\bf 27} 
+ 2 \chi_{{\bf 35}} + 2 \chi_{\bf 35}^* 
\nonumber \\&&
+ \chi_{\bf 64} 
\nonumber
\end{eqnarray}
Instead of the identities in \Eq{4.43}, one can alternatively expand \Eq{3.27}
in powers of $\Omega$, so that everything is expressed in terms of
$\chi_{\bf 3}(\Omega^n)$, $\chi_{\bf \bar{3}}(\Omega^n)$, and
$\chi_{\bf 8}(\Omega^n)$. The two latter
expressions are reduced to fundamental characters through
\begin{equation}
\chi_{\bf \bar{3}}(\Omega^n) = \chi_{\bf 3}(\Omega^{-n})
, \quad
\chi_{\bf 8}(\Omega^n) = \chi_{\bf 3}(\Omega^n)\chi_{\bf 3}(\Omega^{-n}) -1
.
\end{equation}
This allows to repeatedly apply the $\SU(3)$ identity
\begin{equation}
\Omega^2 = \chi_{\bf 3}(\Omega) \Omega - \chi_{\bf \bar 3}(\Omega) + \Omega^{-1}
,\quad
\Omega\in\SU(3)
, 
\end{equation}
as well as its adjoint one. In this way everything is expressed as a
polynomial of the two variables $\chi_{\bf 3}(\Omega)$ and $\chi_{\bf
  3}^*(\Omega)$ and \Eq{B7} applies.

Still another route is to insert $\chi_\mu(\Omega)$, express all characters in
terms of eigenvalues of $\Omega$ and carry out the integrations by residues.

\section{Numerical results for the bag model with selected configurations}
\label{app:C}

\begin{figure}[t]
\begin{center}
\epsfig{figure=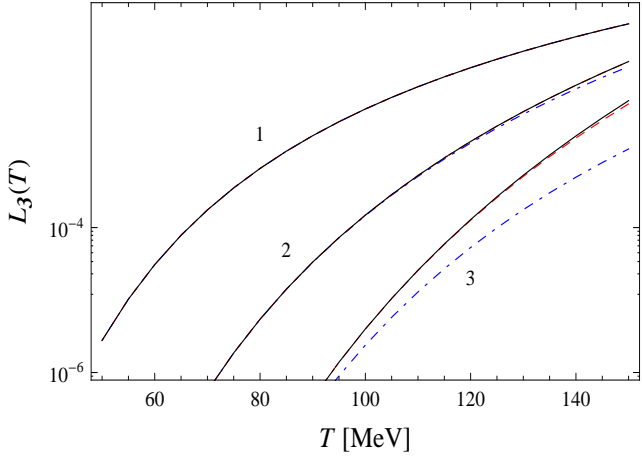,height=60mm,width=85mm}
\end{center}
\caption{Convergence of the sum over levels for $L_{\bf 3}(T)$ as the
  cutoff $\Lambda$ is increased. Labels 1, 2, and 3 correspond to sums
  over configurations with one, two and three constituents,
  respectively. Dot-dashed, dashed and solid lines correspond to
  $\Lambda=2$, $3$, and $4\GeV$, respectively. Convergence is slower
  for higher temperature and/or a larger numbers of constituents, due
  to the rapid increase in the number of active configurations.}
\label{fig:7nb1}
\end{figure}

\begin{figure}[t]
\begin{center}
\epsfig{figure=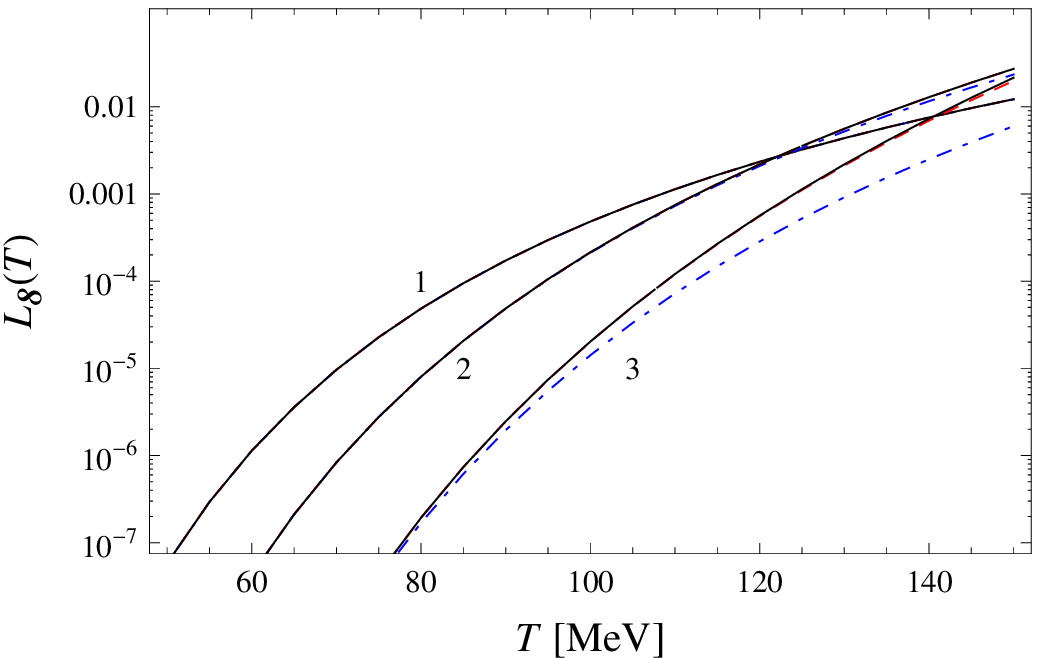,height=65mm,width=85mm}
\end{center}
\caption{
As in Fig.~\protect{\ref{fig:7nb1}} for the adjoint representation.
}
\label{fig:7nb2}
\end{figure}

\begin{figure}[t]
\begin{center}
\epsfig{figure=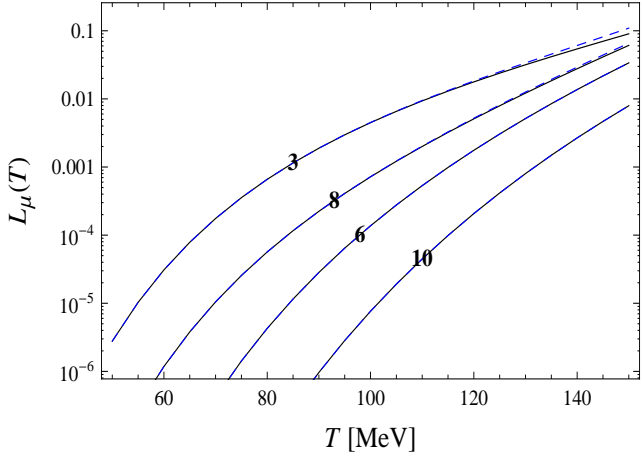,height=60mm,width=85mm}
\end{center}
\caption{Effect of removing reducible color configurations in $L_\mu(T)$ for
  $\mu={\bf 3}$, ${\bf 8}$, ${\bf 6}$, ${\bf 10}$, and including up to three
  constituents. Dashed lines: irreducible plus reducible color
  configurations. Solid lines: irreducible configurations only. }
\label{fig:7nb3}
\end{figure}

\begin{figure*}[h]
\begin{center}
\epsfig{figure=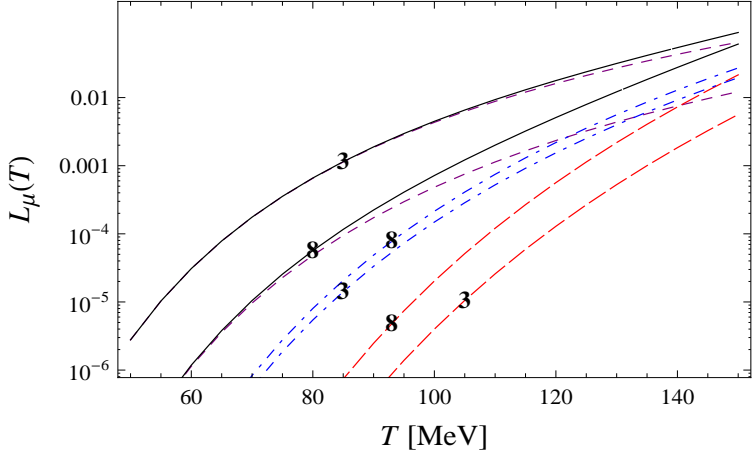,height=60mm,width=100mm}
\end{center}
\caption{Separate contributions to $L_{\bf 3}$ and $L_{\bf 8}$ from 1
  (dashed), 2 (dot-dashed), and 3 (long-dashed) constituents. Solid line:
  $1+2+3$ constituents.}
\label{fig:7nb4}
\end{figure*}

\begin{figure*}[h]
\begin{center}
\epsfig{figure=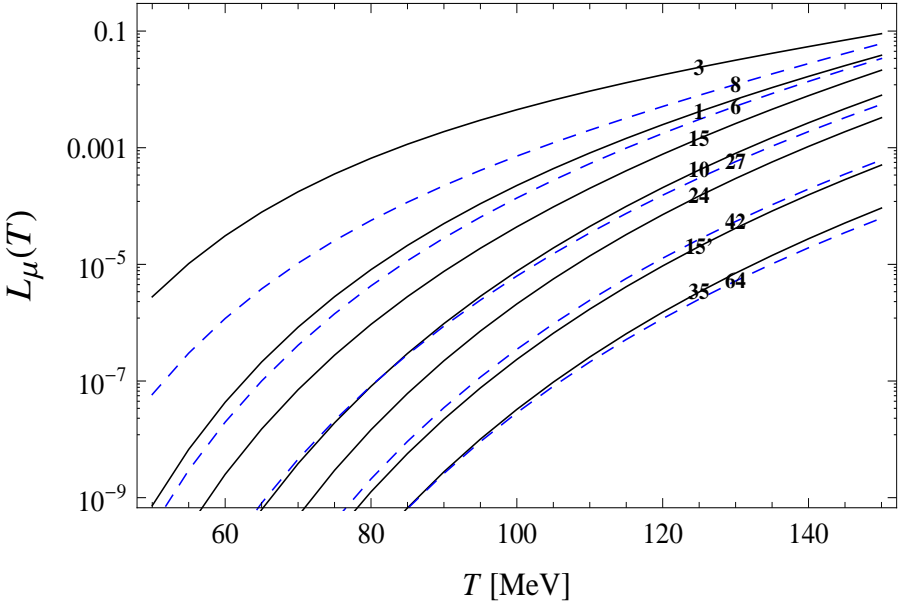,height=80mm,width=120mm}
\end{center}
\caption{$L_\mu(T)$ as a function of the temperature for several irreps and
  including up to three constituents. For the singlet, $Z_{\bf 1}(T)-1$ is
  plotted, it includes singlet reducible states.}
\label{fig:7nb10}
\end{figure*}

\begin{figure*}[h]
\begin{center}
\epsfig{figure=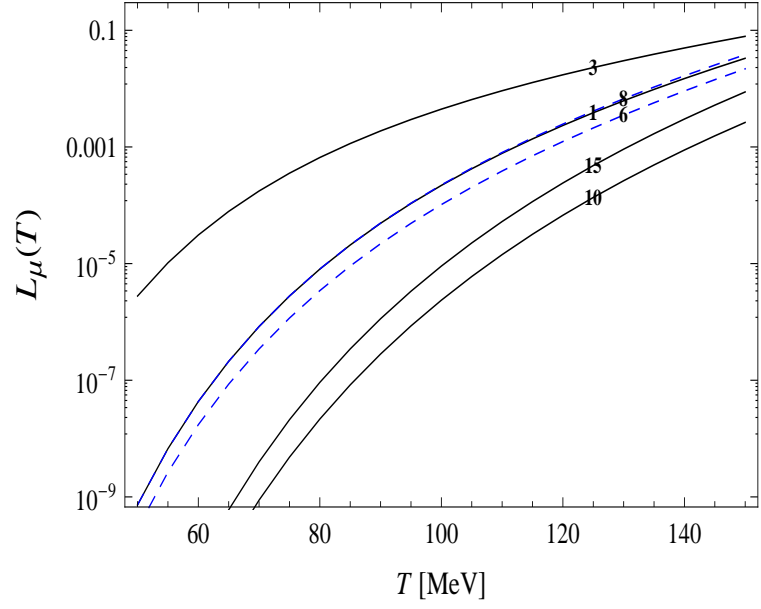,height=80mm,width=100mm}
\end{center}
\caption{As Fig.~\ref{fig:7nb10} but including only quarks and antiquarks.
  Some irreps are not shown because they are too small for the temperature
  range.}
\label{fig:7nb11}
\end{figure*}

\begin{figure*}[h]
\begin{center}
\epsfig{figure=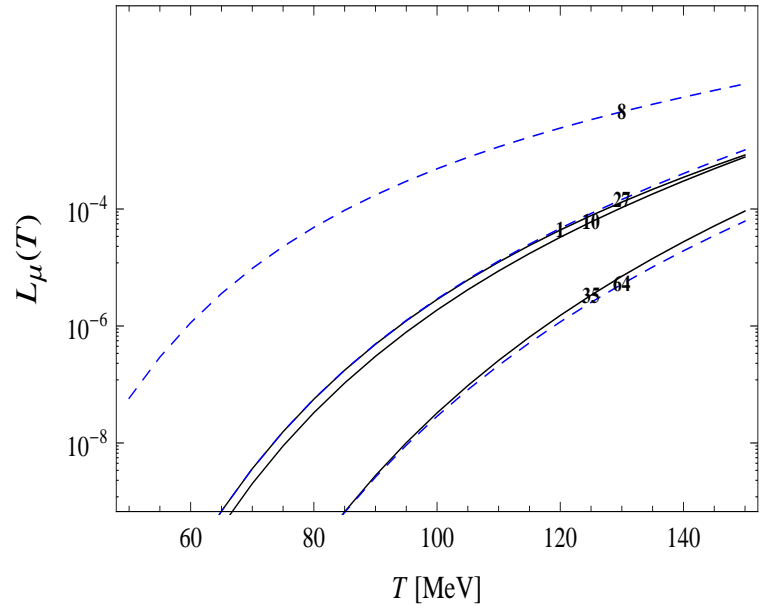,height=80mm,width=100mm}
\end{center}
\caption{As Fig.~\ref{fig:7nb10} but using only gluons.
}
\label{fig:7nb12}
\end{figure*}

\begin{figure*}[h]
\begin{center}
\epsfig{figure=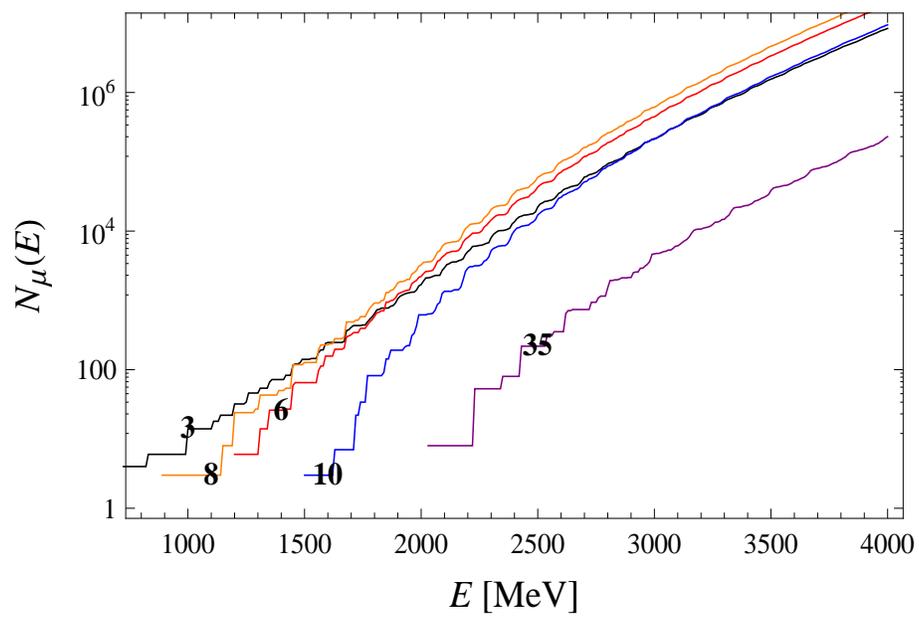,height=80mm,width=120mm}
\end{center}
\caption{Cumulative number of states for some irreps.}
\label{fig:7nb13}
\end{figure*}

Numerical results for Sec. \ref{sec:3} are displayed in Figs. \ref{fig:7nb1},
\ref{fig:7nb2}, \ref{fig:7nb3}, \ref{fig:7nb4}, \ref{fig:7nb10},
\ref{fig:7nb11}, \ref{fig:7nb12} and \ref{fig:7nb13}. Lowest bag states are
displayed in Tables \ref{tab:4}, \ref{tab:5} and \ref{tab:6}.

\begin{table}[t]
\caption{Lowest-lying bag one-constituent configurations with their
  degeneracies for each irrep. ``$0+0$'' entries have been omitted. $\ell$,
  $s$, and $g$ stands for light quark ($u,d$), strange quark, and gluon. $n$
  $j$ and parity of the constituents are indicated.  Irreducible plus
  reducible degeneracies are displayed.}
\label{tab:4}
\begin{tabular}{clcccccccccccc}
\hline 
$\Delta (\MeV)$ & config. 
& {\bf 1} & {\bf 3} & {\bf 6} & {\bf 8} & {\bf 10} & {\bf 15$^\prime$} 
& {\bf  15} & {\bf 24} & {\bf 27} & {\bf 35} & {\bf 42} & {\bf 64}
\cr
\hline
$712.11$ &
 ${\bar\ell}(1\frac{1}{2}^-) $
 & 
 & $4+0 $
 & 
 & 
 & 
 & 
 & 
 & 
 & 
 & 
 & 
 & 
\cr
$821.11$ &
 $\bar s(1\frac{1}{2}^-) $
 & 
 & $2+0 $
 & 
 & 
 & 
 & 
 & 
 & 
 & 
 & 
 & 
 & 
\cr
$888.49$ &
 $g(11^+) $
 & 
 & 
 & 
 & $3+0 $
 & 
 & 
 & 
 & 
 & 
 & 
 & 
 & 
\cr
$998.$ &
 ${\bar\ell}(1\frac{3}{2}^+) $
 & 
 & $8+0 $
 & 
 & 
 & 
 & 
 & 
 & 
 & 
 & 
 & 
 & 
\cr
$1107.$ &
 $\bar s(1\frac{3}{2}^+) $
 & 
 & $4+0 $
 & 
 & 
 & 
 & 
 & 
 & 
 & 
 & 
 & 
 & 
\cr
$1136.77$ &
 ${\bar\ell}(1\frac{1}{2}^+) $
 & 
 & $4+0 $
 & 
 & 
 & 
 & 
 & 
 & 
 & 
 & 
 & 
 & 
\cr
$1149.85$ &
 $g(12^-) $
 & 
 & 
 & 
 & $5+0 $
 & 
 & 
 & 
 & 
 & 
 & 
 & 
 & 
\cr
$1245.77$ &
 $\bar s(1\frac{1}{2}^+) $
 & 
 & $2+0 $
 & 
 & 
 & 
 & 
 & 
 & 
 & 
 & 
 & 
 & 
\cr
$1250.26$ &
 ${\bar\ell}(1\frac{5}{2}^-) $
 & 
 & $12+0 $
 & 
 & 
 & 
 & 
 & 
 & 
 & 
 & 
 & 
 & 
\cr
$1286.07$ &
 $g(11^-) $
 & 
 & 
 & 
 & $3+0 $
 & 
 & 
 & 
 & 
 & 
 & 
 & 
 & 
\cr
$1359.26$ &
 $\bar s(1\frac{5}{2}^-) $
 & 
 & $6+0 $
 & 
 & 
 & 
 & 
 & 
 & 
 & 
 & 
 & 
 & 
\cr
$1387.89$ &
 $g(13^+) $
 & 
 & 
 & 
 & $7+0 $
 & 
 & 
 & 
 & 
 & 
 & 
 & 
 & 
\cr
$1419.04$ &
 ${\bar\ell}(1\frac{3}{2}^-) $
 & 
 & $8+0 $
 & 
 & 
 & 
 & 
 & 
 & 
 & 
 & 
 & 
 & 
\cr
$1475.48$ &
 ${\bar\ell}(2\frac{1}{2}^-) $
 & 
 & $4+0 $
 & 
 & 
 & 
 & 
 & 
 & 
 & 
 & 
 & 
 & 
\cr
$1482.28$ &
 ${\bar\ell}(1\frac{7}{2}^+) $
 & 
 & $16+0 $
 & 
 & 
 & 
 & 
 & 
 & 
 & 
 & 
 & 
 & 
\cr
$1528.04$ &
 $\bar s(1\frac{3}{2}^-) $
 & 
 & $4+0 $
 & 
 & 
 & 
 & 
 & 
 & 
 & 
 & 
 & 
 & 
\cr
$1550.15$ &
 $g(12^+) $
 & 
 & 
 & 
 & $5+0 $
 & 
 & 
 & 
 & 
 & 
 & 
 & 
 & 
\cr
$1584.48$ &
 $\bar s(2\frac{1}{2}^-) $
 & 
 & $2+0 $
 & 
 & 
 & 
 & 
 & 
 & 
 & 
 & 
 & 
 & 
\cr
$1591.28$ &
 $\bar s(1\frac{7}{2}^+) $
 & 
 & $8+0 $
 & 
 & 
 & 
 & 
 & 
 & 
 & 
 & 
 & 
 & 
\cr
$1609.85$ &
 $g(14^-) $
 & 
 & 
 & 
 & $9+0 $
 & 
 & 
 & 
 & 
 & 
 & 
 & 
 & 
\cr
$1620.93$ &
 $g(21^+) $
 & 
 & 
 & 
 & $3+0 $
 & 
 & 
 & 
 & 
 & 
 & 
 & 
 & 
\cr
\hline
\end{tabular}\\
\end{table}

\begin{table*}[t]
\caption{As Table \ref{tab:4} for two-constituent configurations.}
\label{tab:5}
\begin{tabular}{clcccccccccccc}
\hline 
$\Delta (\MeV)$ & config. 
& {\bf 1} & {\bf 3} & {\bf 6} & {\bf 8} & {\bf 10} & {\bf 15$^\prime$} 
& {\bf  15} & {\bf 24} & {\bf 27} & {\bf 35} & {\bf 42} & {\bf 64}
\cr
\hline
$1197.62$ &
 $\ell(1\frac{1}{2}^+) $
 $\ell(1\frac{1}{2}^+) $
 & 
 & $10+0 $
 & 
 & 
 & 
 & 
 & 
 & 
 & 
 & 
 & 
 & 
\cr
$1197.62$ &
 $\ell(1\frac{1}{2}^+) $
 ${\bar\ell}(1\frac{1}{2}^-) $
 & $0+16 $
 & 
 & 
 & $16+0 $
 & 
 & 
 & 
 & 
 & 
 & 
 & 
 & 
\cr
$1197.62$ &
 ${\bar\ell}(1\frac{1}{2}^-) $
 ${\bar\ell}(1\frac{1}{2}^-) $
 & 
 & 
 & $6+0 $
 & 
 & 
 & 
 & 
 & 
 & 
 & 
 & 
 & 
\cr
$1306.62$ &
 $\ell(1\frac{1}{2}^+) $
 $s(1\frac{1}{2}^+) $
 & 
 & $8+0 $
 & 
 & 
 & 
 & 
 & 
 & 
 & 
 & 
 & 
 & 
\cr
$1306.62$ &
 $\ell(1\frac{1}{2}^+) $
 $\bar s(1\frac{1}{2}^-) $
 & $0+8 $
 & 
 & 
 & $8+0 $
 & 
 & 
 & 
 & 
 & 
 & 
 & 
 & 
\cr
$1306.62$ &
 $s(1\frac{1}{2}^+) $
 ${\bar\ell}(1\frac{1}{2}^-) $
 & $0+8 $
 & 
 & 
 & $8+0 $
 & 
 & 
 & 
 & 
 & 
 & 
 & 
 & 
\cr
$1306.62$ &
 ${\bar\ell}(1\frac{1}{2}^-) $
 $\bar s(1\frac{1}{2}^-) $
 & 
 & 
 & $8+0 $
 & 
 & 
 & 
 & 
 & 
 & 
 & 
 & 
 & 
\cr
$1348.67$ &
 $g(11^+) $
 $\ell(1\frac{1}{2}^+) $
 & 
 & 
 & $12+0 $
 & 
 & 
 & 
 & 
 & 
 & 
 & 
 & 
 & 
\cr
$1348.67$ &
 $g(11^+) $
 ${\bar\ell}(1\frac{1}{2}^-) $
 & 
 & $12+0 $
 & 
 & 
 & 
 & 
 & $12+0 $
 & 
 & 
 & 
 & 
 & 
\cr
$1415.62$ &
 $s(1\frac{1}{2}^+) $
 $s(1\frac{1}{2}^+) $
 & 
 & $3+0 $
 & 
 & 
 & 
 & 
 & 
 & 
 & 
 & 
 & 
 & 
\cr
$1415.62$ &
 $s(1\frac{1}{2}^+) $
 $\bar s(1\frac{1}{2}^-) $
 & $0+4 $
 & 
 & 
 & $4+0 $
 & 
 & 
 & 
 & 
 & 
 & 
 & 
 & 
\cr
$1415.62$ &
 $\bar s(1\frac{1}{2}^-) $
 $\bar s(1\frac{1}{2}^-) $
 & 
 & 
 & $1+0 $
 & 
 & 
 & 
 & 
 & 
 & 
 & 
 & 
 & 
\cr
$1444.73$ &
 $\ell(1\frac{1}{2}^+) $
 $\ell(1\frac{3}{2}^-) $
 & 
 & $32+0 $
 & 
 & 
 & 
 & 
 & 
 & 
 & 
 & 
 & 
 & 
\cr
$1444.73$ &
 $\ell(1\frac{1}{2}^+) $
 ${\bar\ell}(1\frac{3}{2}^+) $
 & $0+32 $
 & 
 & 
 & $32+0 $
 & 
 & 
 & 
 & 
 & 
 & 
 & 
 & 
\cr
$1444.73$ &
 $\ell(1\frac{3}{2}^-) $
 ${\bar\ell}(1\frac{1}{2}^-) $
 & $0+32 $
 & 
 & 
 & $32+0 $
 & 
 & 
 & 
 & 
 & 
 & 
 & 
 & 
\cr
$1444.73$ &
 ${\bar\ell}(1\frac{1}{2}^-) $
 ${\bar\ell}(1\frac{3}{2}^+) $
 & 
 & 
 & $32+0 $
 & 
 & 
 & 
 & 
 & 
 & 
 & 
 & 
 & 
\cr
$1457.67$ &
 $g(11^+) $
 $s(1\frac{1}{2}^+) $
 & 
 & 
 & $6+0 $
 & 
 & 
 & 
 & 
 & 
 & 
 & 
 & 
 & 
\cr
$1457.67$ &
 $g(11^+) $
 $\bar s(1\frac{1}{2}^-) $
 & 
 & $6+0 $
 & 
 & 
 & 
 & 
 & $6+0 $
 & 
 & 
 & 
 & 
 & 
\cr
$1494.26$ &
 $g(11^+) $
 $g(11^+) $
 & $0+6 $
 & 
 & 
 & $9+0 $
 & $3+0 $
 & 
 & 
 & 
 & $6+0 $
 & 
 & 
 & 
\cr
$1553.73$ &
 $s(1\frac{1}{2}^+) $
 $\ell(1\frac{3}{2}^-) $
 & 
 & $16+0 $
 & 
 & 
 & 
 & 
 & 
 & 
 & 
 & 
 & 
 & 
\cr
\hline
\end{tabular}\\
\end{table*}

\begin{table*}[t]
\caption{As Table \ref{tab:4} for three-constituent configurations.}
\label{tab:6}
\begin{tabular}{clcccccccccccc}
\hline 
$\Delta (\MeV)$ & config. 
& {\bf 1} & {\bf 3} & {\bf 6} & {\bf 8} & {\bf 10} & {\bf 15$^\prime$} 
& {\bf  15} & {\bf 24} & {\bf 27} & {\bf 35} & {\bf 42} & {\bf 64}
\cr
\hline
$1623.25$ &
 $\ell(1\frac{1}{2}^+) $
 $\ell(1\frac{1}{2}^+) $
 $\ell(1\frac{1}{2}^+) $
 & $0+20 $
 & 
 & 
 & $20+0 $
 & 
 & 
 & 
 & 
 & 
 & 
 & 
 & 
\cr
$1623.25$ &
 $\ell(1\frac{1}{2}^+) $
 $\ell(1\frac{1}{2}^+) $
 ${\bar\ell}(1\frac{1}{2}^-) $
 & 
 & 
 & $40+0 $
 & 
 & 
 & 
 & 
 & 
 & 
 & 
 & 
 & 
\cr
$1623.25$ &
 $\ell(1\frac{1}{2}^+) $
 ${\bar\ell}(1\frac{1}{2}^-) $
 ${\bar\ell}(1\frac{1}{2}^-) $
 & 
 & $0+64 $
 & 
 & 
 & 
 & 
 & $24+0 $
 & 
 & 
 & 
 & 
 & 
\cr
$1623.25$ &
 ${\bar\ell}(1\frac{1}{2}^-) $
 ${\bar\ell}(1\frac{1}{2}^-) $
 ${\bar\ell}(1\frac{1}{2}^-) $
 & $0+20 $
 & 
 & 
 & $20+0 $
 & $4+0 $
 & 
 & 
 & 
 & 
 & 
 & 
 & 
\cr
$1732.25$ &
 $\ell(1\frac{1}{2}^+) $
 $\ell(1\frac{1}{2}^+) $
 $s(1\frac{1}{2}^+) $
 & $0+20 $
 & 
 & 
 & $32+0 $
 & 
 & 
 & 
 & 
 & 
 & 
 & 
 & 
\cr
$1732.25$ &
 $\ell(1\frac{1}{2}^+) $
 $\ell(1\frac{1}{2}^+) $
 $\bar s(1\frac{1}{2}^-) $
 & 
 & 
 & $20+0 $
 & 
 & 
 & 
 & 
 & 
 & 
 & 
 & 
 & 
\cr
$1732.25$ &
 $\ell(1\frac{1}{2}^+) $
 $s(1\frac{1}{2}^+) $
 ${\bar\ell}(1\frac{1}{2}^-) $
 & 
 & 
 & $32+0 $
 & 
 & 
 & 
 & 
 & 
 & 
 & 
 & 
 & 
\cr
$1732.25$ &
 $\ell(1\frac{1}{2}^+) $
 ${\bar\ell}(1\frac{1}{2}^-) $
 $\bar s(1\frac{1}{2}^-) $
 & 
 & $0+64 $
 & 
 & 
 & 
 & 
 & $32+0 $
 & 
 & 
 & 
 & 
 & 
\cr
$1732.25$ &
 $s(1\frac{1}{2}^+) $
 ${\bar\ell}(1\frac{1}{2}^-) $
 ${\bar\ell}(1\frac{1}{2}^-) $
 & 
 & $0+32 $
 & 
 & 
 & 
 & 
 & $12+0 $
 & 
 & 
 & 
 & 
 & 
\cr
$1732.25$ &
 ${\bar\ell}(1\frac{1}{2}^-) $
 ${\bar\ell}(1\frac{1}{2}^-) $
 $\bar s(1\frac{1}{2}^-) $
 & $0+20 $
 & 
 & 
 & $32+0 $
 & $12+0 $
 & 
 & 
 & 
 & 
 & 
 & 
 & 
\cr
$1760.63$ &
 $g(11^+) $
 $\ell(1\frac{1}{2}^+) $
 $\ell(1\frac{1}{2}^+) $
 & 
 & $48+0 $
 & 
 & 
 & 
 & 
 & $48+0 $
 & $18+0 $
 & 
 & 
 & 
 & 
\cr
$1760.63$ &
 $g(11^+) $
 $\ell(1\frac{1}{2}^+) $
 ${\bar\ell}(1\frac{1}{2}^-) $
 & $0+48 $
 & 
 & 
 & $96+48 $
 & $48+0 $
 & 
 & 
 & 
 & $48+0 $
 & 
 & 
 & 
\cr
$1760.63$ &
 $g(11^+) $
 ${\bar\ell}(1\frac{1}{2}^-) $
 ${\bar\ell}(1\frac{1}{2}^-) $
 & 
 & 
 & $48+0 $
 & 
 & 
 & 
 & 
 & 
 & 
 & 
 & 
 & 
\cr
$1841.25$ &
 $\ell(1\frac{1}{2}^+) $
 $s(1\frac{1}{2}^+) $
 $s(1\frac{1}{2}^+) $
 & $0+12 $
 & 
 & 
 & $16+0 $
 & 
 & 
 & 
 & 
 & 
 & 
 & 
 & 
\cr
$1841.25$ &
 $\ell(1\frac{1}{2}^+) $
 $s(1\frac{1}{2}^+) $
 $\bar s(1\frac{1}{2}^-) $
 & 
 & 
 & $16+0 $
 & 
 & 
 & 
 & 
 & 
 & 
 & 
 & 
 & 
\cr
$1841.25$ &
 $\ell(1\frac{1}{2}^+) $
 $\bar s(1\frac{1}{2}^-) $
 $\bar s(1\frac{1}{2}^-) $
 & 
 & $0+16 $
 & 
 & 
 & 
 & 
 & $4+0 $
 & 
 & 
 & 
 & 
 & 
\cr
$1841.25$ &
 $s(1\frac{1}{2}^+) $
 $s(1\frac{1}{2}^+) $
 ${\bar\ell}(1\frac{1}{2}^-) $
 & 
 & 
 & $12+0 $
 & 
 & 
 & 
 & 
 & 
 & 
 & 
 & 
 & 
\cr
$1841.25$ &
 $s(1\frac{1}{2}^+) $
 ${\bar\ell}(1\frac{1}{2}^-) $
 $\bar s(1\frac{1}{2}^-) $
 & 
 & $0+32 $
 & 
 & 
 & 
 & 
 & $16+0 $
 & 
 & 
 & 
 & 
 & 
\cr
$1841.25$ &
 ${\bar\ell}(1\frac{1}{2}^-) $
 $\bar s(1\frac{1}{2}^-) $
 $\bar s(1\frac{1}{2}^-) $
 & $0+12 $
 & 
 & 
 & $16+0 $
 & $4+0 $
 & 
 & 
 & 
 & 
 & 
 & 
 & 
\cr
$1848.83$ &
 $\ell(1\frac{1}{2}^+) $
 $\ell(1\frac{1}{2}^+) $
 $\ell(1\frac{3}{2}^-) $
 & $0+80 $
 & 
 & 
 & $128+0 $
 & 
 & 
 & 
 & 
 & 
 & 
 & 
 & 
\cr
\hline
\end{tabular}\\
\end{table*}



\end{document}